\documentclass[12pt,preprint]{aastex}

\slugcomment{to appear in PASP}

\shorttitle{All-Sky spectrally matched Tycho2 magnitudes}
\shortauthors{Pickles \& Depagne}

\begin{document}


\title{All-Sky spectrally matched UBVRI-ZY and u'g'r'i'z' magnitudes for stars in the Tycho2 catalog}


\author{A. Pickles and \'E. Depagne}
\affil{Las Cumbres Observatory Global Telescope, Goleta, CA 93117, USA}

\email{apickles@lcogt.net}




\begin{abstract}

We present fitted UBVRI-ZY and $u'g'r'i'z'$ magnitudes, spectral types and
distances for 2.4\,M stars, derived from synthetic photometry of a
library spectrum that best matches the Tycho2 $B_TV_T$, NOMAD $R_N$
and 2MASS $JHK_{2/S}$ catalog magnitudes. We present similarly synthesized
multi-filter magnitudes, types and distances for 4.8\,M stars with
2MASS and SDSS photometry to $g<16$ within the Sloan survey region,
for Landolt and Sloan primary standards, and for Sloan Northern (PT)
and Southern secondary standards.

The synthetic magnitude zeropoints for $B_TV_T$, $UBVRI$, $Z_VY_V$, $JHK_{2/S}$,
$JHK_{MKO}$, Stromgren $uvby$, Sloan $u'g'r'i'z'$ and $ugriz$ are
calibrated on 20 {\em calspec} spectrophotometric standards.  The
$UBVRI$ and $ugriz$ zeropoints have dispersions of 1--3\%, for
standards covering a range of color from $-0.3 < V-I < 4.6$; those
for other filters are in the range 2--5\%.

The spectrally matched fits to Tycho2 stars provide estimated
$1\sigma$ errors per star of $\sim$0.2, 0.15, 0.12, 0.10 and 0.08 mags
respectively in either $UBVRI$ or $u'g'r'i'z'$; those for at least
70\% of the SDSS survey region to $g<16$ have estimated $1\sigma$
errors per star of $\sim$0.2, 0.06, 0.04, 0.04, 0.05 in $u'g'r'i'z'$
or $UBVRI$.

The density of Tycho2 stars, averaging about 60 stars per square
degree, provides sufficient stars to enable automatic flux
calibrations for most digital images with fields of view of 0.5 degree
or more.  Using several such standards per field, automatic 
flux calibration can be achieved to a few percent in any filter, at
any airmass, in most workable observing conditions, to facilitate
inter-comparison of data from different sites, telescopes and instruments.

\end{abstract}

\keywords{Stars, Data Analysis and Techniques}


\section{Introduction} \label{intro}

Reliable flux calibration is important for accurate photometry, and to
compare observations taken by different observers, at different times or
different sites, with different equipment and possibly different filter
bandpasses.

Ground based optical calibration is traditionally achieved by
observing with the same equipment both standard stars and at least
some stars in the field of interest, during periods known to be
photometric.  In principle this permits calibration of program stars
on a standard photometric system to better than 1\%, but in practice
filter and instrumental mismatches, atmospheric and other variations
during this process often limits effective calibration to 2\% or
worse.

Internal relative flux calibration of single or repeated data sets are
routinely achieved to 0.2\% or better, including during
non-photometric conditions, by reference to non-variable stars in the
observed field.  But significant questions arise about effective
cross calibration of observations from different epochs.  The
situation is particularly complicated for time-domain science, where
multiple sites, telescope apertures, filters and sets of
instrumentation become involved, or when time constraints or observing
conditions preclude traditional calibrations. Offsets between
otherwise very accurate data sets can be much larger than expected.
This can introduce significant uncertainty in multi-observation
analysis, or obscure real variations.

Cross comparisons can be facilitated by parallel wide-field
observations along the line of sight to each image, as in the CFHT
skyprobe facility
\citep{cui04}\footnote{{http://www.cfht.hawaii.edu/Instruments/Skyprobe/}},
or other ``context'' camera systems.  But having multi-filter
standards within each digital image offers many simplifying
advantages. The ideal scenario for calibration of optical imaging data
would be if there were all-sky stars of adequately known brightness on
standard photometric systems, and present in sufficient numbers to
provide a few in every digital image.

Precursors to such standards include i) the Sloan Digital Sky Survey
(SDSS) of 360 million objects, observed with a 2.5m telescope to about
22\,mag in {\em ugriz} filters and covering about $1/4$ of the sky
\citep{gun98}\footnote{{\tt http://www.sdss.org/dr7/}}, ii) the 2MASS
whole sky survey of about 300 million stars observed with two 1.3m
telescopes to about 15\,mag in $JHK_{2/S}$ filters
\citep{cut98}\footnote{{\tt
http://www.ipac.caltech.edu/2mass/overview/about2mass.html}}, and iii)
the USNO NOMAD catalog \citep{zac04}\footnote{{\tt
http://www.nofs.navy.mil/nomad/}}, which contains over 1 billion stars
covering the whole sky. It is often used for automatic astrometric
fits, and has calibrated photographic $R_N$-band photometry (on the
Landolt system) with a dispersion of about 0.25\,mag.

The Tycho2 catalog \citep{hog00} provides a consistent set of all-sky
optical standards, observed with the {\em Hipparcos}\footnote{{\tt
http://www.rssd.esa.int/index.php?project=HIPPARCOS}} satellite. It
provides $\sim$2.5\,million stars to about 13.5 \& 12.5\,mag in $B_T$
\& $V_T$ filters.  While this catalog contains less than 1\% of the
stars in the 2MASS catalog, and has large photometric errors at the
faint end, it forms the basis along with 2MASS and NOMAD for an
effective and consistent all-sky photometric catalog, that can be used
now by many optical telescopes to achieve reasonable automatic flux
calibration.  

\begin{figure}
\epsscale{.80}
\plotone{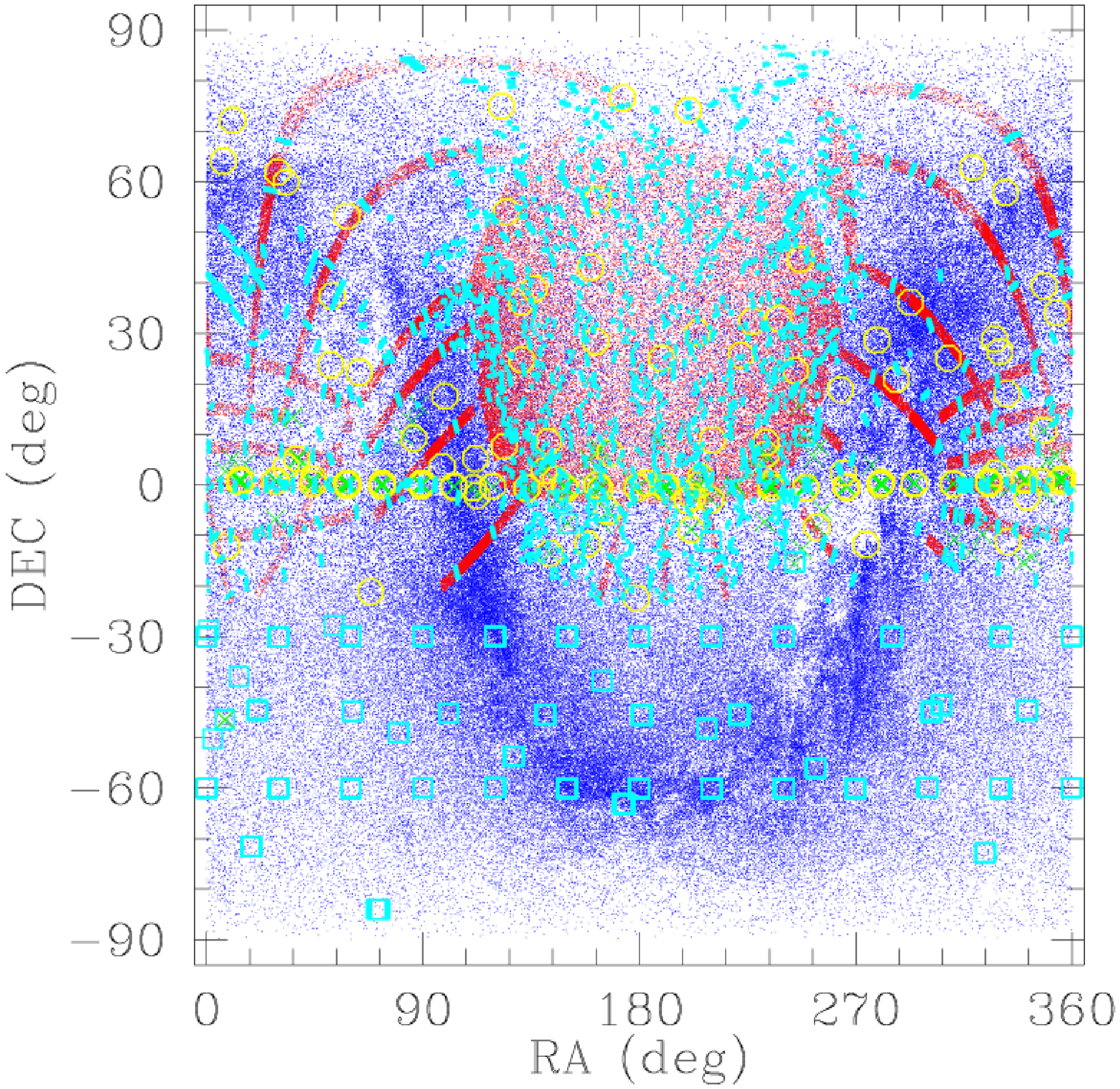}

\caption{The footprint in equatorial coordinates of the various
catalogs discussed here.  The Tycho2 catalog (blue dots) covers the
whole sky to $B_T\sim$~13.5\,mag.  The Sloan survey grey (red in electronic version) dots
covers less area but to a much fainter limiting magnitude, shown here
to g$<$16\,mag.  The Landolt primary standards are marked with green
crosses, the Sloan primary standards with yellow circles, the Southern
Sloan secondary standards are marked with cyan squares, where each
square represents many standards, and the Northern SDSS PT stars
are marked with cyan dots. }

\label{footprint}
\end{figure}

The average density of Tycho2 stars varies from
$\sim$150 stars per square degree at galactic latitude $b = 0\deg$,
through $\sim$50 stars per square degree at $|b| = 30\deg$ to $\sim$25
stars per square degree at $|b| = 90\deg$ \citep{hog00}, so it is
typically possible to find 5---15 Tycho2 stars within a 30-arcmin
field of view. These numbers obviously decrease for smaller fields of
view, and become uninteresting for fields of view much smaller than
15-arcmin.  The footprints in equatorial coordinates of several
catalogs discussed here are shown in figure \ref{footprint}.

\citet{ofe08} described how to produce synthetic {\em g'r'i'z'}
magnitudes for about 1.6\,M Tycho2 stars brighter than 12\,mag in
$V_T$ and 13\,mag in $B_T$. The present paper extends this to 2.4\,M
Tycho2 stars, with synthetic magnitudes on both the Landolt and Sloan
standard photometric systems, and to other filter systems calibrated
here such as Stromgren and UKIRT $ZY$ \& $JHK_{MKO}$\footnote{{\tt
http://www.jach.hawaii.edu/UKIRT/astronomy/calib/phot\_cal/}}.  The
methodology permits {\em post facto} extrapolation to other filter systems of
interest, and can be applied to future all-sky catalogs of greater
depth and accuracy.

In section~\ref{synthetic} we describe our calculations of
synthetic magnitudes and fluxes from flux calibrated digital spectra.

In section~\ref{filters} we describe the zeropoint calibration of our
synthetic photometry against the {\em de facto} standard: 20
spectrophotometric standards with photometric data and covering a
significant color range, taken from the Hubble Space Telescope (HST)
{\em calspec}\footnote{{\tt
http://www.stsci.edu/hst/observatory/cdbs/calspec.html}} project.

In section~\ref{library} we describe the spectral matching library and
calibrated magnitudes in different filter systems: $B_TV_T$, $UBVRI$,
$J_2H_2K_{2/S}$, UKIRT $ZY$ and $JHK_{MKO}$, Stromgren $uvby$ and
Sloan filter sets, both primed and unprimed system bandpasses.

In section \ref{matching} we describe the spectral matching process,
chi-square optimization and distance constraints.

In section~\ref{primary} we illustrate the spectral matching and flux
fitting methodology with results fitted with different combinations of
optical and infrared colors for Landolt and Sloan primary standards.

In section~\ref{SDSS-2} we further illustrate the strengths and
limitations of the spectral fitting method with $\sim$16000 Southern
Sloan standards \citep{smi05}, $\sim$1\,M SDSS PT secondary patch standards
\citep{tuc06,dav07} with 2MASS magnitudes, and $\sim$11000 of those 
which also have Tycho2 $B_TV_T$ magnitudes and NOMAD $R_N$ magnitudes.

In section~\ref{catalogs} we describe online catalogs with fitted
types, distances and magnitudes in UBVRI-ZY and u'g'r'i'z' for: i)
Landolt and Sloan primary standards, ii) Sloan secondary North and
South standards, iii) 2.4\,M Tycho2 stars and iv) 4.8\,M stars within
the Sloan survey region to $g<16$.

\section{Synthetic Photometry} \label{synthetic}

\citet{bess05} describes the preferred method for convolving an $F(\lambda)$
digital spectrum with filter/system bandpass sensitivity functions to obtain
synthetic magnitudes that take account of the photon-counting nature of modern detectors.

$ N_{photons} = \int (F(\lambda)/h\nu).S_b(\lambda).d\lambda = \int (\lambda.F(\lambda)/hc).S_b(\lambda).d\lambda $

which can be normalized by the filter/system bandpass to form the weighted mean photon flux density

$ <\lambda.F(\lambda)_b> = \int \lambda.F(\lambda).S_b(\lambda).d\lambda / \int (\lambda.S_b(\lambda).d\lambda) $

As \citet{bess05} notes, this weights the fluxes by the wavelength,
and shifts the effective wavelength of the bandpass to the red.  This
is the basic convolution methodology used in the HST {\em synphot}\,\footnote{{\tt
http://www.stsci.edu/resources/software\_hardware/stsdas/synphot}} package.
From the Synphot User's Guide we can form:

$mag_\lambda(b) = -2.5*log_{10} <\lambda.F(\lambda)_b> -\ 21.1 $

where the numerical factor for the nominal $F(\lambda)$ at V=0 brings
the resultant magnitudes close to standard values\footnote{$21.0999 =
-2.5 * log_{10}(3631e^{-12} erg.cm^{-2}.s^{-1}.A^{-1})$}.
Additionally, as discussed in the ``Principles of Calibration''
section of the Synphot User's Guide, we can form the effective
wavelength

$<efflam> = \int F(\lambda).S_b(\lambda).\lambda^2.d\lambda / \int F(\lambda).S_b(\lambda).\lambda.d\lambda $

and the source-independent pivot wavelength

$ \lambda_{pivot} = \sqrt{ \int S_b(\lambda).\lambda.d\lambda \over{ \int S_b(\lambda).d\lambda/\lambda} } $  

and form a magnitude system based on $F(\nu)$ as
\begin{equation}
 mag_\nu(b) = mag_\lambda(b) - 5*log_{10}\lambda_{pivot} + 18.692 - ZeroPoint_b
\label{eq-1}
\end{equation}
where the numeric constant, as derived in \citet{sir05},
has the advantage of bringing $mag_\nu$ close to the AB79 system of \citet{og83}. 
The bandpass zeropoints in equation \ref{eq-1} are to be determined.

\citet{pick98} adopted a different approach based on mean flux per
frequency in the bandpass $<F(\nu)>$.  The differences between this
and the {\em synphot} methods for calculating magnitudes are typically
of order 0.02 to 0.04~mag over most of the color range, but for very
red stars with non-smooth spectra can become as large as 0.1~mag in
$R$, which has an extended red tail.  Importantly, the {\em synphot}
approach produces less zeropoint dispersion for stars covering a wide
range of type and color, and is adopted here.

\section{Filters and zeropoints based on spectrophotometric standards} \label{filters}

All the filter-bands discussed here (except the medium-band Stromgren
filters) are illustrated in figure \ref{figfilt}.  Only the wavelength
coverage and particularly the {\em shape} of the system bandpass
matters, not the height, which is taken out in the filter
normalization.

Our approach is similar to that presented by \cite{hb06} but
deliberately attempts to calibrate a large number of filters with
standards covering a wide range of color and spectral type.  The small
zeropoint dispersions achieved here demonstrate the validity of this
approach over a wide variety of spectral types and colors. They
reinforce the advances that have been made in accurate flux
calibration, as the result of careful work by
\citet{boh96,boh97,boh01,boh10}.

\subsection{Calspec standards} \label{calspec}

There are 13 standards with STIS\_NIC\_003 calibrated
spectrophotometry covering 0.1 to 2.5\,$\mu$, which include the latest HST {\em calspec}
calibration enhancements, and the 2010 corrections to
STIS gain settings.  The spectra are mainly of white dwarfs, but
include four G dwarfs and VB8(=LHS\,429, a late M dwarf) so provide
significant color range: $-0.3 < V-I < 4.6$.

There are two K giants (KF08T3, KF06T1) with low reddening which were observed with
NICMOS to provide IRAC calibration \citep{rea05}.  There is little
optical standard photometry for these two K giants, so they are only
included in the 2MASS $JHK_{2/S}$ zeropoint averages, but optical colors
estimated from their types are shown to be consistent with the
derived optical zeropoints. 

There are three additional {\em calspec} white dwarfs with coverage to
$\sim1\mu$ (G93-48, GD\,50, Feige\,34), and two subdwarfs (G158-100, BD\,+26\,2606) observed by \citet{oke90} and
\citet{og83} for which fairly extensive photometric data are available
in the literature. The latter 5 spectra calibrated from the uv to
$\sim1\mu$, were extended to $2.5\mu$ for illustrative
purposes. Their synthetic infrared magnitudes are computed and shown
for comparison, but only their optical zeropoints are combined in the averages.

\subsection{Standard Catalog Magnitudes} \label{catmag}

The matching ``catalog'' photometric data for {\em calspec} standards comes from i) the
Tycho2 catalog for $B_TV_T$, ii) the 2MASS catalog for
$J_2H_2K_{2/S}$, iii) $UBVRI$ data from \citet{land09} for
GD\,71 (DA1), G93-48 (DA3) \& GD\,50 (DA2), and \citet{lanu07} for
G191-B2B (DA0), BD\,+17\,4708 (sdF8), BD\,+26\,2606 (sdF),
AGK\,+81\,266 (sdO), GRW\,+70\,5824 (DA3), LDS\,749B (DBQ4),
Feige\,110 (DOp), Feige\,34 (DA) \& G158-100 (sdG), iv) from
\citet{bess91} for optical and infrared photometry of VB8 (M7\,V), v)
from the UKIRT standards listed on the JAC/UKIRT website\footnote{{\tt
http://www.jach.hawaii.edu/UKIRT/astronomy/calib/phot\_cal/}} for
$JHK_{MKO}$ and their WFCAM $ZY_{MKO}$ data, and vi) from \citet{weg83},
\citet{lac81} and \citet{hau98} for Stromgren standards (white
dwarfs). All the above catalog data are on the ``vega'' system where
the magnitudes of Vega are nominally zero in all bands.

Sloan catalog (vii) data on the AB system are from \citet{smi02} for
primary standards, from the SDSS\_PT catalog \citep{tuc06,dav07} and
Southern SDSS standards \citep{smi05}, from the SDSS DR7
database\footnote{{\tt http://www.sdss.org/dr7/}} for LDS\,749B, VB8,
GD\,50, and in 2 cases (G191-B2B \& GD\,71) from \citet{hb06} for
$u'g'r'i'z'$ magnitudes.

Additional $UBVRI$ and $ugriz$ photometric data for GD\,153 (DA0) are from
\citet{hb06}.  Additional $BV$ data for P041C, P177D, P330E (G0\,V),
KF08T3 (K0.5\,III) and KF06T1 (K1.5\,III) are from the {\em calspec}
website.

Photometric data for the calspec standards are summarized in tables
\ref{tbl-stda} and \ref{tbl-stdb}.  

In electronic table {\em AllFlx} we have computed mean fluxes and
synthetic magnitudes by equation \ref{eq-1} in the system bandpasses
of all the filters for up to 20 standards which have both accurate
digital spectra available from {\em calspec}, and measured photometry
from the literature.  Table {\em AllFlx} lists for each spectrum and
filter all the digitally measured spectrophotometric data including:
average wavelengths $<\lambda>$, effective wavelengths $<efflam>$,
pivot wavelengths (which are the same for each star), mean fluxes
$<F(\lambda)>$, $<\lambda.F(\lambda)>$ and $<F(\nu)>$, and catalog
magnitudes from the literature where available, with quoted errors.

For each standard and each filter, magnitudes have been computed via
equation \ref{eq-1}, and the zeropoints calculated to match synthetic
to observed magnitudes.  The calculated values are
listed in electronic table {\em SynZero}.

\subsection{Zeropoint means and dispersions} \label{zpmean}

In table \ref{tbl-zp} we summarize for each filter the synthesized
pivot wavelengths, zeropoint means and dispersions, adopted
zeropoints, measured magnitudes and fluxes of the STIS\_005
calspec spectrum of Vega, and our derived values of $<F(\nu)>$ for
0-magnitude in all filters.  

Figure \ref{vega} shows the filter bands over-plotted on the the
STIS\_005 {\em calspec} spectrum of Vega, plotted as $F(\nu)$
vs. wavelength.  The spectrum illustrates the nominal 0-mag definition
for filters on the {\em vega} system, and the horizontal line at 3631\,Jy
illustrates the AB=0 mag reference. The points show our synthesized
Vega magnitudes from column 7 of table \ref{tbl-zp}, and the error
bars show our zeropoint $\pm 2\sigma$ dispersion errors
from column 5 of table \ref{tbl-zp}. The electronic versions of these
figures are in color.

Some uncertain zeropoint values in electronic table {\em SynZero} are
marked with a colon ``xxx:'' to indicate they are derived from catalog
photometry (in tables \ref{tbl-stda}, \ref{tbl-stdb} \& electronic
table {\em AllFlx}) which is uncertain by 0.1 mag or more.  Zeropoint
values enclosed in parentheses indicate values which are {\em not}
used to form the final zeropoints.  Either they are photometric
estimates, or literature values which are out of range, as discussed
in the text.  The number of standards included in each filter
zeropoint calculation ranges from 5 for $V_T$ to 17 for $B$ and $V$.

The 2MASS and SDSS coordinates of VB8 differ by 7.4-arcsec, corresponding
to the large proper motion (-0.77, -0.87 arcsec/year) for this star
between the epochs of the two surveys. We obtain good VB8 zeropoint fits
for $BVRI$, $J_2H_2K_{2/S}$ and $HK_{MKO}$ and $grz$, but not
for $J_{MKO}$, $u$ or $i$ bands (see also section \ref{sloan}).
There is no U-band photometry for VB8.

\subsection{Choice of Landolt synthetic filter bandpasses} \label{choice}

For $UBVRI$ we tested several possible synthetic system bandpass
profiles from the literature.  We have made an empirical choice of the best system
bandpass(es) that minimize zeropoint scatter in the fitted mean
zeropoints for each filter, over the full color range.

In table \ref{tbl-zp}, we list zeropoints for $UBVRI$ using
both Landolt (system) filter response functions convolved with a
typical atmosphere and detector response \citep{coh03a}, and synthetic
system bandpasses for $U_MB_MV_M$
from \citet{ma06} and $V_CR_CI_C$ from \citet{bess79}.

It may seem that the Landolt system response curves would provide the
optimum synthetic matches to Landolt photometry, but this is not
necessarily the case, and not shown by our results in table
\ref{tbl-zp} and illustrated in figure \ref{UIzero}.

Both Landolt and Kron-Cousins measurements seek to emulate the
original Johnson system for UBV. Both apply calibrating steps in terms
of color to their instrumental magnitudes, to bring them into
correspondence with standard system values extending back several
decades. These steps are summarized in \citet{land07} for example, and
in \citet{land83,land92a,land92b,land09} to illustrate how
equipment changes over time have required slightly different color
corrections to maintain integrity with the original system definition.
These calibration steps are further reviewed in \citet{sb00}.

Real photometry is done with bandpasses that can vary with evolving
instrumentation.  We could measure synthetic magnitudes in the Landolt
bandpasses and apply color corrections to achieve ``standard'' values.
But synthetic photometry of flux calibrated spectra has the advantage
of being able to select bandpass profiles that minimize dispersion and
color effects.  We have not attempted to optimize any bandpasses here,
but have selected amoung those available from the literature. For UBVRI,
the results in table \ref{tbl-zp} indicate that these are best
provided by the $U_MB_MV_CR_CI_C$ bandpasses.

The zeropoint dispersions are 0.027, 0.020, 0.008, 0.014 and 0.016 mag
for $U_MB_MV_CR_CI_C$ respectively in table \ref{tbl-zp}, for the full
color range from White Dwarfs to VB8 ($-0.3 < V-I < 4.6$).  The
effective wavelengths vary from 355 to 371\,nm, 432 to 472\,nm, 542 to
558\,nm, 640 to 738\,nm and 785 to 805\,nm for $U_MB_MV_CR_CI_C$
respectively, between White Dwarfs and VB8.  Figure \ref{UIzero}
illustrates that the selected bands show much less zeropoint
dispersion than do $U_LB_LV_LR_LI_L$, with negligible trend with
color.  In figure \ref{UIzero} (and for other zeropoint figures) the
ordinate is inverted so that zeropoints that appear vertically higher
result in larger (fainter) magnitudes.

This is {\em not} a criticism of Landolt system response curves, which
enable accurate photometry with appropriate calibration and color
corrections, but indicates that the selected $U_MB_MV_CR_CI_C$ system
profiles are best for deriving synthetic spectrophotometry of flux
calibrated spectra to properly match Landolt photometry.

The zeropoint dispersions for $U_3B_3$ from \citet{as69,bus78} are
worse, at 0.050 and 0.024 mag respectively.  The dispersion for $V_M$ is
marginally worse than for $V_C$, where $V_C$ has a slightly
more elongated red tail than $V_M$.

The zeropoint dispersions for $UBVRI$ and $ugriz$ (primed and
unprimed) in table \ref{tbl-zp}, typically closer to 0.02 mag than
0.01\,mag, indicate both the accuracy and the limitations of comparing
synthetic photometry of well calibrated spectra with good standard
photometry.  Tighter fits can be obtained by restricting the selection
of comparison stars, for instance to only WD standards, but such
zeropoints can then be a function of color and lead to synthesized
magnitude errors for other stellar types much larger than the nominal
dispersion for a restricted set of standards.

We are gratified that these comparisons match so well over a large
range of color and standard types, confirming the increasing
correspondence (currently at the 1--3\% level) between spectrophotometric
and photometric standards. This sets the basis for enabling synthetic
magnitudes of an extended spectral library to be calibrated on
standard photometric systems.

\subsection{Other synthetic filter zeropoints on the Vega system} \label{zp}

The system transmission functions for $B_TV_T$ have been taken from
\citet{ma06}. The zeropoint dispersions for $B_TV_T$ are 0.045 and
0.020 respectively, which is acceptable given the typical errors in
the photometric values for fainter stars, and several stars included here.  There are a total of 7
values covering a color range from white dwarfs to G dwarfs for $B_T$,
but only 5 with accurate $V_T$ information, with HD209458 (G0\,V --
out of planet occultation) being the reddest comparison standard for $B_TV_T$.

The filter transmission functions for $ZY$ have been taken from
the {\em Vista} website\footnote{{\tt
http://www.vista.ac.uk/Files/filters}}. In what follows we refer to (upper case)
$Z_VY_V$ for these filter bandpasses, where we are using the subscript
``V'' to refer to both the VISTA/UKIDSS consortium and the fact that
these are {\em vega} based magnitudes.  The UKIRT WFCAM detector QE is not
included in the system bandpass but, unlike a CCD, is roughly flat over these
wavelengths\footnote{{\tt
http://www.ukidss.org/technical/technical.html}}.  The dispersions for
$Z_VY_V$ zeropoints, compared to only 5 and 4 UKIRT standards measured
with the WFCAM filters, are 0.038 and 0.031 mag respectively. The
$Y_V$ zeropoint for G158-100 is suspect as its {\em calspec} spectrum is not well
defined at 1$\mu$.  $Z_VY_V$ zeropoint determinations may improve as
more photometric standards in common with spectrophotometric standards
are measured.  The zeropoint results and somewhat restricted color ranges for
$B_TV_TZ_VY_V$ are illustrated in figure \ref{BVZYzero}.

Having a wider color range here for comparison would clearly be
advantageous, but there are several mitigating factors.  The $B_T$ and
$Z_VY_V$ bandpasses are more ``rectangular'' in shape than are the $B$
and $z$ bandpasses for example. They have effective wavelengths that
vary less with color, and are therefore less susceptible to color
effects suffered by filters with extended ``tails''.  The range of
effective wavelengths from white dwarfs to VB8 are 416.3 to 440.4\,nm,
524.3 to 543.1\,nm, 875.1 to 884.9\,nm, and 1018.8 to 1023.2\,nm for
$B_TV_TZ_VY_V$ respectively.  Also the $B_T-V_T$ colors derived from
our synthesized {\em LibMags} flux library (see section \ref{library})
follow the standard Tycho2 conversion formulas\footnote{{\tt
http://heasarc.nasa.gov/W3Browse/all/tycho2.html}}, with standard
deviations less than 0.02 \& 0.03\,mag respectively over a large color
range.

$V = V_T -0.09*(B_T-V_T)$

$B-V = 0.85*(B_T-V_T)$ 

Similarly, as shown in section \ref{zy}, the synthesized $Z_V$ and
$z'$ data fit well over a wide color range, indicating an extended range
of validity beyond that illustrated in figure \ref{BVZYzero}.

The $Y_V$ data are included because several survey cameras
(PanStarrs\footnote{{\tt http://pan-starrs.ifa.hawaii.edu/public/}},
SkyMapper\footnote{{\tt http://rsaa.anu.edu.au/skymapper/}}, Dark
Energy Survey (DES)\footnote{{\tt https://www.darkenergysurvey.org/}},
UKIDSS/VISTA\footnote{{\tt http://www.vista.ac.uk}} and our
LCOGT\footnote{{\tt http://lcogt.net}} monitoring cameras are using or
plan to use a Y filter at $\sim$1 micron. This is further discussed in
section \ref{zy}.

The transmission functions of the 2MASS filters, including detector
and typical atmosphere, have been taken from the {\em IPAC}
website\footnote{{\tt
http://www.ipac.caltech.edu/2mass/overview/about2mass.html}}.  The
zeropoint dispersions for $J_2H_2K_{2/S}$, indicated graphically in
figure \ref{JKzero}, are about 0.02 mag, quite comparable to the
typical (low) 2MASS errors for these stars, and the zeropoints show no trend with color.

The transmission functions for the JHK filters on the Mauna Kea
Observatory (MKO) system have been taken from \citet{tok02}.  Unlike
the 2MASS system bandpasses, these do not include detector QE or
atmosphere. The zeropoint dispersions for $JHK_{MKO}$ are 0.02, 0.02
and 0.04 respectively, for relatively few standards, but they do show
trends with color. We indicate the $J_{MKO}$ zeropoint for VB8 in
figure \ref{JKzero} but, due to uncertainty with its catalog
magnitude, have not included it in the $J_{MKO}$ zeropoint average.

The transmission functions for the Stromgren filters (not illustrated)
are from \citet{ma06}.  The zeropoints derived in table \ref{tbl-zp}
are $-0.29 \pm 0.04$, $-0.32 \pm 0.01$, $-0.18 \pm 0.02$ and
$-0.04 \pm 0.03$ for Stromgren $uvby$ respectively.  These are
averaged over 7 {\em calspec} white dwarfs for which Stromgren
photometry was found in the literature. The validity of these
zeropoints for redder stars is not demonstrated, but the
effective wavelength variations with color are
small for medium band filters with ``rectangular'' profiles, so
zeropoint variations with color should not be large for the Stromgren
filter bandpasses.

\subsubsection{Mould R-band}

The $R_p$ filter system profile shown in figure \ref{figfilt} is a
``Mould'' R-band interference filter with rectangular profile, of the
type used at many observatories to measure R on the Landolt system.
In this case it is the CFH12K 7603 filter originally from
CFHT\footnote{{\tt http://www.cfht.hawaii.edu/Instruments/Filters/}}
and now used by PTF\footnote{{\tt http://www.astro.caltech.edu/ptf/}}.
The $R_p$ system profile includes the CCD response and atmospheric
transmission appropriate to 1.7\,km and Airmass\,1.3, although both of
these are roughly flat in this region. The zeropoint (table
\ref{tbl-zp}) for this filter profile excludes VB8, since the lack of
a red-tail results in magnitudes which are too faint for stars redder
than $R-I > 0.6$.

Figure \ref{Rpplot} shows (upper panel) the comparison of $R_p-R_C$
with Landolt $R-I$ color, which results in a tight 2-segment curved
fit.  The lower two panels show $R_p-R_C$ and $R_p-r'$ against the
Sloan $r'-i'$ color, both of which show reasonable 2-segment linear
fits, confirming that rectangular-shaped ``Mould'' R observations can be
reliably converted to standard R magnitudes.

$R_p \sim R_C - 0.022 - 0.092*(r'-i') ~~~~   (r'-i') < 0.56 $

$R_p \sim R_C - 0.202 - 0.227*(r'-i') ~~~~   (r'-i') > 0.56 $

$R_p \sim r' - 0.198 - 0.318*(r'-i') ~~~~   (r'-i') < 0.36 $

$R_p \sim r' - 0.227 - 0.220*(r'-i') ~~~~   (r'-i') > 0.36 $

\subsubsection{NOMAD R-band} \label{nomad}

The USNO NOMAD catalog \citep{zac04} contains BVR and JHK data for
many stars.  The NOMAD $BV$ data for brighter stars are typically
derived from Tycho2 $B_TV_T$ using standard conversion formulas
referenced in section \ref{zp}.  It is possible to compare these
converted values with synthesized BV magnitudes, but it is better and
more accurate to compare as we have done the Tycho2 $B_TV_T$ values
directly with their synthesized values in the $B_TV_T$ system bandpasses. The
NOMAD JHK magnitudes are from 2MASS, so do not provide additional
information.

The NOMAD R-band data are derived from the USNO-B catalog
which was photometrically calibrated against Tycho2 stars at the
bright end, and two fainter catalogs as described in \citet{mon03},
with a typical standard deviation of 0.25 mag.

Figure \ref{Rnomad} illustrates our comparison of NOMAD R-band data
against Landolt standards, and 16 {\em calspec} standards.  The data
are plotted as the differences $R_{Nomad}-R_{Landolt}$
vs. $R_{Landolt}$.  A $3\sigma$ clip has been applied which excludes
about 10 stars plotted as black crosses, but retains more than 98\%
plotted as grey (red in electronic color version) crosses.  The dashed
histogram of all the stars shows the distribution of these differences
plotted with increasing number to the right against the magnitude
delta on the ordinate. This shows a good one-to-one fit, with zero
mean offset and a $1\sigma$ dispersion of 0.25\,mag for sigma clipped
Landolt stars. The equivalent dispersion is 0.17\,mag for the {\em
calspec} stars excluding VB8.  The location of the red M dwarf VB8 is
indicated: the photographic R band, like the ``Mould'' type R-band
above lacks the red-tail of the Kron/Cousins/Landolt R filter, so
measures magnitudes which are too faint for extremely red stars where
the flux is rising rapidly through the bandpass.

It would be preferable to compare digitized photographic R-magnitudes
with synthesized values in the appropriate system bandpass, but this is
clearly impossible, and probably would not improve this
calibration. The NOMAD R-band calibration is linear over a significant
magnitude range and the dispersion is roughly constant with magnitude,
probably indicating the difficulties of measuring digitized plates
rather than any linearity or systemic calibration issues.

In section \ref{matching} we include the NOMAD $R_N$ band data
together with Tycho2 and 2MASS magnitudes to provide 6-band spectral
matching of the Tycho2 stars, where the catalog $R_N$ is matched to
our synthesized $R_C$ magnitude, because inclusion does improve the
fits slightly over just Tycho2/2MASS fits.  We adopt a photometric
error of 0.25\,mag for all NOMAD $R_N$ magnitudes.

\subsection{Sloan filters on the AB system} \label{sloan}

The system transmission functions of the unprimed $ugriz$ filters used in the
imaging camera, including typical atmosphere and detector response, have been
taken from the {\em skyservice} website\footnote{{\tt
http://skyservice.pha.jhu.edu}} at Johns Hopkins University (JHU).
Those for the primed $u'g'r'i'z'$ filters used for standard
observations have been taken from the United States Naval Observatory
(USNO) website\footnote{{\tt
http://www-star.fnal.gov/ugriz/Filters/response.html}}, 
for 1.3 airmasses.

In order to compute zeropoints for the imaging survey (unprimed) Sloan
bandpasses $ugriz$, we computed conversion relations between the
primed and unprimed Sloan system values by comparing synthetic
magnitudes of digital library spectra from \citet{pick98}, as listed
in electronic table {\em LibMags} (section \ref{library}). This approach avoids many
complexities detailed in \citet{tuc06,dav07}, but produces tight
correspondence between synthetically derived Sloan primed and unprimed
magnitudes, over a wide color range.

With the exception of two zeropoint adjustments for $g$ and $r$,
these relations are identical to those listed on the SDSS
site\footnote{{\tt
http://www.sdss.org/dr5/algorithms/jeg\_photometric\_eq\_dr1.html}},
and are summarized here.

$u = ~~~~~~~~~~~ u' ~~~~~~~~~~~~~~~~~~~~~~~~~~~~~~~~~~~ (\sigma = 0.011)$

$g = 0.037 + g' + 0.060 * (g'-r'-0.53) ~~~~ (\sigma = 0.012)$

$r = 0.010 + r' + 0.035 * (r'-i'-0.21) ~~~~ (\sigma = 0.007)$

$i = ~~~~~~~~~~~ i' + 0.041 * (r'-i'-0.21) ~~~~ (\sigma = 0.015)$

$z = ~~~~~~~~~~ z' ~~~~~~~~~~~~~~~~~~~~~~~~~~~~~~~~~~~~ (\sigma = 0.003)  $

These relations were used to convert SDSS standard values on the
primed system to unprimed values (and {\em vice versa} when only DR7
data was available, eg. for LDS~749B, P177D, P330E \& GD\,50) then
compared with synthetic magnitudes computed in the unprimed bandpasses
to derive their zeropoints and dispersions.

The SDSS DR7 u-band data for VB8 appears too bright, likely because of
the known red leak.  The DR7 i-band magnitude is close to the r-band
magnitude, and appears too faint for such a red star.  As mentioned in
section \ref{catmag} we have omitted the $u$ and $i$ band data for VB8
from our zeropoint averages.

The zeropoint dispersions for $u'g'r'i'z'$ and $ugriz$ listed in table
\ref{tbl-zp} are about 0.02 mag (slightly worse for the g-band) for up
to 14 standards covering a wide color range.

There is little evidence for non-zero zeropoints in {\em griz}; they
are zero to within our measured dispersions, and we have chosen to set
them zero.  We find a zeropoint for both $u'$ and $u$ of $-0.03 \pm
0.02$ mag.  The zeropoint results for $u'g'r'i'z'$ and $ugriz$ are
illustrated in figure \ref{uzzero}, and show essentially no trend with
color.

We find synthetic spectrophotometry with the published Sloan system
transmission functions ($u'g'r'i'z'$ \& $ugriz$) matches Sloan
standard values well, with the quoted zeropoints and no color
terms. We note however that there are commercial ``Sloan'' filters
available that have excellent throughput, but which do not match the
bandpasses of the Sloan filters precisely, for instance with an
$i'/z'$ break close to 800\,nm.  We checked synthetic
spectrophotometry of our library spectra with a measured set of such
filters (and standard atmosphere \& CCD response) against standard
Sloan values. We obtain good fits to standard values for these
commercial filters but with the addition of color terms, particularly
in $i'$.

\subsection{V-band magnitude of Vega} \label{Vega-V}

We note that our $V_C$ zeropoint derivation based on 17 standards
(including VB8) results in a Landolt V magnitude for Vega of $0.021
\pm 0.008$, with Vega colors of $U-B= -0.007 \pm 0.031$, $B-V= -0.009
\pm 0.022$, $V-R= -0.002 \pm 0.016$ and $V-I= -0.010 \pm 0.018$. These
values, the same as in \citet{pick10}, are based on 13--17 standards
and fall within quoted errors of those derived in \citet{ma07}.

We further note that using the \citet{coh03a} filter definition
results in a zeropoint ($ZP_{VL}=-0.014 \pm 0.021$) which leads to
$V_L(Vega)=0.013 \pm 0.021$ using all 17 standards, or a zeropoint
based solely on the first 3 DA white dwarfs ($ZP_{VL}=-0.024 \pm
0.002$), which leads to $V_L(Vega)=0.023 \pm 0.002$. The latter value
is closer to the recently adopted {\em calspec} value of Landolt
$V(Vega) = 0.025$ mag. However the $V_L$ zeropoint is clearly a
function of color, as seen in figure \ref{UIzero}.  In figure
\ref{UIzero} (and for other zeropoint figures) the ordinate is
inverted so that zeropoints that appear vertically higher result in
larger (fainter) magnitudes, so $V_L$ magnitudes are indicated fainter
for bluer $V-I$ color.  The effect is fairly small, but for the
appropriate color for Vega ($V-I=-0.01$ vs. $V-I\sim-0.3$ for WDs),
the color-corrected zeropoint ($ZP_{VL}=-0.022$), leads to
$V_L(Vega)=0.021$, ie. the {\em same} as our $V_C(Vega)=0.021$
derivation when color effects are taken into account.

We argue that our $V_C(Vega)$ magnitude of $0.021 \pm 0.008$
represents a realistic mean value and error for the Landolt V
magnitude of the STIS\_005 spectrum of Vega. The zeropoint dispersion could
probably be improved by standard photometric measurements of more
red stars, including the IRAC K giants, which are shown in the diagram but
not used here, as they lack optical standard photometry.

The further question of the small differences between Landolt,
Kron-Cousins (SAAO) and Johnson V magnitudes are discussed in
\citet{land83,land92a} and \citet{men91}, and summarized again in
\citet{sb00}, but is sidestepped here. For $UBVRI$ we are comparing
synthetic spectrophotometry of {\em calspec} standards to available
published photometry on the Landolt system.

\subsection{Vega and Zero magnitude fluxes}

Columns 7 and 8 of table \ref{tbl-zp} show for each bandpass: our
synthesized magnitudes of Vega with the adopted zeropoints and the
$<F(\nu)>$ fluxes measured on the STIS\_005 spectrum of Vega.  Column
9 shows and our inferred fluxes ($F(\nu)~in~Jy$) for a zero-magnitude
star (ie. zero {\em vega} mag for $B_TV_T, UBVRI, Z_VY_V, JHK$,
Stromgren $uvby$; and zero AB mag for Sloan $u'g'r'i'z'$ and $ugriz$
filters).

The zero magnitude fluxes can be compared with other values given for
instance by \citet{bess79,bess88}, \citet{coh03b}, \citet{hew06}, and
the IPAC\footnote{{\tt
http://www.ipac.caltech.edu/2mass/releases/allsky/doc/sec6\_4a.html}},
SPITZER\footnote{{\tt
http://casa.colorado.edu/~ginsbura/filtersets.htm}} 
and Gemini\footnote{{\tt
http://www.gemini.edu/?q=node/11119}} websites. The
largest discrepancy is for U, where our 0-mag U-band flux is about 4\%
less than the IPAC value for example, compared to our measured sigma of 2.7\%.

We obtain zero magnitude fluxes for $griz$ and $g'r'i'z'$ close to 3631\,Jy,
and close to 3680\,Jy for $u$ and $u'$.

With our adopted Stromgren zeropoints, we derive Vega
magnitudes of 1.431, 0.189, 0.029 and 0.046 in Stromgren $uvby$
respectively, compared to values of 1.432, 0.179, 0.018 and 0.014
derived in \citet{ma07}.

\subsection{Absolute and Bolometric magnitudes}

Electronic table {\em LibMags} (section \ref{library}) lists in the last column a nominal
absolute magnitude $M_V$ for each spectrum, mainly taken from
\citet{pick98}. The absolute magnitudes of the white dwarf spectra, in
the range $M_V$=12.2 to 12.6, were estimated from the tables of cooling
curves in \citet{chab00}, using their $V-I$ colors and hence
temperature. Absolute magnitudes are used to calculate distances from
spectral fits, as described in section \ref{distances}.  Absolute
magnitudes in other bands can be formed as (eg) $M_K$=$M_V-(V-K)$.

Also in table {\em LibMags} and table \ref{tbl-zp} we have included a digital
``bolometric'' magnitude Emag, with unit throughput over the full
spectral range (0.115\,$\mu$ to 2.5\,$\mu$).  The ``E'' zeropoint,
on the {\em vega} system, was adjusted to give approximately correct
bolometric corrections for solar type stars.  Bolometric corrections
in the V and K bands can be computed as $BC_V=E-V$ and $BC_K=E-K$, and
bolometric magnitudes can be formed as $M_{bol}=M_V-(V-E)=M_V+BC_V$ or
$M_{bol}=M_K-(K-E)=M_K+BC_K$. Thus it is
possible to derive HR diagrams for the fitted catalog stars in different
magnitude vs. color combinations.

\section{Spectral Matching Library and calibrated Magnitudes} \label{library}

Electronic table {\em LibMags} shows the magnitudes computed by the
{\em synphot} $<\lambda.F(\lambda)>$ method for all the filters
described, using the zeropoints listed in table \ref{tbl-zp}, for 141
digital spectra: Vega, 131 library spectra from \citet{pick98}, eight
additional {\em calspec} spectra, and one
{\em DA1/K4V} double star spectrum, discussed in section \ref{landolt}

The additional spectra from {\em calspec} added to extend coverage
are: G191-B2B (DA0), GD\,153 (DA0), GD\,71 (DA1), GRW\,+70\,5824
(DA3), LDS\,749B (DBQ4), Feige\,110 (D0p), AGK\,+81\,266 (sdO) and VB8
(M7\,V).

In table {\em LibMags} the magnitudes listed for Vega are those
derived from the STIS\_005 spectrum, but scaled to zero mag in V.  All the
flux calibrated library spectra, which are available from the quoted sources,
have been {\em multiplied} by the scaling factor indicated to produce
synthetic $V_C$ magnitudes of zero, ie. the \citet{pick98} spectra
have been scaled down and the {\em calspec} spectra have been scaled
up to achieve this normalization to $V_C=0$.

In later sections, library spectra are referred to by number and
name, together with the V-mag scaling necessary for the matched
spectrum to fit the program star photometry. Magnitudes in other
bandpasses, including $JHK_{MKO}$ or Stromgren $uvby$ can be derived
by adding the appropriate filter magnitude from table {\em
LibMags} for the fitted library spectrum, to the fitted V-mag for the
program star.

\section{Spectral Matching} \label{matching}

\subsection{Spectral scaling and Chi-square optimization}

For each star with catalog magnitudes (CM) to be fitted, and for each
library spectrum (0...140) with measured synthetic magnitudes (SM), we
form the weighted magnitude difference between the catalog and program
magnitudes:

$wmag = \Sigma (w_i * (CM_i-SM_i)) / \Sigma (w_i)$

$w_i = 1/(eCM_i * fac_i)$

where the weights are the inverse of the catalog errors ($eCM$) in
magnitudes, scaled by factors determined empirically to optimize the
fits.  After experimentation with data sets where cross checks are
available, we chose factors of \{6,4,3,4,4\} for $UBVRI$,
\{6,4,2,3,3\} for $u'g'r'i'z'$, and \{1,1,1.5,1,1.3,1.3\} for
$B_TV_TR_NJ_2H_2K_{2/S}$. All the filter bands display broad, shallow
minima of $synthetic-catalog$ error with factor, and there is some
cross-talk between bands. The results are {\em not} critically
dependent on these factors, they are much more dependent on the
catalog photometric errors themselves.  But the factors were selected
within these broad minima to minimize the RMS errors in each band and
to improve the consistency of spectral type fitting between different
fits.

Smaller factors imply higher weights, although the weighting also
depends on the catalog magnitude errors, so the weighting for NOMAD
$R_N$ (dispersion 0.25\,mag, factor 1.5) is usually least.  Larger
factors are necessary for data with small intrinsic errors, because
the library spectra cannot achieve milli-mag fits to the data.  The
factors are not completely independent, but are inter-connected by the
library spectra themselves.  Fitted magnitudes in U/u' or B/g' for
instance can be improved slightly by reducing the scale factor for
these bands, but at the expense of slightly worse fits in redder
bands. This tends to indicate spectral library calibration errors in the U/u' region.

For each filter band to be fitted, we then form:

$dM_i = (CM_i - SM_i - wmag) $

and form a chi-square normalized by the number of filters:

$chi2 = \Sigma (w_i * dM_i)^2 / nbands $

The resultant $chi2$ will be dominated by those bands where $dM_i$
exceeds $eCM_i$.  

Additionally we can form RMS magnitude error values for the 6~Tycho2/NOMAD~$R_N$/2MASS bands,
the 5~Landolt, 5~Sloan and three~2MASS bands when they are available, as:

$RTM = \sqrt{ \Sigma(dM_{Tycho2/R_N/2MASS}^2) / 6 } $

$RUI = \sqrt{ \Sigma(dM_{Landolt}^2) / 5 } $

$Ruz = \sqrt{ \Sigma(dM_{Sloan}^2) / 5 } $

$R2M = \sqrt{ \Sigma(dM_{2MASS}^2) / 3 } $.

RMS values without the influence of U/u' can also be formed, eg. $RBI$ and $Rgz$.

The spectral matching process also computes a distance modulus using
the adopted absolute V magnitude from table {\em LibMags} and the fitted (apparent) V
magnitude, and hence a distance estimate in parsec for each fit.

$D(pc) = 10^{(1 - 0.2*(M_V - SM_V))} $

For each program star we get an ordered list of 141 values of $chi2$,
one for each spectrum, where the optimum spectrum has the lowest
$chi2$.  Good fits normally have $chi2 \le~1$ because of the
normalization by photometric errors and number of filters, but higher
values are possible when there are significant differences between
catalog and synthetic magnitudes in one or more bands

Several fits at the top of the ordered list may have close,
and good fits.  These typically represent uncertainty between close
spectral types, metal-rich spectra of slightly earlier type,
metal-weak spectra of later type, or confusion between dwarf,
giant or even supergiant spectra of similar colors. 

Additional constraints used to discriminate between these cases are
described below.  The predicted magnitudes of fits with similar 
$chi2$ values are themselves similar to within 1---3\%.  The predicted
distances however can vary widely, particularly when there are close
dwarf and giant matches.

\subsection{Distance Constraints} \label{distances}

We can use the (J-H, H-K) color-color diagram as
in fig. 5 of \citet{bess88} to discriminate between late-type dwarfs
and giants. We have adopted the simple criterion that an M-star fitted
type must be a dwarf if the 2MASS colors $(H_2-K_2)~>~0.23$ and $(J_2-H_2)~<~0.75$.

Luminosity class discrimination can also be based on any
available distance information, including Proper Motions listed in the
Tycho2 catalog. The Hipparcos catalog displays a fairly good relation between parallax and Proper Motion:

$Parallax(mas) \sim 0.065 * PM(mas.yr^{-1}) ~~~~~~~~~~ \sigma = 130~mas$

where $PM = \sqrt{PMra^2 + PMdec^2} ~~~~~~ mas.yr^{-1}$

Expressed as a limit, where we adopt the outer boundary of the relation, this is

$Parallax(mas) > 0.015 * PM(mas.yr^{-1})$ ~~~ or ~~~  $Distance (pc) < 67000/PM$

The proper motions listed in the Tycho2 catalog can therefore be used,
with some caution, to check derived distances. Small measured proper
motions do not provide a significant limit to distance, but large
measured proper motions can be used to discriminate between giant fits
at many kpc or nearer subgiants or dwarfs. We did not use parallax
measurements directly to determine distance, as they are not available
for most catalog stars.

For catalog stars with low or no known proper motion, we set a somewhat
arbitrary Galactic upper distance limit of 20\,kpc.  Usually this excludes
supergiants, except where the catalog star is bright, but does not
exclude giant fits.  We do not make any allowance for reddening in our
spectral fits, or in our distance estimates.

We also adopt a lower distance limit of 3\,pc, which excludes some
very nearby dwarf fits (eg. at a distance of 0\,pc) in favor of
brighter candidates.  There are however several nearby stars in the
Tycho2 catalog, including Barnard's star (BD\,+04~3561a fitted as an
M3\,V at a distance of 5\,pc; an M4.2\,V fit at 2.9\,pc was excluded by
our lower distance limit), Lalande\,21185 (BD\,+36~2147 fitted as an
M1.9\,V at 4\,pc), Ross\,154 (fitted as an M4.2\,V at 5\,pc), Epsilon
Eridani (BD\,-09~697 fitted erroneously as a G5\,III at 12\,pc; a
K2\,V fit at 3\,pc occurs at a lower rank), Groombridge 34 (GJ\,15
fitted as an M\,1.9V at 5\,pc), Epsilon Indi (CPD\,-57~10015 fitted as
a K4\,V at 3\,pc), Luyten's star (BD\,+05~1668 fitted as an M3\,V at
6\,pc), Kapteyn's star (CD\,-45~1841 fitted as an M2.5\,V at 7\,pc),
Wolf 1061 (fitted as an M4.2\,V at 4\,pc), Gliese 1 (fitted as an
M1.9\,V at 6pc), GJ\,687 (BD\,+68~946 fitted as an M3\,V at 5pc),
GJ\,674 (fitted as an M3\,V at 5\,pc), GJ\,440 (WD\,1142-645 fitted as
an L749-DBQ4 type WD at 5\,pc), Groombridge 1618 (GJ\,380 fitted as an
M1\,V at 3\,pc), and AD Leonis (fitted as an M3\,V at 5\,pc).

We always accept the highest ranked spectrum (lowest $chi2$) unless we
have proper motion, distance limit or color information to exclude a
top-ranked giant or supergiant fit.

In the case that the top-ranked spectral fit contravenes one of the
above four criteria, our process descends the list of rank-ordered
chi-squared fits to the first fit compliant with our distance constraints. This
additional test increases the $chi2$ value, but often not by very
much. In checks of Tycho2/NOMAD/2MASS (TNM) fits where we also have Sloan or
Landolt information, it increases the $RTM$ values, but can reduce the
$RUI$ or $Ruz$ values. 

Our checks indicate that the (relatively noisy) TNM fits can select
giant fits over dwarf fits more often than do fits with better optical
data.  Nevertheless for TNM fits we have set our proper
motion/distance limit very conservatively, modifying the selected fit only
when the predicted distance is clearly too large. There is therefore
some residual bias in the TNM fits towards giants over dwarfs.  The
rank of each accepted fit is listed in the results, and the predicted
distances can be checked against the Tycho2 proper motions.

We have used SM\footnote{{\tt
http://www.astro.princeton.edu/~rhl/sm/}} and Enthought
Python\footnote{{\tt http://www.enthought.com/}} for most of the data
processing; these scripts are available on request. By judicious use
of {\em numpy} arrays we can fit $\sim$1M stars in eight bands with
141 sets of library magnitudes in about 10 minutes on a typical single CPU, so processing time is not
a limiting factor.

\subsection{Accuracy of the Spectral Matching process} \label{accuracy}

The Library spectra typically have relatively high S/N, but are
limited by their own flux calibration uncertainties particularly in
the U-band which, unlike the uv, was observed from the ground.  In
approximate order of decreasing uncertainty, the accuracy of our
spectrally fitted magnitudes are limited by the accuracies of:
\begin{enumerate}
\item the input catalog magnitudes,
\item the spectral matching process, discussed below, 
\item the adopted system transmission functions, discussed in section \ref{filters},
\item our derived zeropoints listed in table \ref{tbl-zp}, discussed in sections \ref{choice} to \ref{sloan},
\item our library spectra particularly in U and, to a much lesser extent,
\item the {\em calspec} spectra.
\end{enumerate}

The first item is usually largest but it is
noteworthy that our fits to catalog values, where they can be properly
checked, approach (and sometimes exceed) the accuracy of catalogs themselves. 

The methodology presented here can not (yet) approach the goal of sub
1\% photometry, and is fundamentally limited by the accuracy of the
{\em calspec} spectra, currently to about 2\%. In some ways it avoids
the end-to-end calibration of telescopes and detectors, discussed for
instance in \citet{st06}, but complements it by offering an easy way
to monitor total system throughput by enabling automatic pipelined
measurements for wider-field images.

The issues of atmospheric variability in nominally photometric
conditions are discussed in \citet{sb00} and \citet{mcg10}.  The
methodology presented here is currently less accurate than the
expected atmospheric changes for field-to-field comparisons. But using
standards in each digital field, rather than transferring them, offers
the advantage of both monitoring and consistently removing such
variability within the same field, even during known non-photometric
conditions.

\section{Primary Standards} \label{primary}

\subsection{Landolt Standards with 2MASS} \label{landolt}

We illustrate the spectral matching process by
matching library spectra to 594 Landolt standards with accurate UBVRI
magnitudes, for which we also have 2MASS $JHK_{2/S}$ magnitudes.  

We took the updated Landolt coordinates, magnitudes, uncertainties and
proper motions from \citet{land09}. We used the {\tt find2mass} tool
in the {\em cdsclient} package from Centres de Donn\'ees
(CDS)\footnote{{\tt http://cdsarc.u-strasbg.fr/doc/cdsclient.html}} to
check 2MASS matches within 3 arcsec to these coordinates and found
good correspondence to \citet{land09} for all except SA92-260 and SA109-956.  For these last
two stars we have adjusted the coordinates, and matching 2MASS
magnitudes, checking them with Landolt finding charts and DSS
images. As indicated in \citet{land09} the star 'E' near the T-Phe
variable has a NOMAD entry but no 2MASS entry within 30 arcsec; it has
been excluded from our list here.

The catalog data for the Landolt standards: sexagesimal coordinates,
UBVRI magnitudes and errors, 2MASS magnitudes, errors and coordinates
in degrees, together with UKIRT $ZY$ \& $JHK_{MKO}$ data where
available are listed in the online table {\em LandoltCat}.  Where
\citet{land09} photometric errors were left blank, indicating one or few
standard observations, we have set them to 0.009\,mag for the purpose
of our fitting process.

For each standard, we compute the $chi2$ value for each of 141 library
spectral magnitudes as described in section \ref{matching} for the
eight L2M bands: Landolt $UBVRI$ and 2MASS $JHK_{2/S}$. We then order
them by $chi2$, check against our distance constraints, and select the
one that satisfies the constraints with the best match to the eight catalog magnitudes,

The results are listed in the online table {\em LandoltFit}, which
lists the standard name, coordinates, Landolt magnitudes and colors,
number of observations, and spectral type from \citet{dl79}, then
fitted rank, $chi2$ values, two RMS error in Landolt magnitudes $RUI$ \& $RBI$,
fitted spectral library number and type, and fitted magnitudes in
$B_TV_T$, $JHK_{2/S}$, $UBVRI$, UKIRT $Z_VY_V$, Sloan $u'g'r'i'z'$,
and distance in parsec.
Magnitude values in other bandpasses can be
derived by adding the library magnitude from {\em LibMags} for the
appropriate band and spectral type to the fitted V magnitude.

There is reasonably good correspondence between the objective prism
spectral types from \citet{dl79} and those fitted here; about 123
close matches, {\em versus} 13 discrepant types -- the worst two being
SA110-499 and SA110-450.  

Figure \ref{BVRI_l2m} shows the 8-band L2M fitted magnitudes plotted
as $Fit-Landolt$ magnitudes against their Landolt standard values for
$BVRI$.  The display range excludes a few outliers discussed below.
The dashed histograms of number (to the right) vs. delta magnitude (ordinate)
illustrate that most stars lie close to the zero-delta line. The
clipped $1\sigma$ dispersions are 0.05, 0.03, 0.04 and 0.05 mag for
$BVRI$ respectively. In order to compute these dispersions we formed a
sigma for each of $BVRI$ separately, performed a $3\sigma$ clip, then
repeated a second (tighter) $3\sigma$ clip to arrive at the quoted
$1\sigma$ dispersions. This typically clips 4---8\% of the values, or
25---45 of the values here, but retains about 95\% of the
values.  

Of the clipped values, almost $2/3$ have been observed only once or
twice in table {\em LandoltCat}; the others are mainly white dwarfs,
very red stars, or perhaps double stars.  A few poor fits with large
values of $chi2$ and $RUI$ include some late giants and supergiants
where the spectral library coverage is weak and Feige\,24, a double
star, which is fairly well fit with a double-star spectrum constructed
from a 68:32\% mix at V-band of DA1 (GD\,71) and K4\,V spectra. 
PG1530+057 is listed as a uv-emission source, and is also
best fit with this DA/K4\,dwarf double star spectrum, as is SA107-215.
No effort was made to optimize these coincidental double-spectrum
fits, but they emphasize that double stars are likely to be present
among Landolt standards, as among other catalog stars.  

In many cases of multiple stars within the observing aperture,
including known doubles like BD\,+26\,2606, the spectral flux will be
sufficiently dominated by one spectrum at all wavelengths that fitting
to a single spectral type is valid. But some data do show the
predominant influence of one spectrum in the blue and another in the
red.  Most stars here ($\sim$94\% or 560) are very well fit by the
(single) spectral matching process.

The clipped $1\sigma$ dispersion for $U$ shown in figure \ref{U_l2m}
is 0.16 mags; 0.12 mags to U$<$16 and increasing for fainter magnitudes. 
The four faintest U standards were not observed often in \citet{land09}.
Our fits for these find brighter U values, but have fairly large $chi2$ and $RUI$.

The distribution of spectrally fitted types and luminosities for
Landolt standards is shown graphically as an HR diagram in figure \ref{HRland}.
The fitted types are plotted as absolute V magnitudes vs. $V-I$
color, with different symbols for different Luminosity classes and
metallicities.  The area of each symbol is proportional to the number of fitted stars of that type.
The truncation of the lower main sequence is of course
a brightness selection effect.  The Landolt primary standards cover a wide range
of colors, types and metallicities, with most MK types well represented.

\subsubsection{$ZY$ on Vega and AB systems} \label{zy}

Table {\em LandoltCat} also contains $Z_VY_V$ data from UKIRT for 14
standards, and $JHK_{MKO}$ for 17 Landolt standards (see section
\ref{intro}). The $1\sigma$ dispersions (not shown) for both $Z_V$ and
$Y_V$ are 0.06 mag for 14 stars with standard values (including two
calspec standards: GD\,50 fit with a GD\,153 DA0 type spectrum \&
GD\,71 fit with its own Library GD\,71 DA1 spectrum).

Despite the fact that the $Z_V$ bandpass terminates shortward of $z$
or $z'$, and because of the falling long-wavelength response of CCDs,
the pivot wavelengths of $Z_V$ \& $z'$ in table \ref{tbl-zp} are similar. It is
therefore not surprising that there is a simple one-to-one relation between
fitted values for $Z_V$ on the {\em vega} system and $z'$ and $z$ on the AB
system, which holds for 14 Landolt/Sloan standards and also for 594
fitted values for Landolt Standards and for the spectral library
magnitude data in table {\em LibMags}, both of which cover a wide
color range:

$0.58 + Z_V \simeq z_{AB} \simeq z'_{AB}  ~~~~~~~~~~~ \sigma = 0.014$

where the additive constant is simply the derived zeropoint for $Z_V$
(and that for $z_{AB}$ is zero). This is equivalent to the magnitude difference
between the {\em vega} and AB system 0-mag fluxes at $Z_V$:
 $0.58 = -2.5*log_{10}(2128/3631) $ from table \ref{tbl-zp}.

For AB magnitudes measured with a bandpass similar to $Y_V$ used by
UKIRT/WFCAM, the relation between the $Y_V$ {\em vega} and AB system
magnitudes should also be the {\em vega} to AB zeropoint offset at this bandpass, ie.

$K_Y + Y_V \simeq y_{AB} $

where $K_Y$ should be close to $0.61 - 0.0$, or alternatively to the
difference in magnitudes between the 0-mag flux at $Y_V$ on the {\em
vega} system and 0-mag (3631\,Jy) for the AB system, ie. 
 $0.61 = -2.5*log_{10}(2072/3631) $ from table \ref{tbl-zp}.

\begin{figure}
\epsscale{.80}
\plotone{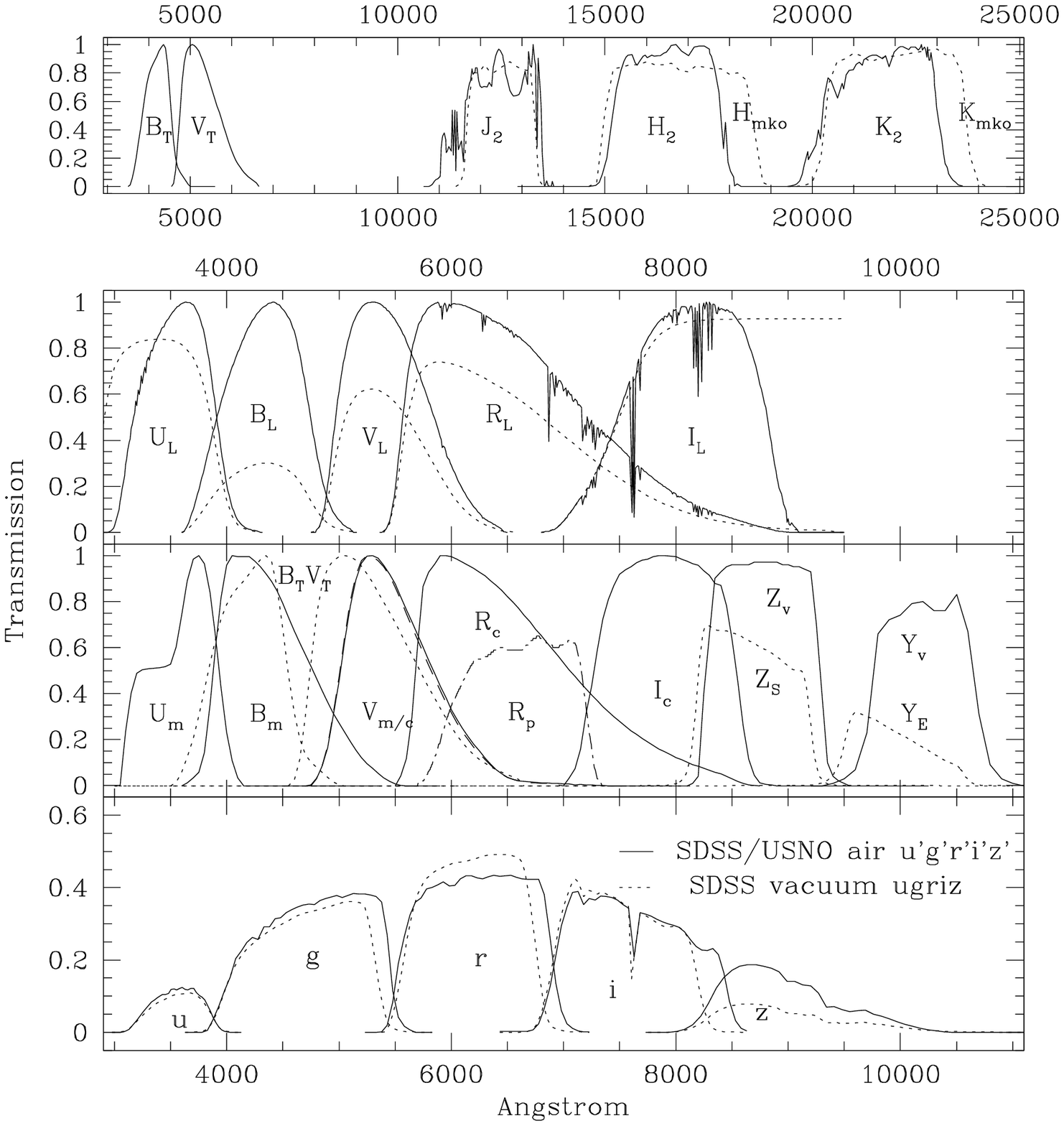}

\caption{The filters discussed here are illustrated.  In the first
panel the Tycho2 $B_TV_T$ filter system transmissions are taken from
\citet{ma06}, and over-plotted (dotted lines) in the third panel for
convenient reference. The 2MASS $J_2H_2K_{2/S}$ system total response
curves are from the IPAC website; the $JHK_{MKO}$ filter transmission curves
(dotted) are from \citet{tok02}.  In the second panel the Landolt
$U_LB_LV_LR_LI_L$ transmission curves for the filters alone are shown
-- dotted, \citet{land92a}; their system response including typical
atmosphere and detector QE are shown -- solid, \citet{coh03a}.  In the
third panel the synthetic $U_MB_MV_M$ filters are from \citet{ma06}
($V_M$ plotted as dashed line); the $V_CR_CI_C$ curves are from
\citet{bess79} and the $Z_VY_V$ (solid) curves are from the UKIRT/VISTA
collaboration.  The PTF ``Mould-type'' $R_p$ system response is also
illustrated, as are the LCOGT $Z_SY_E$ system responses (dots) discussed in
section \ref{zy}.  In the bottom panel the Sloan survey system bandpasses are
shown for both the unprimed survey imaging bandpasses (dotted lines,
where the filter interference red edges move bluer in the vacuum of
the survey camera) and primed bandpasses used in air for PT and
Southern standard observations ({\em u'g'r'i'z'} -- solid lines); they
are from the JHU skyservice and the USNO FNAL websites.}

\label{figfilt}
\end{figure}

\begin{figure}
\epsscale{1.0}
\plotone{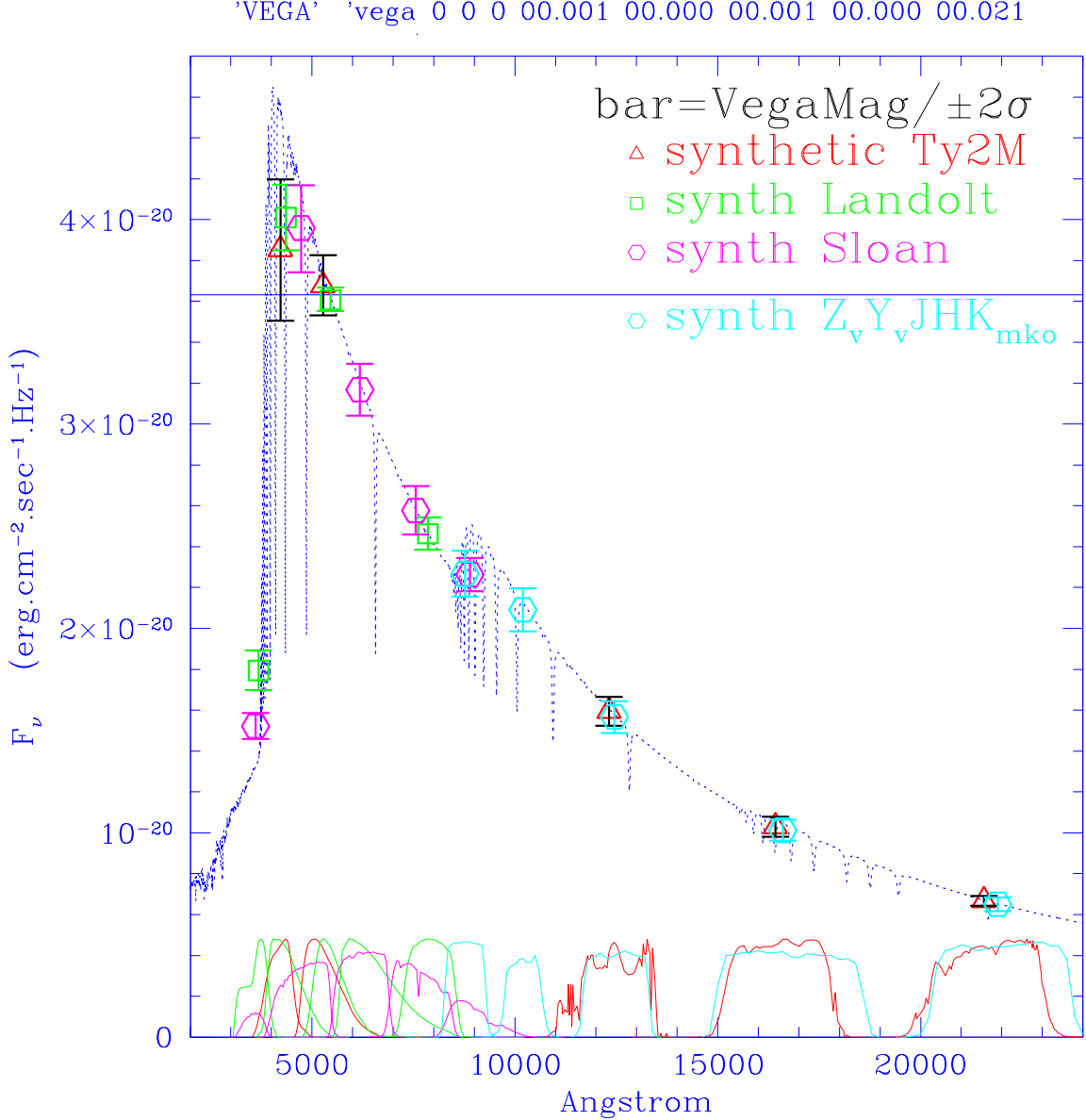}

\caption{The bandpasses, zeropoints and synthetic measurements are
illustrated for the {\em calspec} spectrum STIS\_005 spectrum of Vega
(blue dotted trace), which defines the nominal 0-mag definition for
filters on the {\em vega} system.  The electronic version of this
figure is in color.  The horizontal line at
$3.631*10^{-20}~erg~cm^{-2}~sec^{-1}~Hz^{-1}$ (3631\,Jy) illustrates
the AB=0 mag reference. Tycho2/2MASS $B_TV_TJHK_{2/s}$ bandpasses are
indicated by red traces, synthetic flux measurements by red triangles
and zeropoint dispersion values about them by error bars in black. All
error bars here show $\pm2\sigma$ zeropoint dispersion for better
visibility.  Adopted Landolt $U_MB_MV_CR_CI_C$ bandpasses are shown as
green traces, synthetic flux measurements by green squares and
zeropoint dispersion values by green bars.  Sloan $u'g'r'i'z'$
bandpasses are shown in magenta, synthetic measurements by magenta
circles and zeropoint dispersion values by magenta bars. The UKIRT
$Z_VY_VJHK_{MKO}$ bandpasses are shown as cyan traces, synthetic
measurements as cyan circles and zeropoint dispersion values about
them by cyan bars.}

\label{vega}
\end{figure}

\begin{figure}
\epsscale{1.0}
\plotone{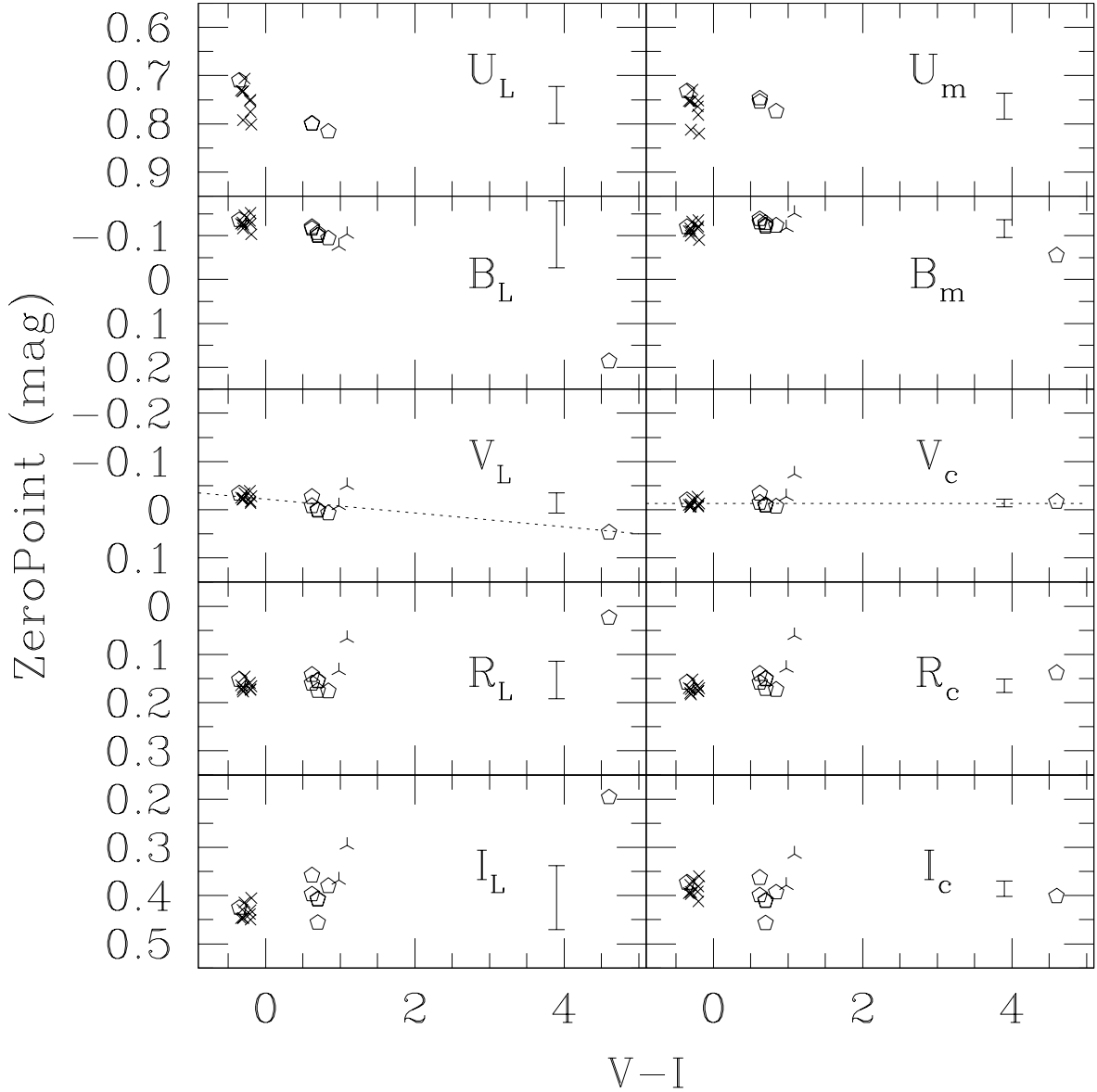}
\caption{Zeropoints are shown for $U_LB_LV_LR_LI_L$ (left) and
$U_MB_MV_CR_CI_C$ system bandpasses (right) as a function of standard $V-I$ color.  
The ordinate scale is set so that vertically higher zeropoints result in
magnitudes which are larger (fainter).  Vertical error bars indicate
the average and $\pm 1\sigma$ dispersion in each panel.  Crosses
indicate White Dwarfs, Pentagons indicate dwarfs.  Three-point crosses indicate
the two K giants, which are not included in these zeropoint averages or
dispersions as they lack UBVRI photometry. The reddest point
corresponding to the M7\,V VB8 lacks U-band data.  The selected UBVRI
system profiles on the right minimize zeropoint scatter and trend with
color. }
\label{UIzero}
\end{figure}

\begin{figure}
\epsscale{1.0}
\plotone{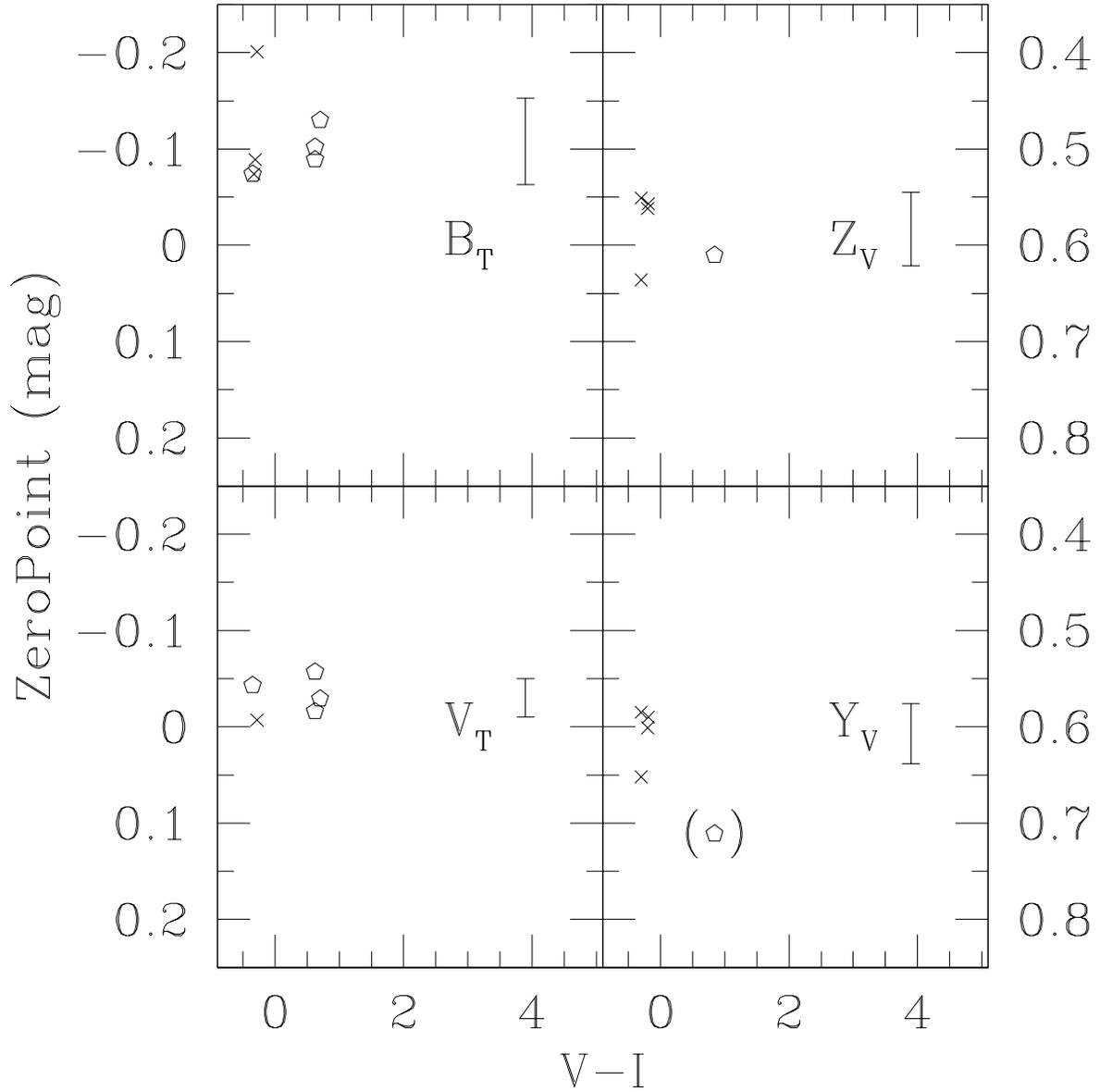}
\caption{Zeropoints are shown for
$B_TV_T$ (left ordinate) and $Z_VY_V$ (right ordinate) as a function of 
standard $V-I$ color, with averages and dispersions as vertical bars. 
The $Y_V$ zeropoint for G158-100 is shown in parentheses but not
included in the average. Symbols as before.}
\label{BVZYzero}
\end{figure}

\begin{figure}
\epsscale{1.0}
\plotone{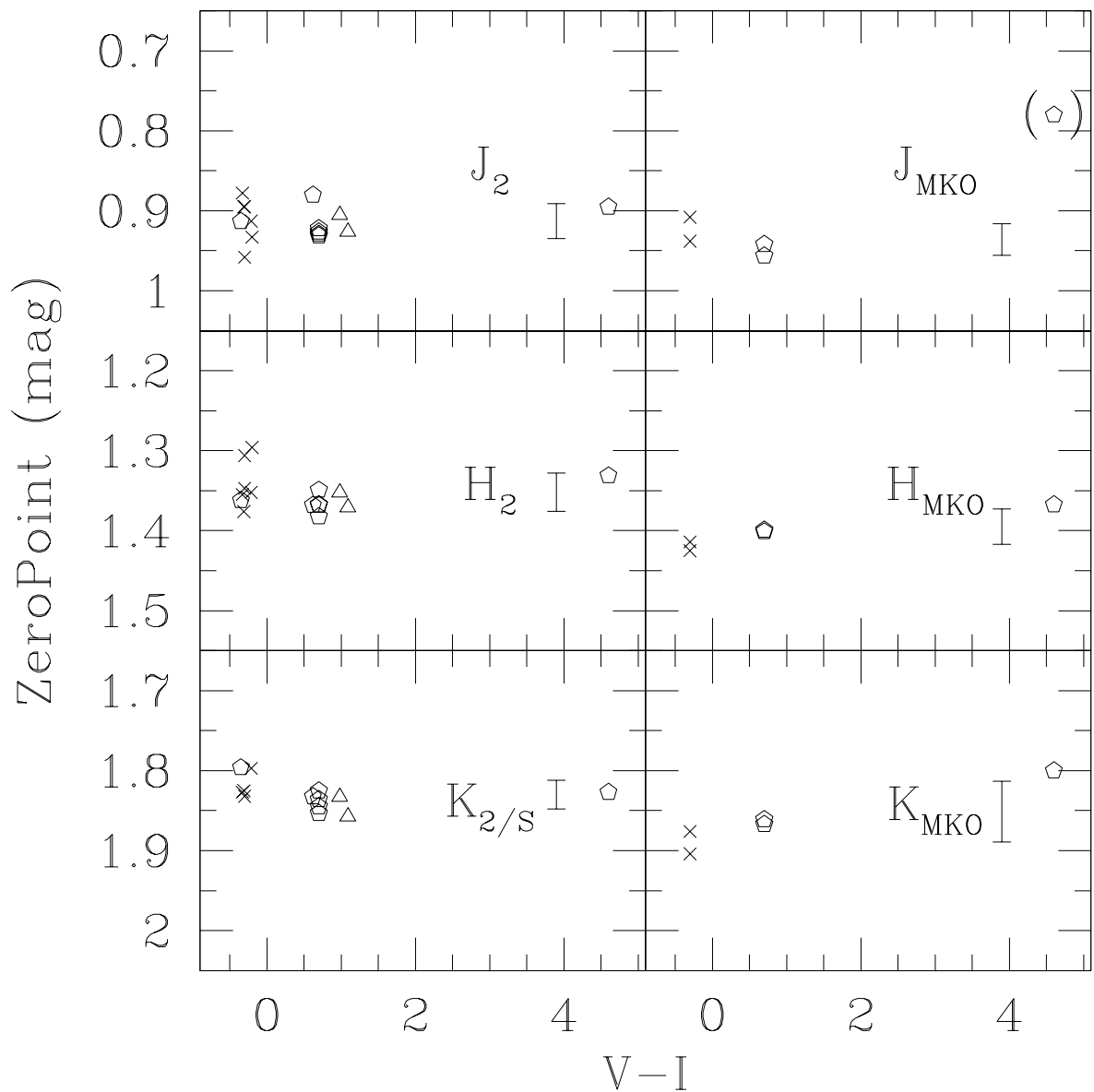}

\caption{Zeropoints are shown for $J_2H_2K_{2/S}$ (left) and
$JHK_{MKO}$ (right) as a function of standard $V-I$ color, with averages and
dispersions.  Symbols as before. The K\,III points (triangles) have
been included for $JHK_{2/S}$. The zeropoints for $JHK_{MKO}$ show some
trend with color, but the $J_{MKO}$ value for VB8 has been excluded
due to uncertainty over its magnitude. }

\label{JKzero}
\end{figure}

\begin{figure}
\epsscale{1.0}
\plotone{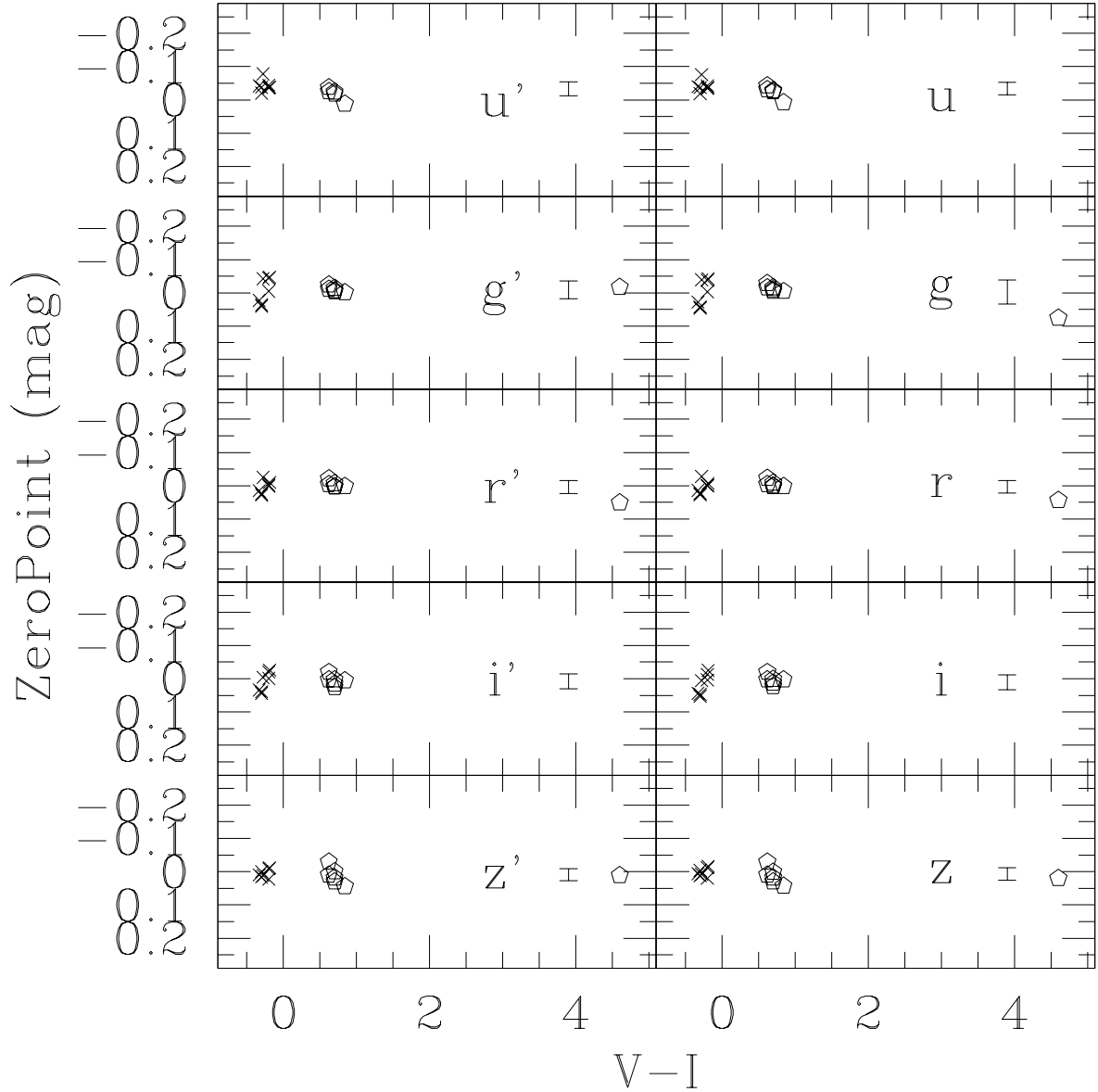}

\caption{Zeropoints are shown for $u'g'r'i'z'$ (left) and $ugriz$ (right) as a
function of standard $V-I$ color, with averages and dispersions.  VB8
is present in the $grz$ plots, but has been omitted in the $ui$
plots. Symbols as before.}

\label{uzzero}
\end{figure}

\begin{figure}
\epsscale{1.0}
\plotone{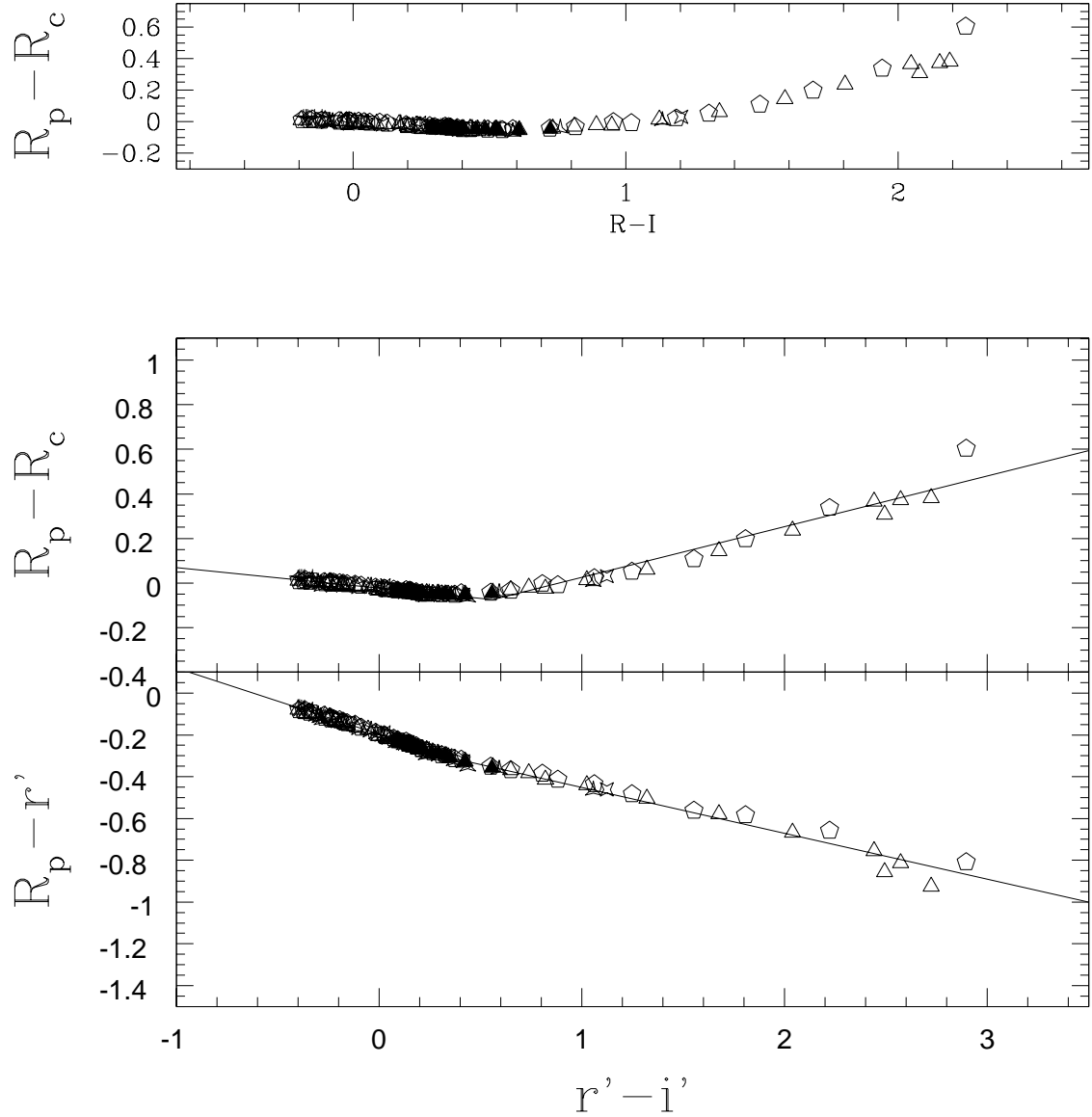}

\caption{Upper panel shows the $R_p - R_C$ color plotted against the
Landolt $R-I$ color. The fit is tight, but curved in both
segments.  Dwarfs are indicated as pentagons, subgiants as
squares, giants as triangles and supergiants as crosses.  The lower
two panels show respectively the $R_p - R_C$ and $R_p - r'$ colors
vs. Sloan $r'-i'$. Same symbols as before. In these cases
there are good two-segment straight-line fits.}

\label{Rpplot}
\end{figure}

\begin{figure}
\epsscale{1.0} 
\plotone{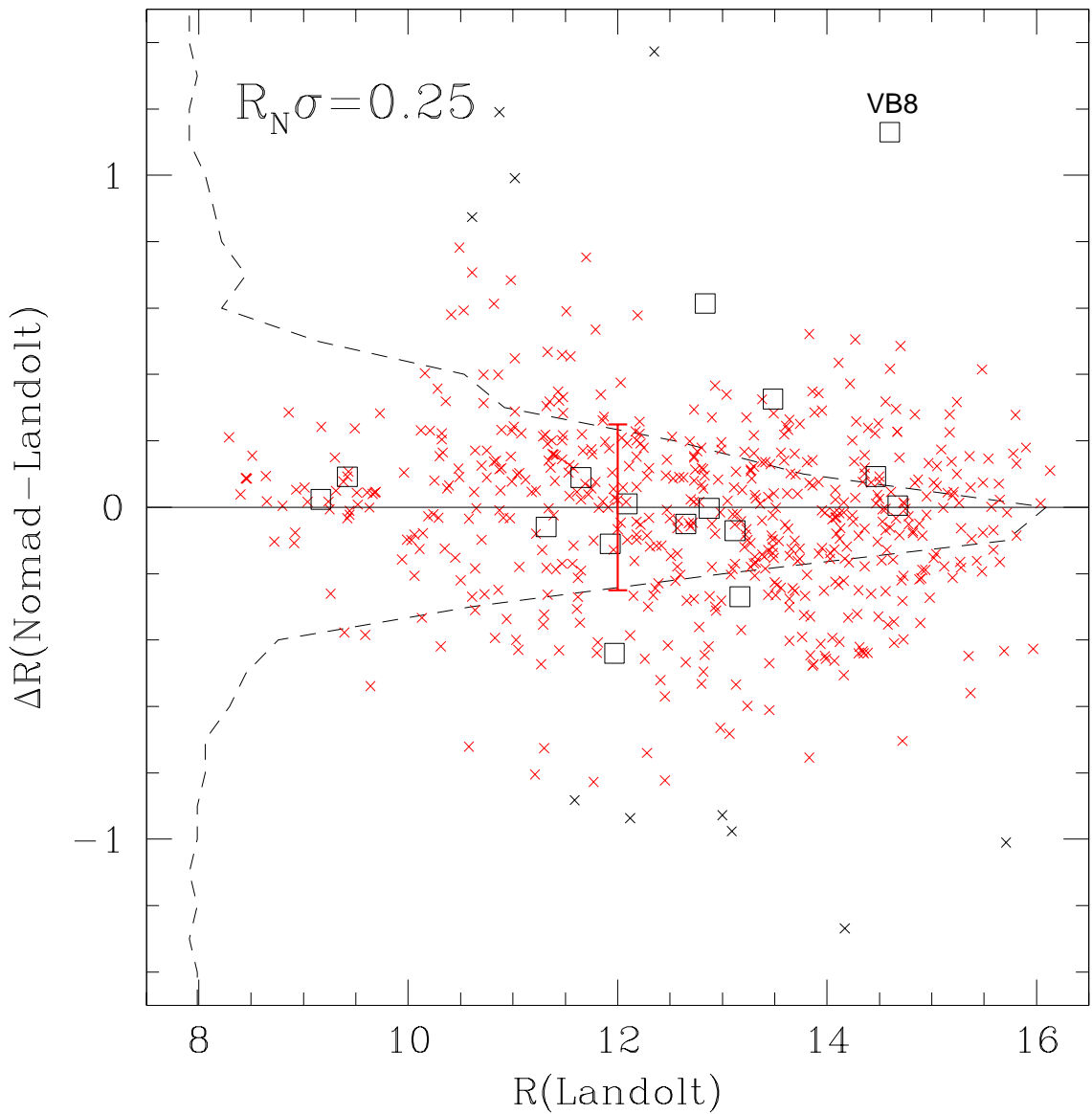}

\caption{Comparison of NOMAD and Landolt R-band data plotted as the
difference $R_N$-$R_{Landolt}$ vs. $R_{Landolt}$ (crosses) and CALSPEC
standards (squares); VB8 is identified. Black crosses show about 10
points rejected by a $3\sigma$ clip, and grey crosses (red in the
color version) those remaining. The resulting $1\sigma$ dispersion of
0.25\,mag is indicated.  
The dashed
histogram of all the stars shows the distribution of these differences
plotted with increasing number to the right against the magnitude
delta on the ordinate.}

\label{Rnomad}
\end{figure}

\begin{figure}
\epsscale{1.0}
\plotone{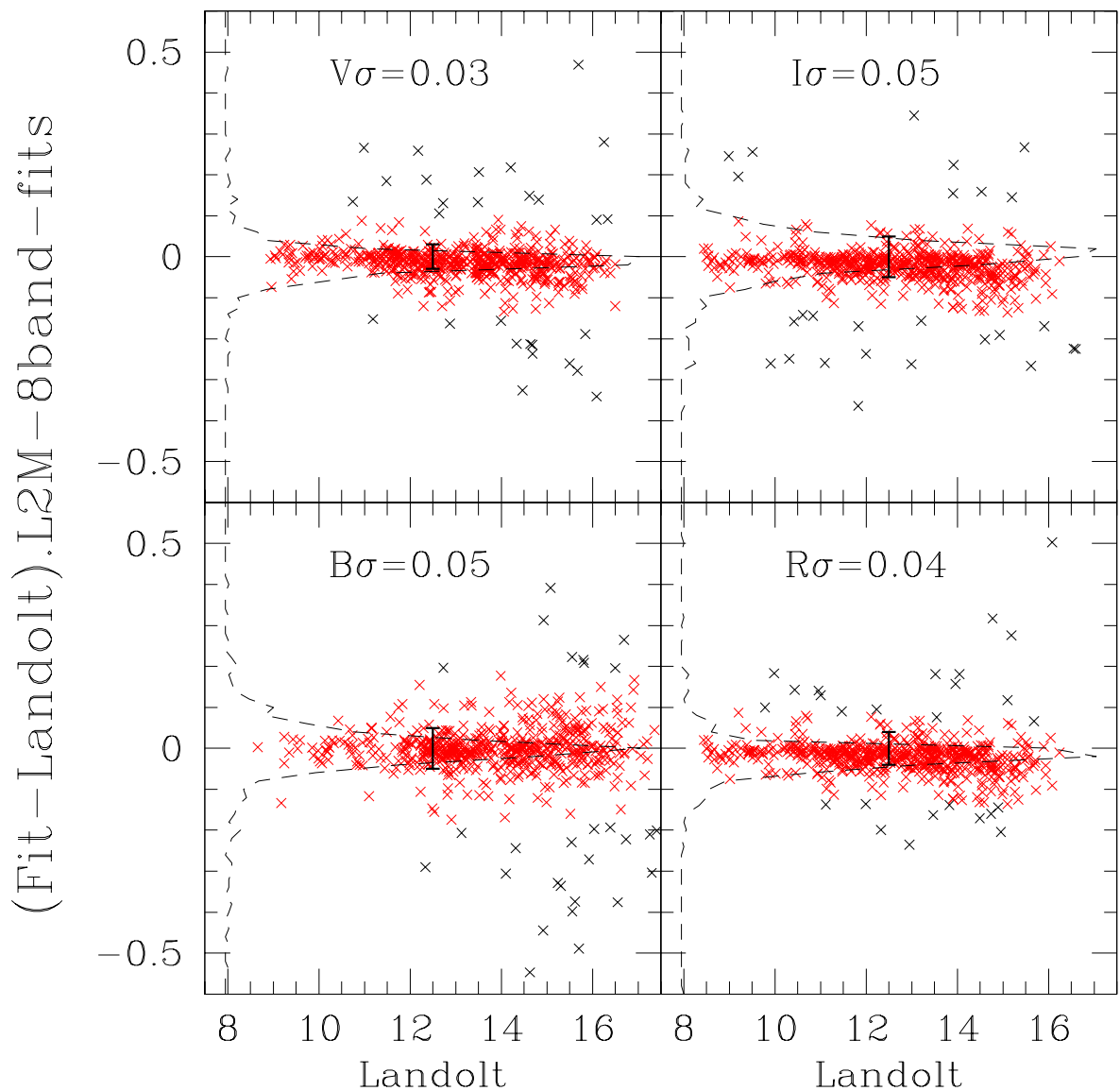}

\caption{$BVRI$ fits with eight L2M $UBVRI-JHK_{2/S}$ bands for 594
Landolt standards.  The data are plotted as magnitude differences
$Fit-Landolt$ on the ordinate vs. Landolt catalog magnitudes on the
abscissa.  Black crosses indicate points excluded by sigma clipping;
the scaling excludes a few outliers discussed in the text.  Grey (red) points
indicate those remaining after sigma clipping. The dashed histograms
plotted vertically in each panel illustrate the number distribution of all the
points about the zero-delta (horizontal) line.  Clipped $\pm1\sigma$
errors are indicated in each panel caption, and by the vertical error
bars.}

\label{BVRI_l2m}
\end{figure}

\begin{figure}
\epsscale{1.0}
\plotone{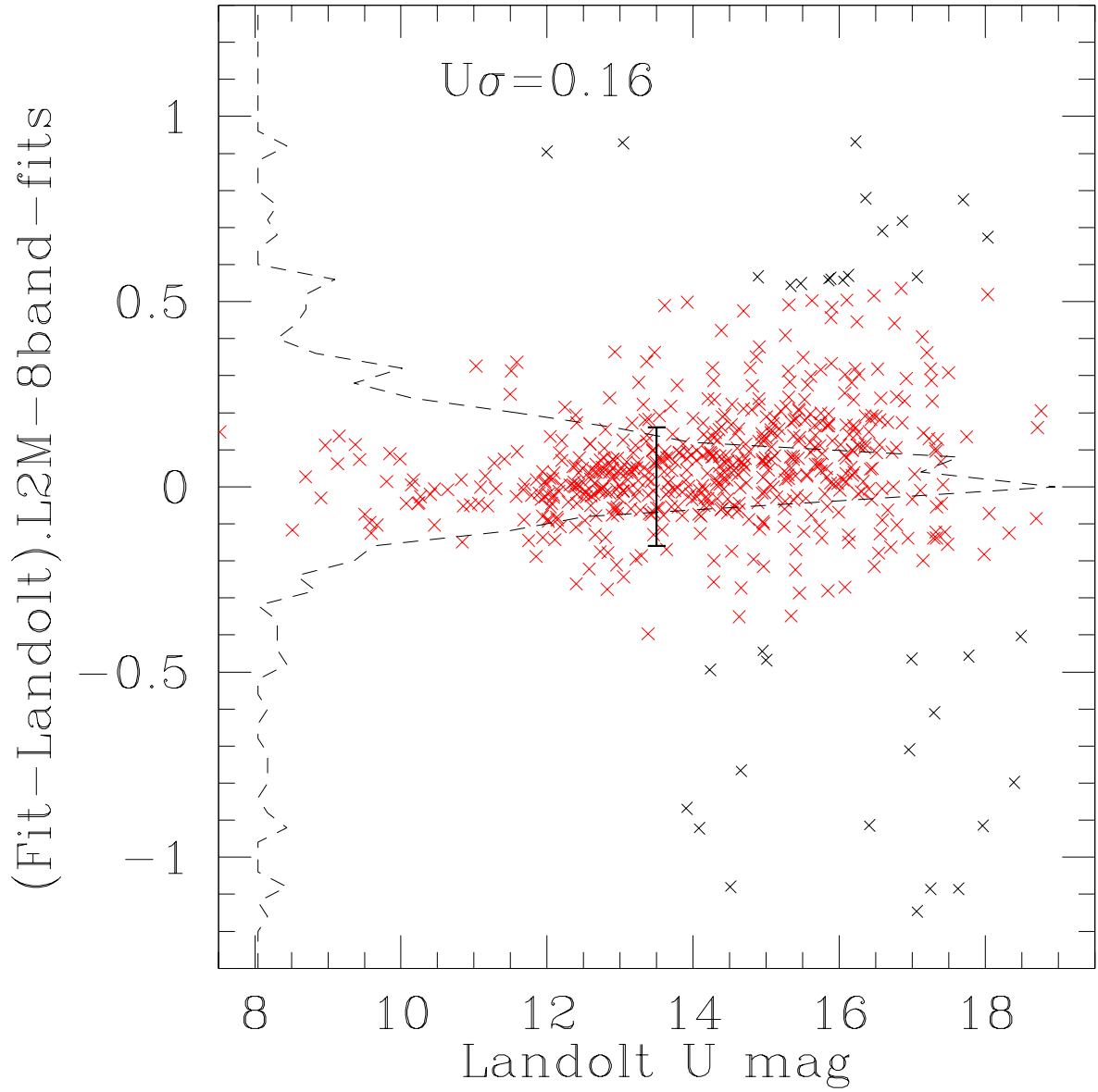}

\caption{U-band fits with eight L2M $UBVRI-JHK_{2/S}$ bands plotted as
$Fit-Landolt$ for 594 Landolt standards. Grey (red) crosses indicate points
retained after sigma clipping, and the dashed histogram illustrates
the number distribution of all the points around the zero-delta
line. The sigma is 0.12 mag to U$<$16, increasing for fainter stars.}

\label{U_l2m}
\end{figure}

\begin{figure}
\epsscale{1.0}
\plotone{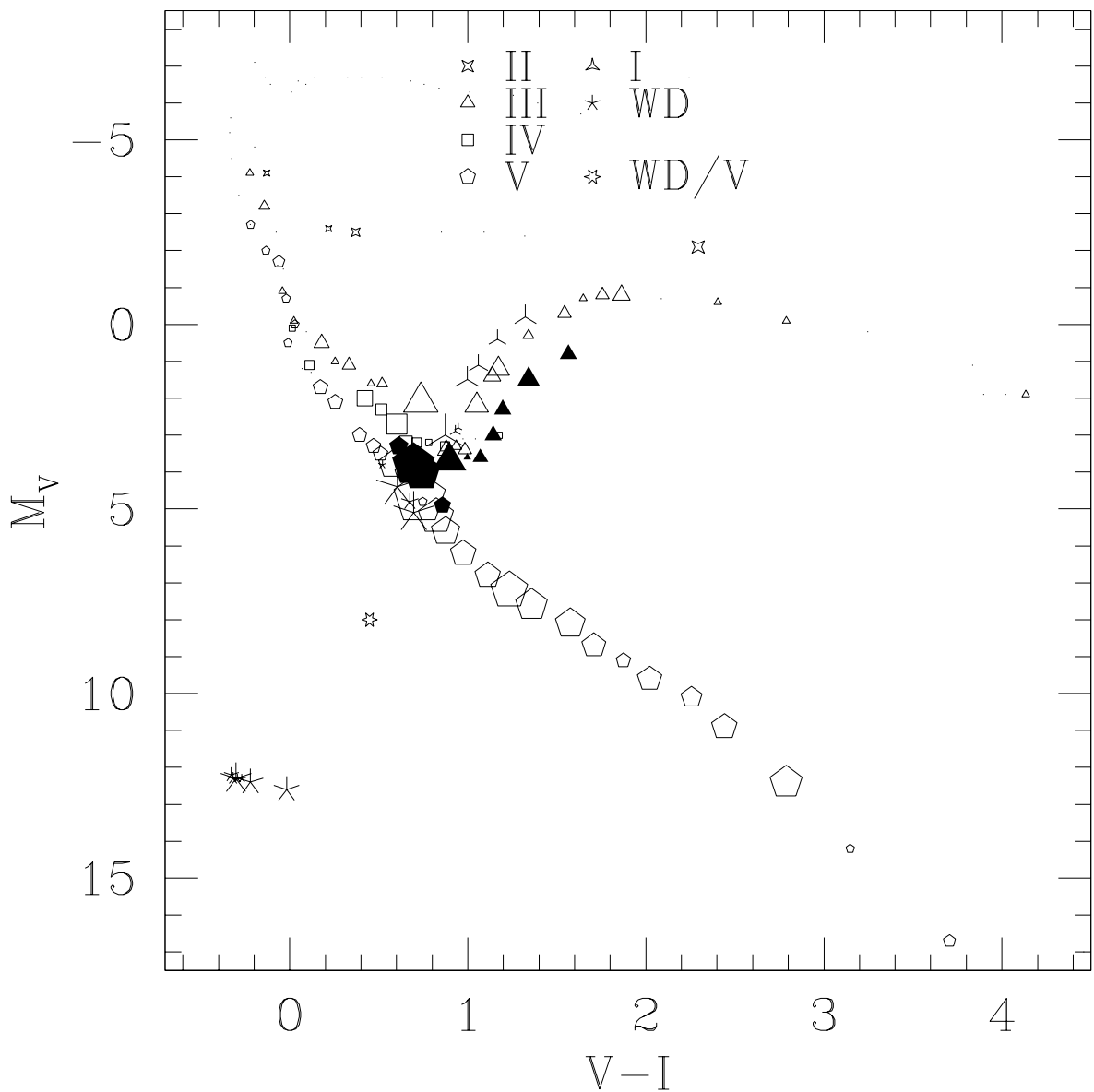}
\caption{HR diagram for 594 Landolt standards shows our fitted types
as absolute V magnitudes vs. spectral library $V-I$ colors.
Dwarfs are represented as pentagons, subgiants as squares, giants as
triangles, Luminosity class II and I as four-point and 3-point stars
respectively.  White dwarfs are shown as 5-point crosses.  Metal-weak
dwarfs and giants are shown as 5 or 3 sided crosses, metal-rich
components as 5 or 3 sided filled symbols.  The area of each symbol is
proportional to the frequency that library spectrum was selected.  The
6-starred symbol at (0.45,8) represents the DA1/K4\,V double spectral type, fitted to
Feige\,24, PG1530+057A and SA107-215.}
\label{HRland}
\end{figure}

Care must be taken however, particularly with Y-band measurements made
with CCDs, because CCD QEs fall very fast in the 1.0--1.1\,$\mu$
region, and CCD Y-bandpass profiles are necessarily quite different
from the rectangular UKIRT WFCAM HgCdTe $Y_V$ bandpass, see figure
\ref{figfilt}.  Many CCDs have little or no sensitivity at
Y-band. Deep depletion CCDs can have significant QE out to the Silicon
bandgap at $1.1 \mu$, but CCDs even of the same type may have varying
QE curves and system bandpasses at Y, and hence also have different
effective and pivot wavelengths.

At LCOGT we use the PanStarrs-type $Z_S$ (for Z-short) and Y filters with our
Merope E2V 42-40 2K CCDs on Faulkes Telescopes North and South (FTN \&
FTS). Our LCOGT $Z_SY_E$ system bandpasses including filter,
atmosphere and detector are illustrated in fig \ref{figfilt}.  We
refer to $Y_E$ to specifically reference the PanStarrs-type Y bandpass with
our E2V system QE curve, and 1.3\,airmasses of extinction at a typical
elevation of 2100m.  The measured pivot wavelengths for $Z_V$ and
$Z_S$ are 877.6 and 865.1\,nm respectively; those for $Y_V$ and $Y_E$
are 1020.8 and 989.9\,nm respectively. The effective wavelengths for
either type of Z and Y bandpasses vary by less than 1\% and 0.5\%
respectively however, from white dwarf to M-dwarf spectra, so Z/Y band
measurements can provide convenient near infrared flux and temperature measurements,
enabling quite good dwarf/giant discrimination for M stars, somewhat
insulated from atmospheric and stellar spectral features.

For the same (few) standards shown in table \ref{tbl-zp} we obtain
zeropoints (on the {\em vega} system): $Z_S = 0.56 \pm 0.04$ and $Y_E
= 0.55 \pm 0.03$.  We can form $Z_SY_E$ filter system magnitudes for
all the spectral library stars, and form linear relationships with
$Z_VY_V$ of the same stars, but because of the bandpass differences
they are no longer simply one-to-one relationships of unit slope.  For
synthetic measurements with LCOGT $Z_SY_E$ of our 141 spectral library
standards in table {\em LibMags} we obtain:

$1.05*(0.56 + Z_S) \simeq z = z'  ~~~~~~~~~~~~~ \sigma = 0.024$, and

$1.04*(Y_E - 0.03) \simeq Y_V    ~~~~~~~~~~~~~~~~ \sigma = 0.022$

Where $Z_SY_E$ \& $Y_V$ are on the {\em vega} system.
These relations are dependent on the particular bandpasses used here,
but indicate that CCD based Y-band magnitudes (on either the {\em
vega} or AB systems) can be successfully tied to the UKIRT/WFCAM
Y-band standard system, with CCD-system-dependent equations.

\subsubsection{$JHK_{MKO}$ fits}

Our fitted values of $JHK_{MKO}$ on the {\em vega} system match 17 Landolt stars with standard UKIRT values\footnote{{\tt
http://www.jach.hawaii.edu/UKIRT/astronomy/calib/phot\_cal/fs\_izyjhklm.dat}},
which further supports our derivation of their zeropoints.

$J_{MKO}(fit) \simeq J_{MKO} ~~~~~~~ \sigma = 0.04 $

$H_{MKO}(fit) \simeq H_{MKO} ~~~~~~~ \sigma = 0.05 $

$K_{MKO}(fit) \simeq K_{MKO} ~~~~~~~ \sigma = 0.07 $

$J-H$ and $H-K$ fitted colors form linear relationships with
$J-H_{MKO}$ \& $H-K_{MKO}$ for these same 17 stars with sigmas of 0.04
\& 0.03\,mag respectively, but with little color range.
There are color terms between the $JHK_{MKO}$ and 2MASS $JHK_{2/S}$ systems.

\subsection{Sloan Standards}

There are 158 Sloan Standards listed in electronic table {\em
SloanCat}, together with UKIRT values of $Z_VY_V$ and $JHK_{MKO}$ where available,
and 2MASS values of $JHK_{2/S}$. These have been spectrally matched
with eight S2M bands of $u'g'r'i'z'J_2H_2K_2$, and their fitted values
of Landolt and Sloan magnitudes and distances in parsec are listed in
electronic table {\em SloanFit}. The comparisons with standard data
are discussed below.

\subsection{Landolt/Sloan/Tycho2/2MASS Standards}

There are 96 standards in common between the Landolt and Sloan lists;
they all have 2MASS $J_2H_2K_{2/S}$ and NOMAD $R_N$ magnitudes; 65 of
these also have Tycho2 $B_TV_T$ magnitudes. This very small
sample with cross-referenced data provides useful illustrations of the
strengths and limitations of the spectral matching process.

Electronic table {\em LanSloCat}, lists Landolt, Sloan, NOMAD, 2MASS
and Tycho2 magnitudes, errors, coordinates and proper motions. In this
case the proper motions are from Tycho2; they are similar to but
slightly different from proper motions listed in electronic table {\em LandoltCat} which are from
\citet{land09}. The proper motions are only used to discriminate distance limits here.

These stellar magnitudes have been fitted four ways: i) they were fitted
with $UBVRI$, $u'g'r'i'z'$ and $B_TV_TJ_2H_2K_{2/S}$, a total of 15
bands, ii) they were fitted with Landolt and 2MASS data (L2M,
8-bands), iii) they were fitted with Sloan and 2MASS data (S2M,
8-bands), and iv) they were fitted with Tycho2, NOMAD $R_N$ and 2MASS
data (TNM, 6 bands).

The fits are listed in electronic table {\em LanSloFit} where the
results are listed in full for the first method, then only for those
stars with different matching spectra for subsequent fits.

In each case the fitting process produces synthesized magnitudes at
all bands, together with library types and distances. The synthesized
magnitudes were compared with the standard values, and the fitted
library types checked for consistency between fits.

\subsubsection{Examples of spectral matching} \label{lscex}

\begin{figure}
\epsscale{1.0}
\plotone{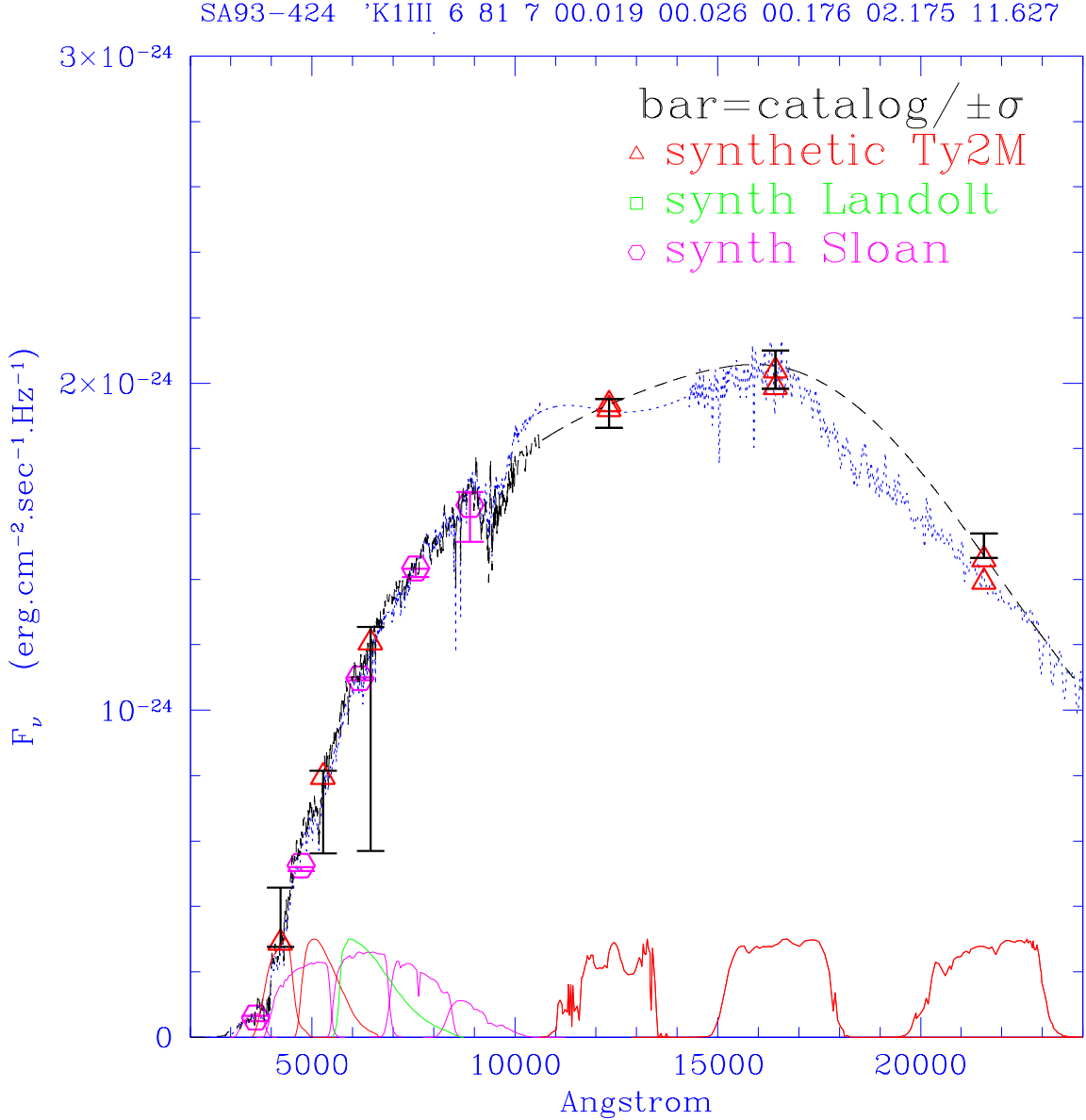}

\caption{The Sloan/2MASS (8-band, S2M) K1\,III fit for SA93-424 is
shown (fainter, blue dotted trace) as $F(\nu)$ vs. wavelength.
Electronic versions of these figures are in color.  The Sloan filter
profiles are shown in magenta, the Tycho2/2MASS profiles in red and
the R profile (matched here to NOMAD $R_N$ in green.  Black, green and
magenta bars identify the Tycho2/2MASS, Landolt (and $R_N$) and Sloan
catalog magnitudes and errors.  Red triangles, green squares and
magenta circles identify fitted synthetic fluxes and magnitudes in the
same systems.  The overlaid black dashed spectrum is the rK0\,III fit
obtained with 6 Tycho2/$R_N$/2MASS (TNM) bands.  The long vertical
black bar represents the $R_N$ catalog error, scaled by a factor 1.5,
on the second fit. The catalog fluxes are at the center of each bar,
and the synthesized fluxes are marked by the symbols.}

\label{SA93-424}
\end{figure}

Figure \ref{SA93-424} shows the S2M 8-band fit for SA93-424, a K1\,III at
768\,pc, with $F(\nu)$ plotted against wavelength.  The same K1\,III
fit is obtained with the 15-band fit and the L2M 8-band fit; ie. the
fits have the same type and the scale is different by less than 0.2\%.
For this fit, RUI, Ruz and RTM rms values are 0.02, 0.03 \& 0.16 mag
respectively.  Over-plotted (black dashed line) is the best-fit
rK0\,III type at 393\,pc obtained with the TNM fit (6 bands).  The
6-band fit is forced a bit brighter in the optical region by the
$B_TV_T$ data, but matches the $K_{2/S}$ point better.  This latter
fit gives synthesized Landolt/Sloan magnitudes with RUI, Ruz and RTM
of 0.16, 0.09 and 0.18 mag respectively. 
\citet{dl79} list a G8\,III spectral type for SA93-424, from objective prism spectra.

\begin{figure}
\epsscale{1.0}
\plotone{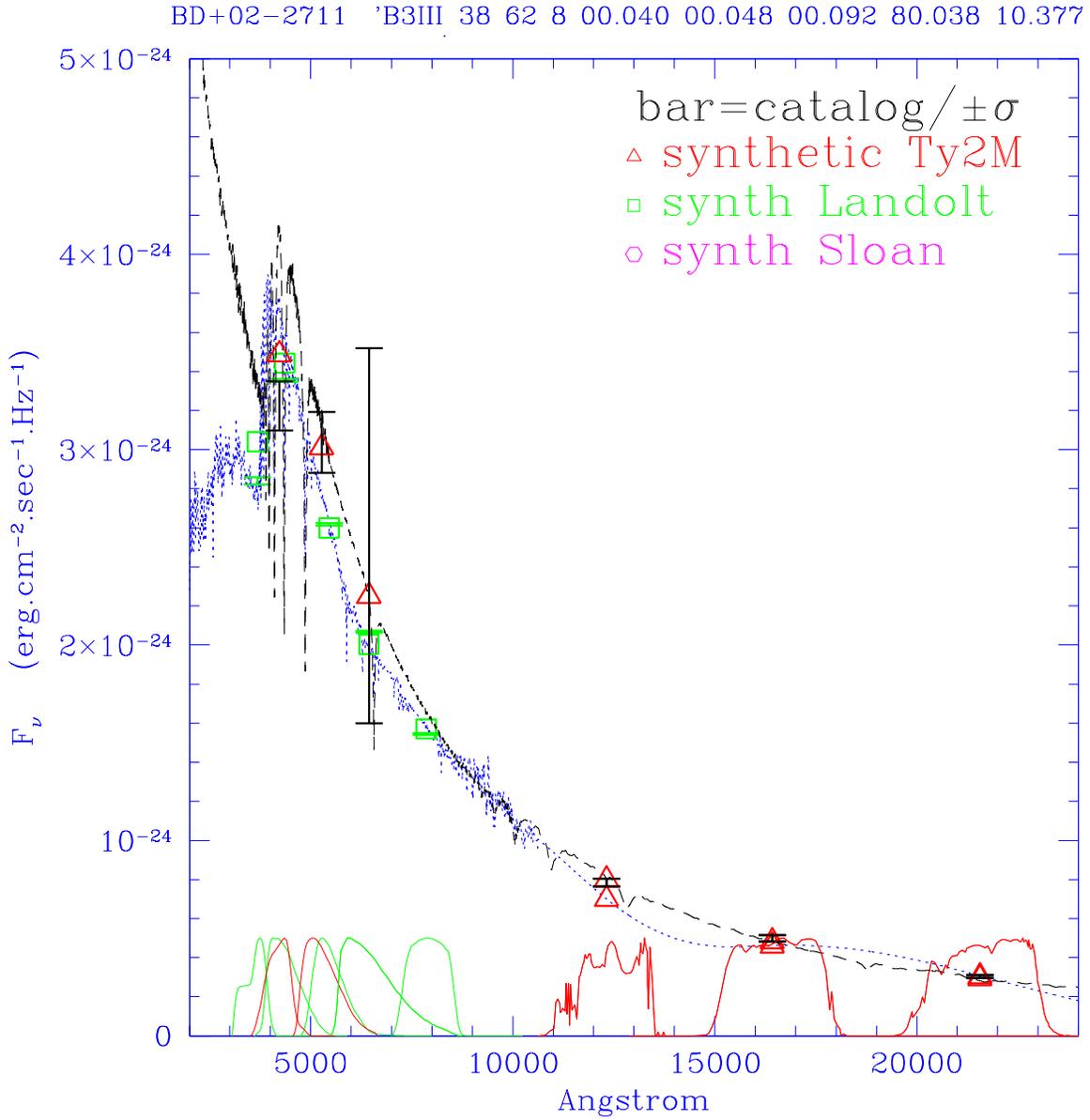}

\caption{The Landolt/2MASS (8-band, L2M) B3\,III fit is shown (lower,
blue dotted line) as $F(\nu)$ vs. wavelength.  The Tycho2/$R_N$/2MASS
(6-band, TNM) DA3-WD fit is shown as the overlaid black-dashed trace.
The long vertical black bar represents the NOMAD $R_N$ magnitude (in
its center) and error for the TNM fit, with red triangle symbol for
the synthesized flux. The green bar (above the fitted green
square) represents the Landolt R catalog magnitude and error for this
star. Symbols as before.}

\label{BD+02}
\end{figure}

Figure \ref{BD+02} shows the L2M fit for BD\,+02\,2711, a B3\,III at
5.2\,kpc, plotted as a blue dotted line, with RUI, Ruz and RTM values
of 0.04, 0.05 and 0.08 mag.  The S2M and 15-band fits are similar. The
6-band TNM fit of a GRW\,+70\,5824 (DA3) type White Dwarf at 4\,pc is
shown as the over-plotted black dashed line, with RUI, Ruz and RTM
values of 0.13, 0.14 and 0.07 mag respectively. The latter fit is a
better match to the Tycho2 bands, but is unconstrained at wavelengths
shorter than $B_T$, and is a poorer fit to the Landolt and Sloan
bands.  There is no objective prism type for BD\,+02\,2711 from
\citet{dl79}, but SIMBAD lists a type of B5\,V at a distance (from its
Hipparcos parallax) of 1.3\,kpc. An early B dwarf type was selected as
a high, but not top-ranked fit by all the spectral matches.

These examples were chosen to illustrate both the successes and
limitations of the spectral matching method.  They emphasize that
color or spectral type are matched better than luminosity class,
particularly when the input photometric errors are larger, but that
good magnitude fits can be obtained in most cases.

\begin{figure}
\epsscale{1.0}
\plotone{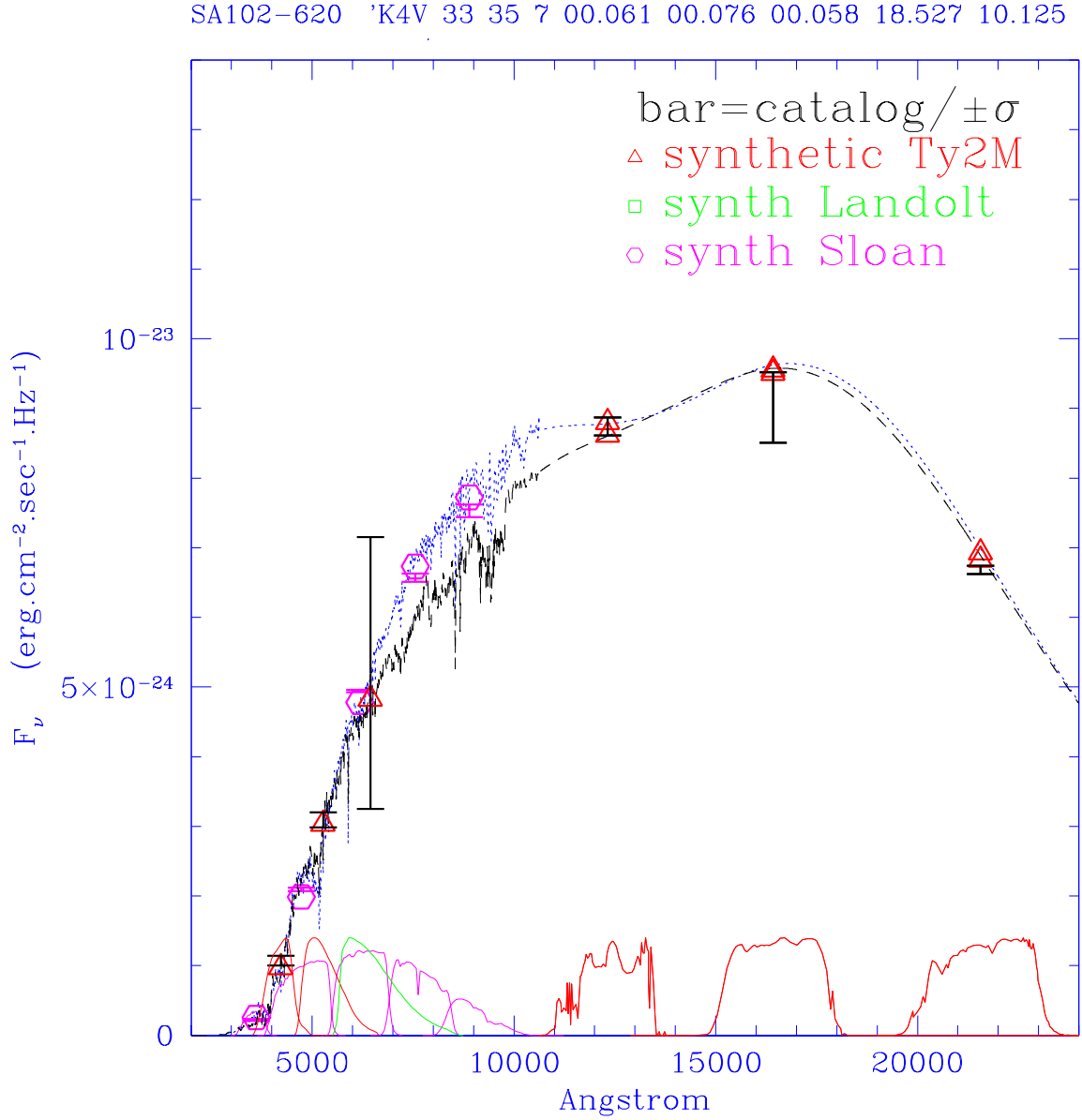}

\caption{The Sloan/2MASS (8-band, S2M) K4\,V fit for SA102-620 is
shown (blue dotted line) as $F(\nu)$ vs. wavelength.  The overlaid
black dashed spectrum is the rK1\,III fit obtained with 6
Tycho2/$R_N$/2MASS bands. Symbols as before.}

\label{SA102-620}
\end{figure}

Figure \ref{SA102-620} shows the first ranked S2M fit for SA102-620, a
K4\,V at 38\ pc. The K4\,V fit is also obtained as the second ranked
fits with the 15 band and the 8-band L2M fits, where a slightly higher
ranked K2\,III fit at 540\,pc was rejected in these two cases as being
beyond a calculated proper-motion-limited distance of 285 pc.  For the
S2M fit, RUI, Ruz and RTM values are 0.06, 0.08 \& 0.05 mag
respectively.  Over-plotted (black dashed line) is the second ranked
6-band TNM fit rK1\,III at 200\ pc, which is a bit bluer in the
optical region because of the $B_TV_TR_N$ data.  This latter fit has
RUI, Ruz and RTM values of 0.11, 0.13 and 0.06 mag respectively;
\citet{dl79} list an M0\,III type for SA102-620. SIMBAD lists a
K5\,III but with a parallax (from Hipparcos) of 22\,mas, implying a
distance of 45\ pc; ie. the star must be a dwarf. The K4\,V L2M and
S2M fits are best, but the TNM rK1\,III fit still matches the
magnitudes quite well.

\begin{figure}
\epsscale{1.0}
\plotone{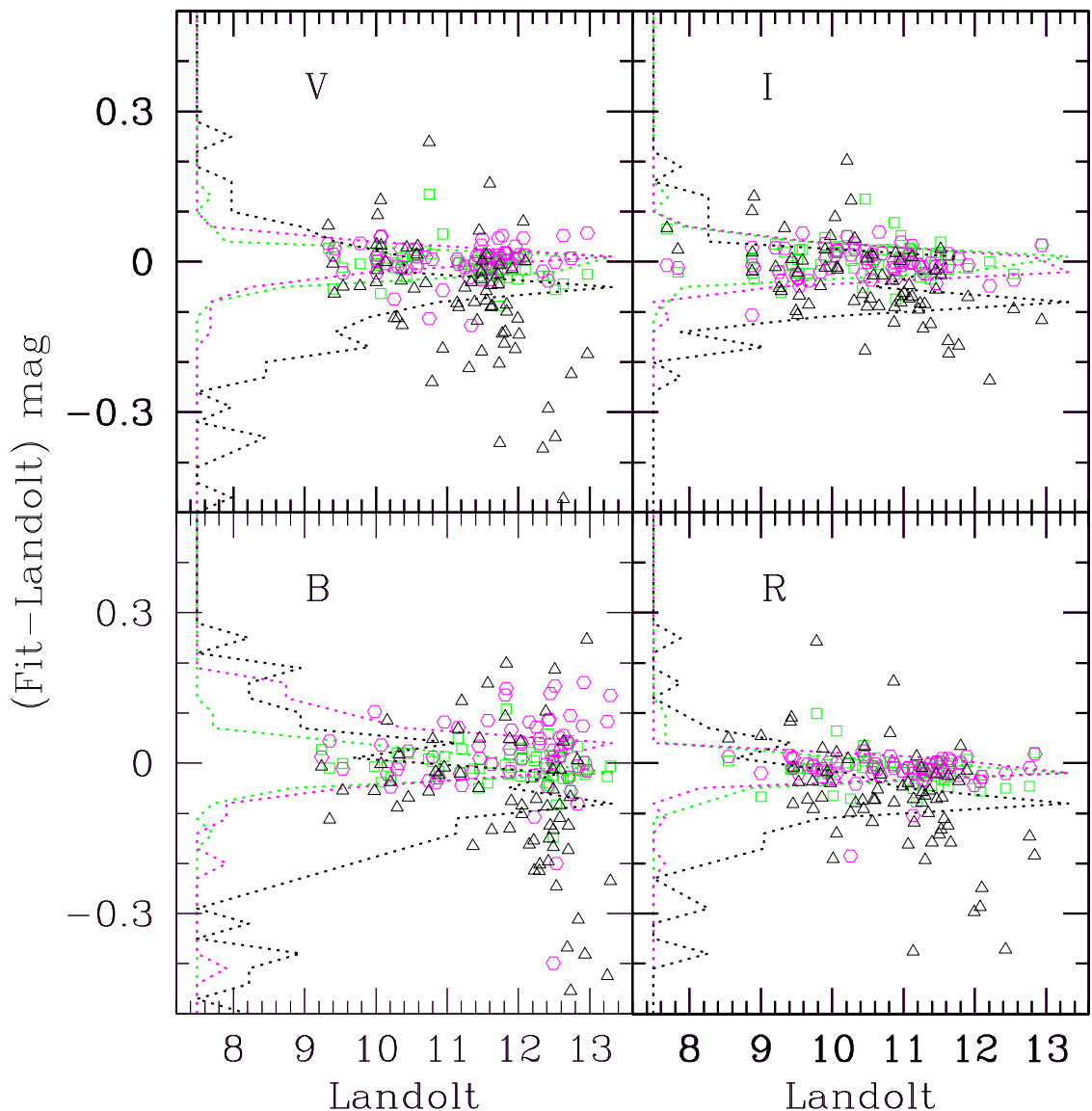}

\caption{Landolt $BVRI$ fits for 65 stars with Landolt, Sloan, Tycho2,
NOMAD and 2MASS standard magnitudes.  The fitted $BVRI$ magnitudes on
the ordinate are plotted as $Fit-Landolt$ vs. the catalog values on
the abscissa. $U$ fits are not shown.  Green squares and histogram indicate the L2M
fits obtained with five Landolt and three 2MASS magnitudes, magenta
circles and histogram indicate the S2M fits obtained with five Sloan and three 2MASS
magnitudes, and black triangles and the wider histogram indicate the TNM fits obtained with
$B_TV_TR_N$ and three 2MASS magnitudes.
The green and magenta histograms are narrow, indicating good fits to the zero-delta line.
The black histogram is wider as expected for these TNM fits. }

\label{BVRI_lstm}
\end{figure}

Figure \ref{BVRI_lstm} compares three types of fit for the {\em
LanSloCat} sample.  The fitted $BVRI$ magnitudes are plotted as
$Fit-Landolt$ on the ordinate vs. Landolt catalog values on the
abscissa. The electronic version of this figure is in color.  A plot
of $Fit-Sloan$ vs. Sloan catalog magnitudes shows similar dispersions.
The clipped sigmas, typically containing more than 92\% of the points,
are: 0.12, 0.03, 0.02, 0.03, 0.03 \& 0.13, 0.03, 0.03, 0.03, 0.04\,mag
for $UBVRI$ \& $u'g'r'i'z'$ bands respectively for the 8-band L2M
fits.  They are: 0.14, 0.06, 0.03, 0.02, 0.03 \& 0.15, 0.03, 0.02,
0.02, 0.03\,mag for $UBVRI$ \& $u'g'r'i'z'$ bands respectively for the
8-band S2M fits, and are: 0.21, 0.15, 0.12, 0.10, 0.08 \& 0.19, 0.15,
0.11, 0.09, 0.07\,mag for $UBVRI$ \& $u'g'r'i'z'$ bands respectively
for the 6-band TNM fits. 
These latter TNM sigmas, combined and
averaged for $UBVRI$ and $u'g'r'i'z'$ as $\sim$0.2, 0.15, 0.12, 0.10
\& 0.08\,mag, are what are reported as the typical $1\sigma$ error per
Tycho2 star per band in the abstract, and can be compared to the TNM
$g'r'i'z'$ fits reported in section \ref{SDSS-2}.

\subsubsection{Accuracy and consistency of the fits} \label{consistent}

It can be seen that there is good correspondence between the $UBVRI$
and $u'g'r'i'z'$ bandpass sigmas where we are able to compare them
directly. This is natural because the fits depend primarily on the
catalog errors and the spectral library errors, not on any mismatch
between Landolt and Sloan systems, see section \ref{accuracy}. In other cases we are only able to
measure the quality of fit in either Landolt or Sloan bands, but can
infer that the quality of fit for the other system will be
quantitatively similar.

Landolt $UBVRI$ magnitudes can also be derived in this simple case
with accurate Sloan $u'g'r'i'z'$ magnitudes, using the formulas from
\citet{jest05}. For these 65 stars this results in sigmas of 0.06,
0.05, 0.03, 0.04 and 0.04 for $UBVRI$ respectively. The spectral
matching method for L2M and S2M methods are therefore comparable to or
better than the \citet{jest05} method for $BVRI$ and $griz$ bands, but
{\em worse} for $U/u$.  The TNM fits are less accurate, but generally fit
better at BVR bands than the $B_TV_TR_N$ errors themselves. They therefore
offer a reliable fitting method in the absence of more accurate optical data.

Note that the Jester formulas are susceptible to errors in just one
filter, that can propagate to several fitted filter bands. The
spectral matching process is not immune to input catalog errors but,
by giving lower weight to discrepant points with larger photometric
errors, can provide more robust synthetic fits in all bands.

\begin{figure}
\epsscale{1.0}
\plotone{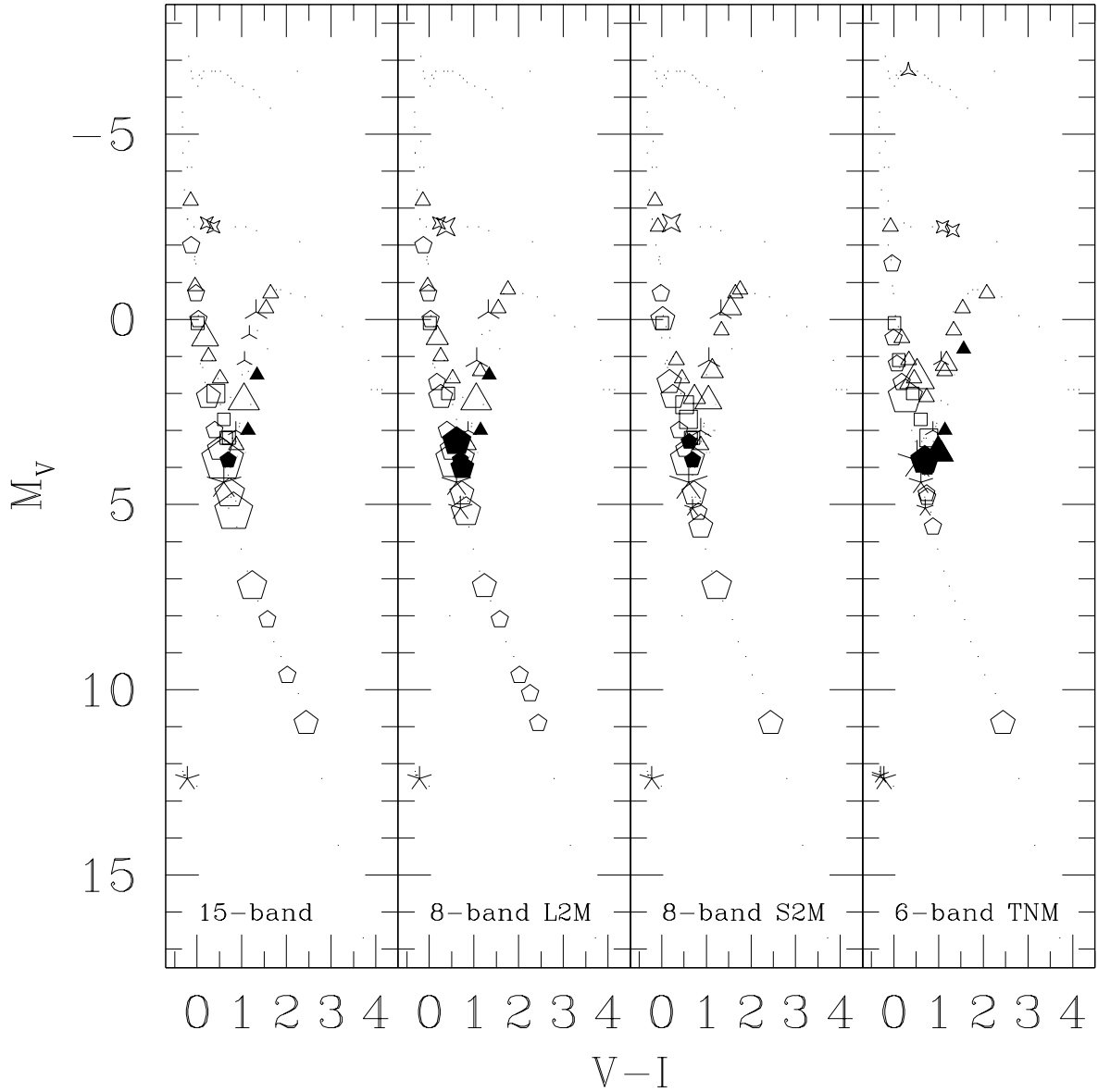}

\caption{HR diagram for 65 Landolt/Sloan standards shows our fitted
types as adopted absolute V magnitudes vs. spectral library $V-I$
colors, for four different fits. Same symbols as fig \ref{HRland}.
The area of each symbol is proportional to the frequency that library
spectrum was selected.}

\label{HRlstm}
\end{figure}

The fitted spectral types can vary according to the input data
matched, but the results are generally consistent.  Figure
\ref{HRlstm} illustrates our derived distributions of stellar types
and luminosities for 65 Landolt/Sloan standards, binned into 141
library spectral types, and plotted as adopted absolute V magnitudes
against their $V-I$ colors.  The same symbols for different
Luminosity classes and different metallicities are used from figure \ref{HRland},
and the symbol area is proportional to the number of stars in
each library type bin.

The first fit on the left is for all 15 Landolt, Sloan \& Tycho2/2MASS
bands.  The second fit is for the 8-band L2M fits, third for the
8-band S2M fits, and fourth on the right for the 6-band TNM fits.

These fits and their distributions of selected spectral types and
luminosities are different in detail, but similar statistically, and
produce closely similar magnitude fits.

The HR diagram on the right illustrates that TNM fits show a slight
preference for giant rather than dwarf types.  The TNM fits also
select one supergiant for SA97-351, an F0\,I at a distance of 19\,kpc
(just inside our distance limit), rather than an A7\,V at 342\,pc
(L2M) or F0\,III at 553\,pc (S2M).  \citet{dl79} list an objective
prism spectral type of A0 for SA97-351.  

There are 27 matches and 19 close matches between the L2M and S2M
fits.  There are 14 matches and 21 close matches between the S2M and
TNM fits.  In all cases the fitted spectra and derived magnitudes are
similar, as shown in figure \ref{BVRI_lstm}.

The difference in optimal fits, depending on the number and accuracy
of filter photometry available, illustrates both the strengths and
limitations of the spectral matching process, and of the spectral
library.  The library quantization is too fine for this purpose in
some places, and insufficiently sampled in others.  In general, the
spectral matching process is better at determining the spectral type
than the luminosity class. 

The differences between accurate and lower quality photometric data
are apparent. Nevertheless the spectral matching process works well
even for poorly determined optical photometry, where the more accurate
2MASS data help constrain the fitted spectra in both type and
magnitude scale.

Better spectral type discrimination is possible when more accurate
filter data are available, but the fitted magnitudes are similar in
all these cases, within the quoted errors. Fitted spectral types are
more accurate than Luminosity class. The fitted distance estimates
can be used as a sanity check on the derived fit.

\section{Secondary SDSS Standards} \label{SDSS-2}

\subsection{SDSS PT observations}

\citet{tuc06} \& \citet{dav07} have published lists of almost 3.2\,M
SDSS observations in $g'r'i'z'$\footnote{{\tt
http://das.sdss.org/pt/}}. These were observed on the US Naval
Observatory photometric telescope (PT), and used to calibrate the SDSS
survey scans\footnote{{\tt
http://www.sdss.org/tour/photo\_telescope.html}}. They are more
accurate than survey data, as evidenced by the consistency of their repeated observations and (small) errors.
We use them here as secondary standard
calibrators. We first combined the lists to produce a list of
$\sim$1\,M repeated observations, with photometric errors being the
maximum of any single observation, or the rms of the average if that
was greater.  These were matched against 2MASS stars within 3-arcsec
to produce a slightly shorter list of stars which reach as faint as
g'$\sim$21\,mag, and are discussed in section \ref{second}.

This list was further matched against the Tycho2 and NOMAD catalogs,
to produce a smaller list of 10,926 SDSS-PT standards with
$B_TV_TR_N$ and reaching as faint as $g' \simeq 13.5$ mag.  These were
fitted two ways: i) S2M with 7-bands $g'r'i'z'$ \& $JHK_{2/S}$, and
ii) TNM with 6-bands $B_TV_TR_NJ_2H_2K_{2/S}$, and their results
compared with the PT standard values.

\begin{figure}
\epsscale{1.0}
\plotone{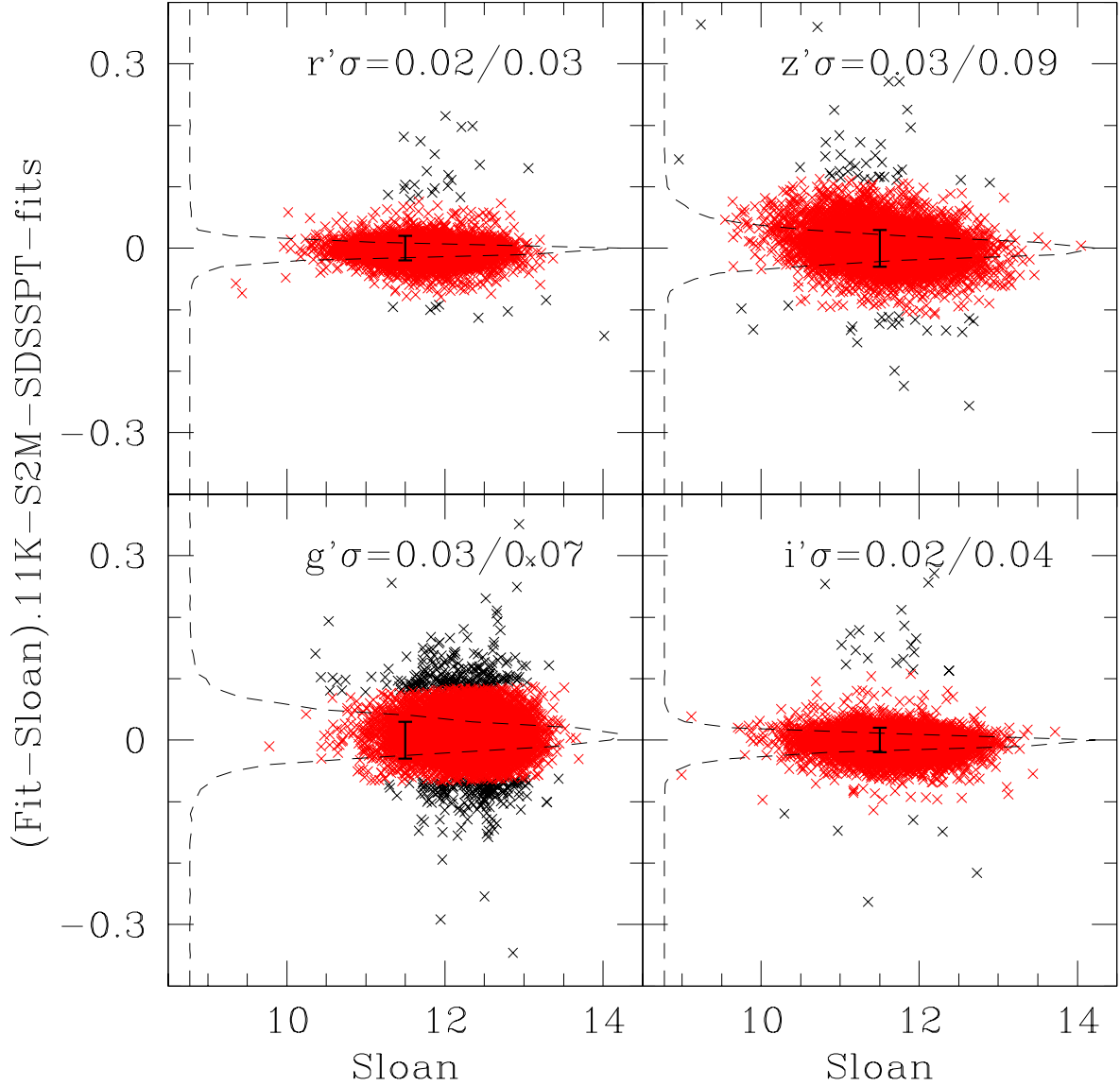}

\caption{S2M fits with seven $g'r'i'z'-JHK_{2/S}$ bands for $\sim$11,000
SDSS PT standards with 2MASS, Nomad \& Tycho2 photometry, plotted as
(Fit-Sloan) vs. Sloan magnitudes. Black crosess are points rejected by
the sigma clipping, and grey (red) points are those remaining after sigma
clipping. The clipped points overlap, but the number distributions
for all the points about the zero-delta line are indicated by the
dashed histograms. The sigmas indicated in the panel captions are for
the clipped / and all the stars in each band, with $\pm 1\sigma$ error
bars drawn for the former. }

\label{griz-spt7}
\end{figure}

Figure \ref{griz-spt7} shows the S2M $g'r'i'z'$ fits to the
$\sim$11,000 SDSS-PT/2MASS stars with Tycho2/Nomad data, with sigmas of
0.03, 0.02, 0.02 and 0.03 in $g'r'i'z'$ respectively, after clipping
outliers, but retaining more than 96\% of the points in each band. The
sigma in $g'$ is 0.07\,mag for $\sim$11,000 stars, but the grey (red) dots
show a sigma of 0.03\,mag after clipping about 400 outliers.  This
illustrates typical $1\sigma$ errors for accurate SDSS data coupled
with 2MASS data.  The histograms illustrate the number distributions
of errors about the mean for all the points where they become
overlapped and blurred.

From section \ref{consistent} we infer that the errors for $BVRI$ are similar. 
The sigmas for fitted $JHK_{2/S}$ are 0.04, 0.04 \& 0.03\,mag respectively.

\begin{figure}
\epsscale{1.0}
\plotone{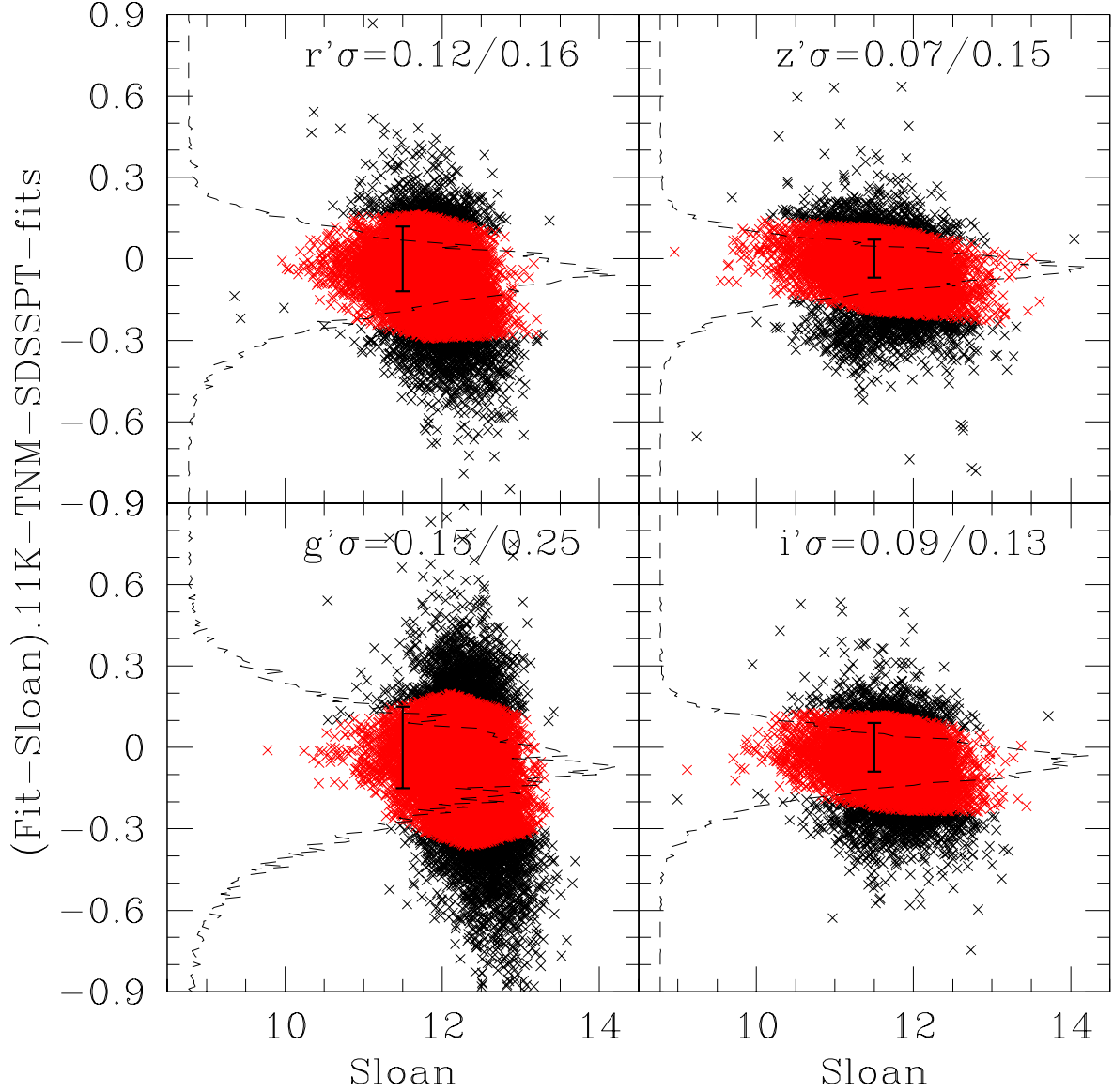}

\caption{TNM fits with six $B_TV_TR_N-JHK_{2/S}$ bands for the same
$\sim$11,000 SDSS-PT standards plotted as (Fit-Sloan) vs. Sloan
magnitudes.  Note the vertical scale change from the previous
figure. Black crosses show all the stars. The grey points in the middle
(red in the electronic version) show 80--90\% of stars remaining after
sigma clipping. Sigmas after and before clipping are indicated.  The
histograms show the number distribution of errors about the mean for
all the points.  There may be some bias towards brighter fitted
magnitudes in at least the $g'r'$ bands, but the errors are dominated
here by the input Tycho2/Nomad catalog accuracies.}

\label{griz-spt10}
\end{figure}

Figure \ref{griz-spt10} shows the corresponding TNM fits to these
stars, with clipped and unclipped sigmas of 0.15/0.25, 0.12/0.16,
0.09/0.13 and 0.07/0.15\,mag in $g'r'i'z'$ respectively, after
clipping outliers (shown in black) but leaving 78\% of the grey (red)
points in g' and more than 88\% of the grey (red) points in the other
three colors. We again infer that these illustrate typical $1\sigma$
errors for both $griz$ and $BVRI$ fits to the Tycho2 catalog when
coupled with NOMAD and 2MASS magnitudes, and note their similarity to
typical TNM errors quoted in section \ref{lscex} and the abstract.  The histograms show the
number distribution of errors about the mean for all the TNM fits to
$\sim$11,000 SDSS-PT/2MASS/Tycho2 stars.

The TNM fit errors quoted here for $r'i'z'$ are similar to those found by
\citet{ofe08}, but for a larger magnitude range.  
The error quoted for $g'$-band is slightly larger than previously
reported, but for a wider fitted magnitude range.  The
sigmas for fitted $JHK_{2/S}$ are 0.03, 0.02 \& 0.02\,mag respectively.

\subsection{Southern SDSS standards}

\begin{figure}
\epsscale{1.0}
\plotone{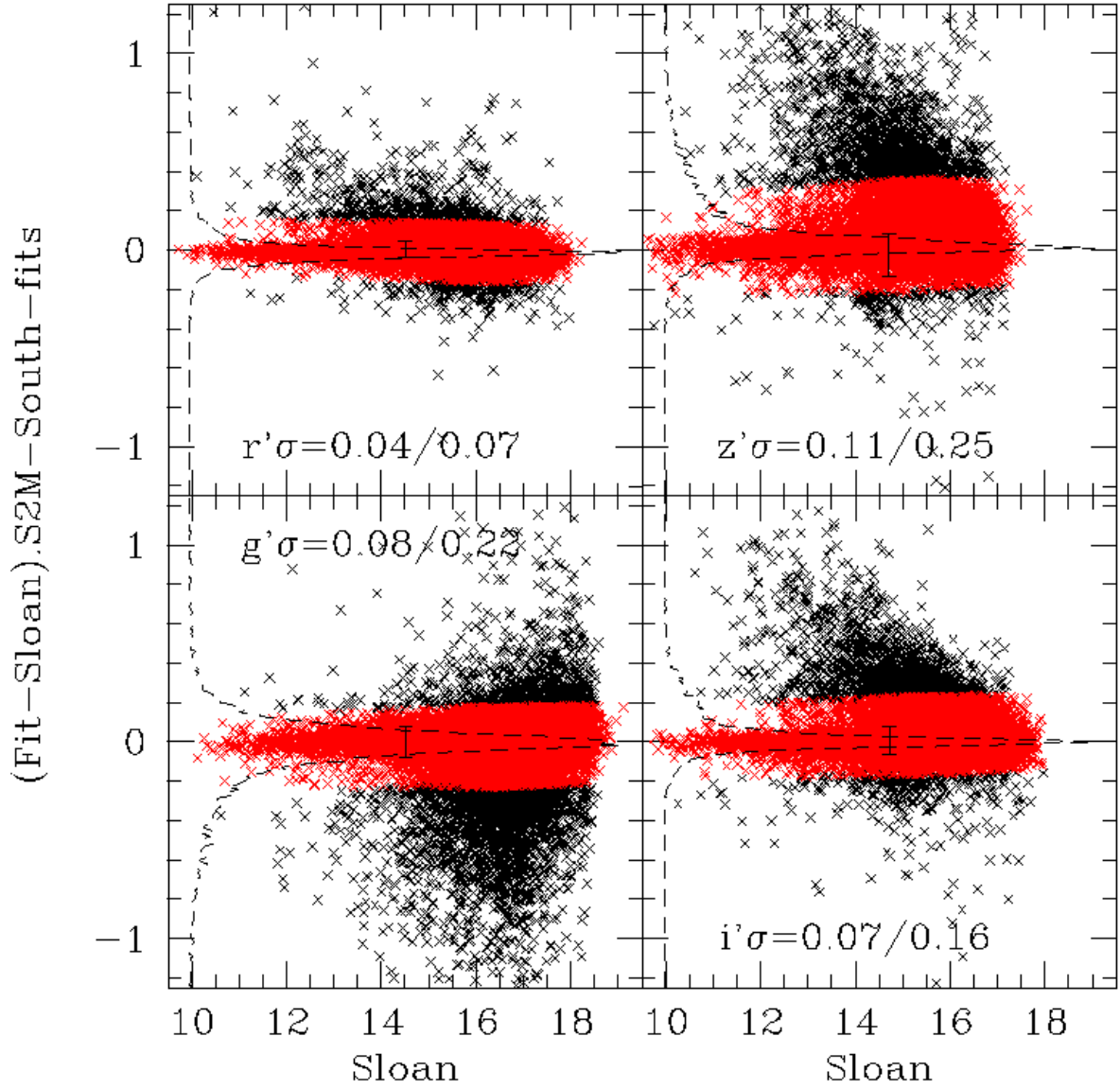}

\caption{S2M fits with eight $ugriz-JHK_{2/S}$ bands for $\sim$16,000
Southern SDSS standards, to g'$\sim$19\,mag. Clipped/unclipped sigmas, histograms and
error bars as before.}

\label{griz-sss7}
\end{figure}

\citet{smi05}\footnote{{\tt http://www-star.fnal.gov/}} have published
a list of $\sim$16,000 southern SDSS standards, which includes some
repetition. We matched 15,673 of these to 2MASS catalog values and
fitted them in eight S2M bands: $u'g'r'i'z'$ and $JHK_{2/S}$. The
clipped and unclipped $1\sigma$ fits are 0.22/0.51, 0.08/0.22,
0.04/0.07, 0.07/0.16 and 0.11/0.25\,mag for $u'g'r'i'z'$ respectively,
for all Sloan Southern stars ranging up to $g \sim 19$\,mag, as shown
in figure \ref{griz-sss7}.  The vertical histograms show the number
distribution of errors about the mean for all these Southern Sloan
stars, and indicate that our fits produce an excess of rejected fits
{\em fainter} than the catalog values in $z'$ and, to a lesser extent,
in $i'$.

The $JHK_2$ fits are not shown, but the $J_2$ fit shows a
corresponding excess of fitted points {\em brighter} than the catalog
values, whereas the $HK_2$ fits are symmetrically distributed about
the mean-delta (zero) line.  Thus our fits to southern Sloan $i'z'$ are being
pushed slightly fainter at the red end by the usually reliable 2MASS data.

There are $\sim$6000 stars with $g'<16$:
for these stars the clipped sigmas are 0.13, 0.07, 0.04, 0.06 and 0.09
in $u'g'r'i'z'$ respectively, ie. not as good as for the SDSS-PT stars
in figure \ref{griz-spt7}, but for a larger magnitude range in this
case.

\begin{figure}
\epsscale{1.0}
\plotone{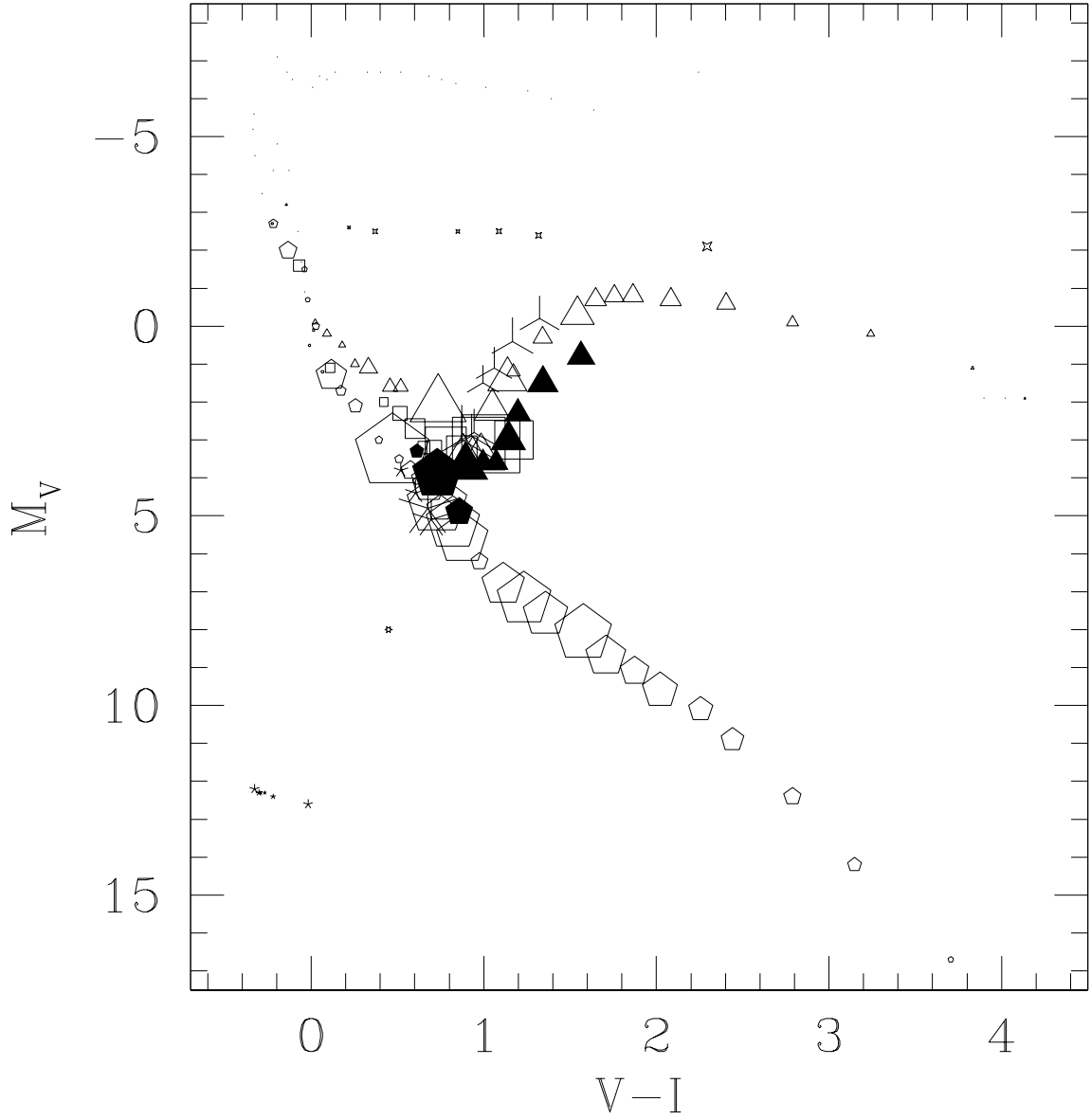}
\caption{HR diagram for $\sim$16,000 SDSS Southern standards. Same symbols as fig \ref{HRland} }
\label{HRsss7}
\end{figure}

The HR diagram for this fit is shown in figure \ref{HRsss7}. Compared
with the Landolt distribution in figure \ref{HRland}, this shows a
slightly less populated lower main sequence, more stars near the main
sequence turnoff, a slightly better defined red giant branch out to
M8\,III, some ``horizontal branch'' type giants but no supergiants.

\begin{figure}
\epsscale{1.0}
\plotone{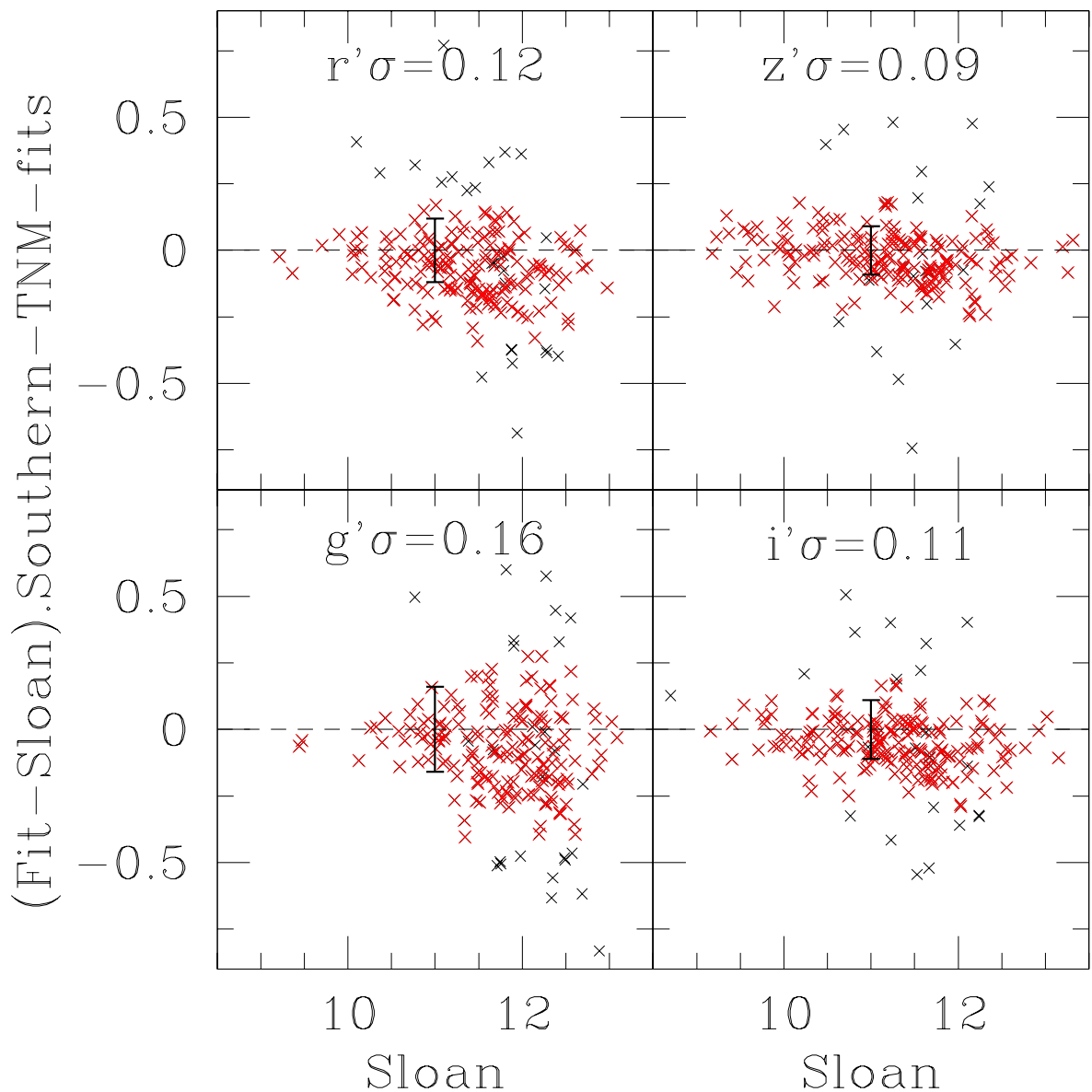}
\caption{TNM fits with six $B_TV_TR_N-JHK_{2/S}$ bands for 201
Southern SDSS standards with Nomad \& Tycho2 magnitudes.  The fits for
$g'r'i'z'$ are plotted as (Fit-Sloan) vs. Sloan magnitudes, with the
clipped data shown as grey (red) crosses, with their clipped sigmas indicated
in the panels, and as error bars. The black crosses were rejected by
the sigma clipping process.}
\label{griz-sss10}
\end{figure}

We have further matched this list to Tycho2 $B_TV_T$ and NOMAD $R_N$
magnitudes, resulting in a list of only 201 Southern stars as most of the
Southern SDSS standards are too faint to have Tycho2 matches.  The
comparison of fitted $griz$ magnitudes to standard values for fits
using 6 TNM bands is shown in figure \ref{griz-sss10},
with sigmas of 0.23, 0.16, 0.12, 0.11 and 0.09\,mags in $ugriz$
respectively, and comparable to the TNM fits to SDSS PT stars above for a similar magnitude range.

\section{Catalogs} \label{catalogs}

\subsection{Landolt and Sloan Fitted data}

Landolt standards including updated coordinates, magnitudes and
errors, $JHK_{2/S}$ 2MASS data and UKIRT $Z_VY_VJHK_{MKO}$ data where
known are listed for 594 stars in electronic table {\em LandoltCat}.
Table {\em LandoltFit} contains the spectrally fitted rank, $chi2$,
RUI, RBI, library number and type, fitted magnitudes
$B_TV_TJ_2H_2K_2UBVRIZ_VY_V$, $u'g'r'i'z$, and distances in parsec.

Similar data are contained in tables {\em SloanCat} and {\em
SloanFit} for Sloan standards, and in tables {\em LanSloCat} and
{\em LanSloFit} for stars common to both lists, as described in
section \ref{primary}.

\subsection{Secondary standard fitted data} \label{second}

Electronic table {\em SDSSPTFit} contains coordinates, standard
magnitudes and errors in $g'r'i'z'$, number of repeated PT
observations, merged $J_2H_2K_2$ 2MASS data, and fitted Rgz, R2M
(2MASS), library types, $UBVRI-Z_VY_V$ and $u'g'r'i'z'$ magnitudes and
distances for $\sim$1\,M SDSS-PT stars.

\begin{figure}
\epsscale{1.0}
\plotone{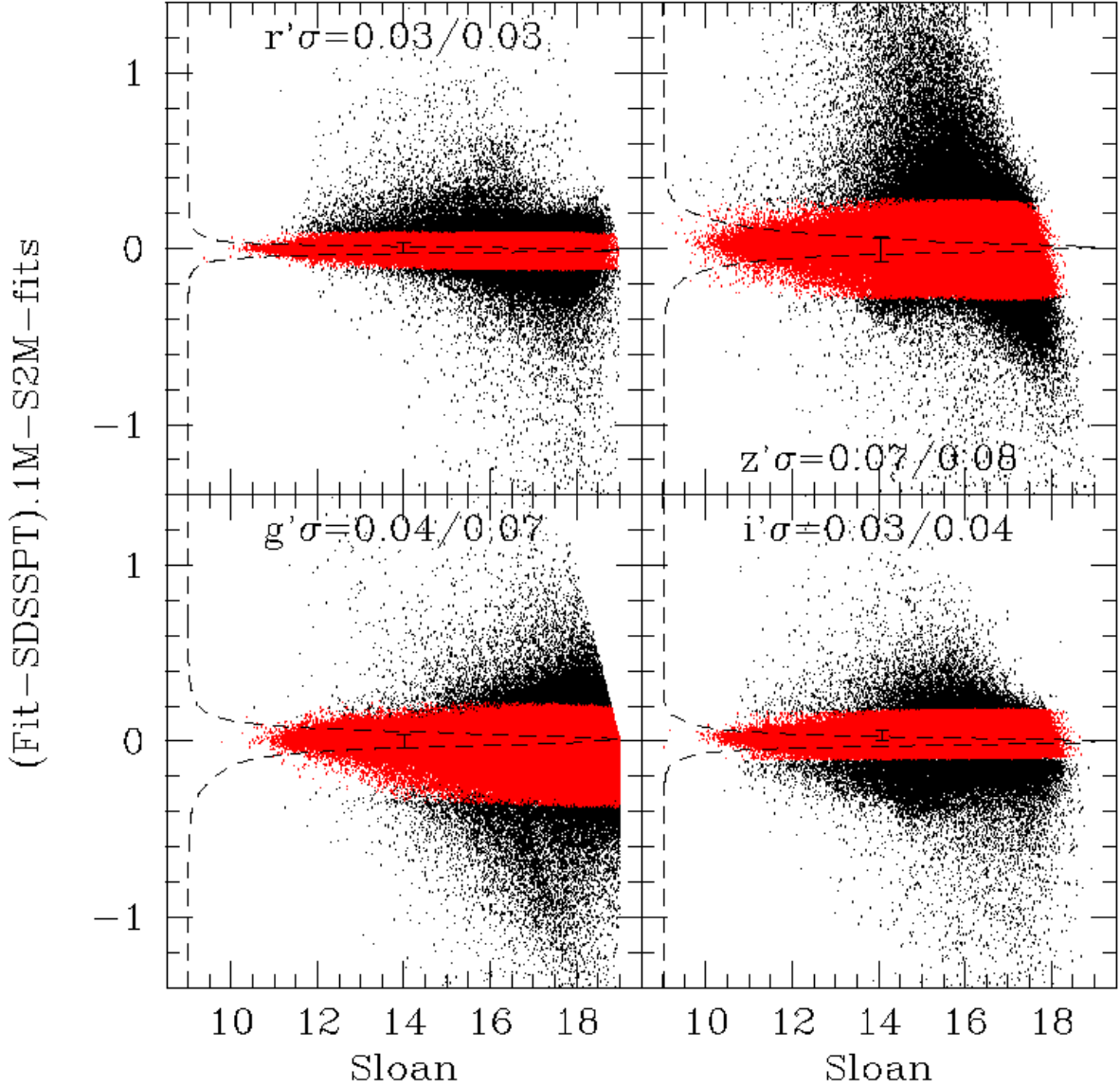}

\caption{S2M fits with seven $g'r'i'z'JHK_{2/s}$ bands for $\sim$1\,M
SDSS-PT standards with 2MASS magnitudes, to g'$\sim$19\,mag.  plotted
as (Fit-SDSSPT) vs. SDSSPT magnitudes in separate panels for
$g'r'i'z'$.  Black dots show all the rejected stars. The grey bands through
the middle (red in the electronic version) shows 91\% of stars left
after sigma clipping in $g'$ and $>$95\% of stars left in $r'i'z'$.
The dashed histograms show the number distribution of errors
about the mean for all the stars.  The annotated sigmas are to a
limiting magnitude of 16 in each band / and for the full magnitude
range.}

\label{griz-spt7-1m}
\end{figure}

Figure \ref{griz-spt7-1m} shows the 7-band S2M fits for these stars.
The outer, black dots show all the rejected stars, and the lighter grey,
narrower band (red in the electronic version) contain more than 90\%
of the stars after sigma clipping. The sigma-clipped bands widen
somewhat with magnitude, particularly for g' and z'.  The number
distributions of these points are strongly peaked to the center (zero)
line, and are shown as dashed histograms.  The sigmas are 0.04, 0.03,
0.03 \& 0.07 to a limiting magnitude of 16 in $g'r'i'z'$
bands respectively, and are 0.07, 0.03, 0.04 \& 0.08, to the sample
limiting magnitudes of about 19, 19, 18.5 and 18 in $g'r'i'z'$
respectively.  Infrared sigmas (not shown) are 0.04, 0.06 and 0.07 for
$JHK_2$ respectively.

\begin{figure}
\epsscale{1.0}
\plotone{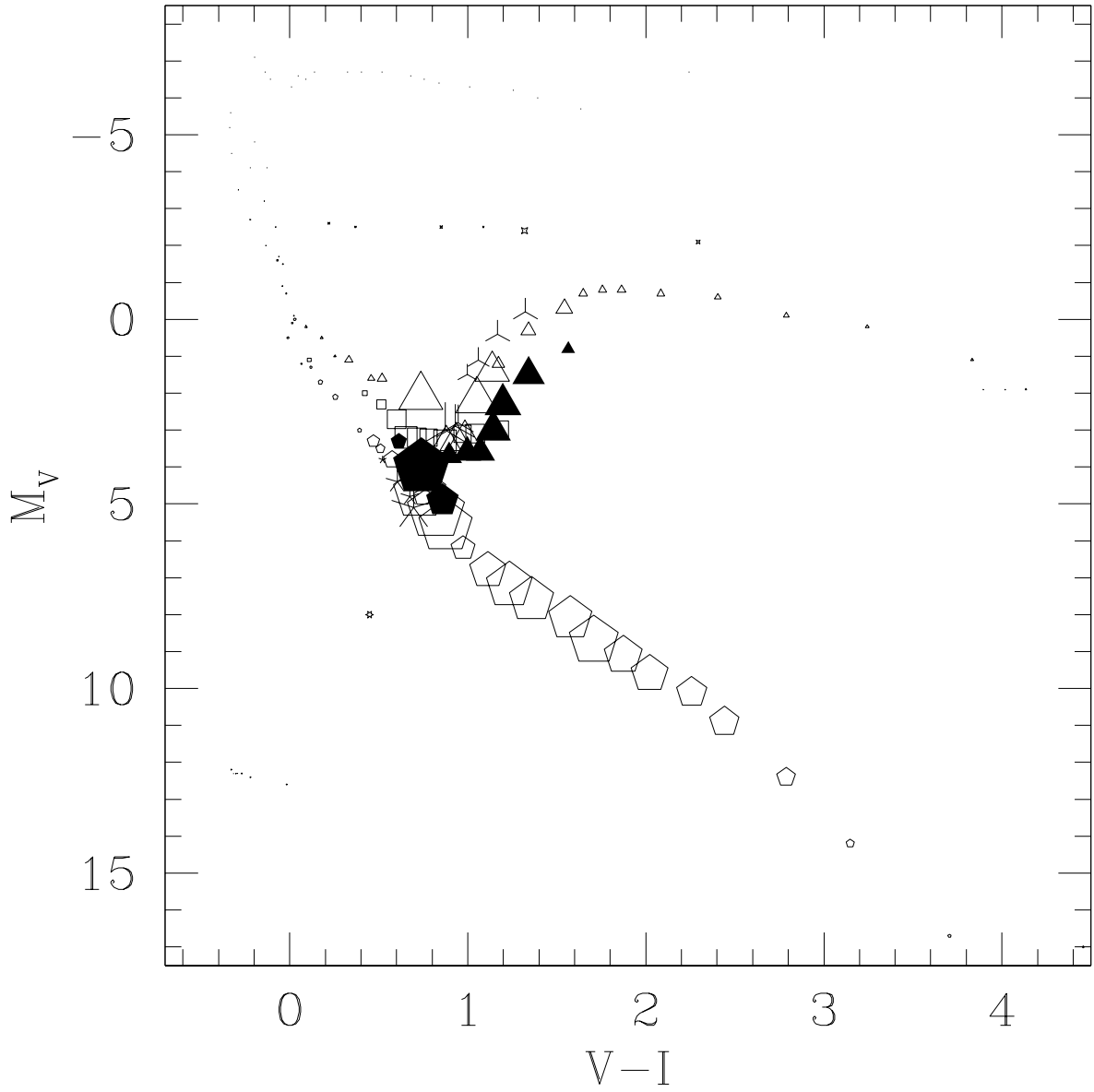}
\caption{HR diagram for $\sim$1\,M SDSS-PT standards. Same symbols as fig \ref{HRland} }
\label{HRspt7-1m}
\end{figure}

Figure \ref{HRspt7-1m} shows the distribution of these stars as
absolute magnitude against V-I color.  The larger SDSS-PT sample shows
a more heavily populated lower main sequence than for the Southern
Sloan standards.

Electronic table {\em SDSSSouthFit} contains similar fitted data for
15,673 Sloan Southern standards, where the name contains the original
sexagesimal coordinates.  Missing input magnitudes are listed as
-9.999 with associated errors of +9.999

\subsection{Spectrally matched Tycho2 catalog}

The catalog of 2.4\,M fitted Tycho2 stars is published in electronic
table {\em Tycho2Fit}.  It contains for each star: Tycho2 coordinates,
proper motions in RA/DEC in mas/yr, 
$B_TV_T$ catalog magnitudes and errors, NOMAD $R_n$
magnitudes and 2MASS $J_2H_2K_{2/S}$ magnitudes, mean errors and Quality Flag,
followed by fitted values for rank, $chi2$, RTM (mag), number and type of
matching spectrum, fitted $B_TV_T$, $UBVRI-ZY$ magnitudes on the {\em vega}
system, fitted $u'g'r'i'z'$ magnitudes on the AB system, and Distances in pc.

Because the Tycho2 catalog covers a wide range of optical magnitudes
and colors, it also covers a wide range of 2MASS magnitudes.  We have
included bright 2MASS sources which have larger errors due to
saturation effects, typically brighter than 5\,mag in $J$, $H$ or
$K_S$\footnote{{\tt
http://www.ipac.caltech.edu/2mass/releases/allsky/doc/sec2\_2.html}}. In
these cases (about 40,000 entries with 2MASS quality flag 'C' or 'D') the fits tend to be dominated by the
optical rather than the infrared bands.  About 11,000 entries lack mean
error information (quality flag 'U'); we assign 0.04, 0.05 and
0.06\,mag mean errors for $JHK_S$ respectively to enable our spectral
matching procedure to work with the quoted 2MASS upper magnitude limits.
About 570 entries lack magnitude information (quality flag 'X') in one
or more 2MASS bands; for these entries our catalog shows -9.999\,mag, with
an error of +9.999. This permits the matching process to proceed with negligible input
from the affected band.  The 2MASS quality flags are included in the
electronic catalogs, and most entries are 'AAA'.

\begin{figure}
\epsscale{1.0}
\plotone{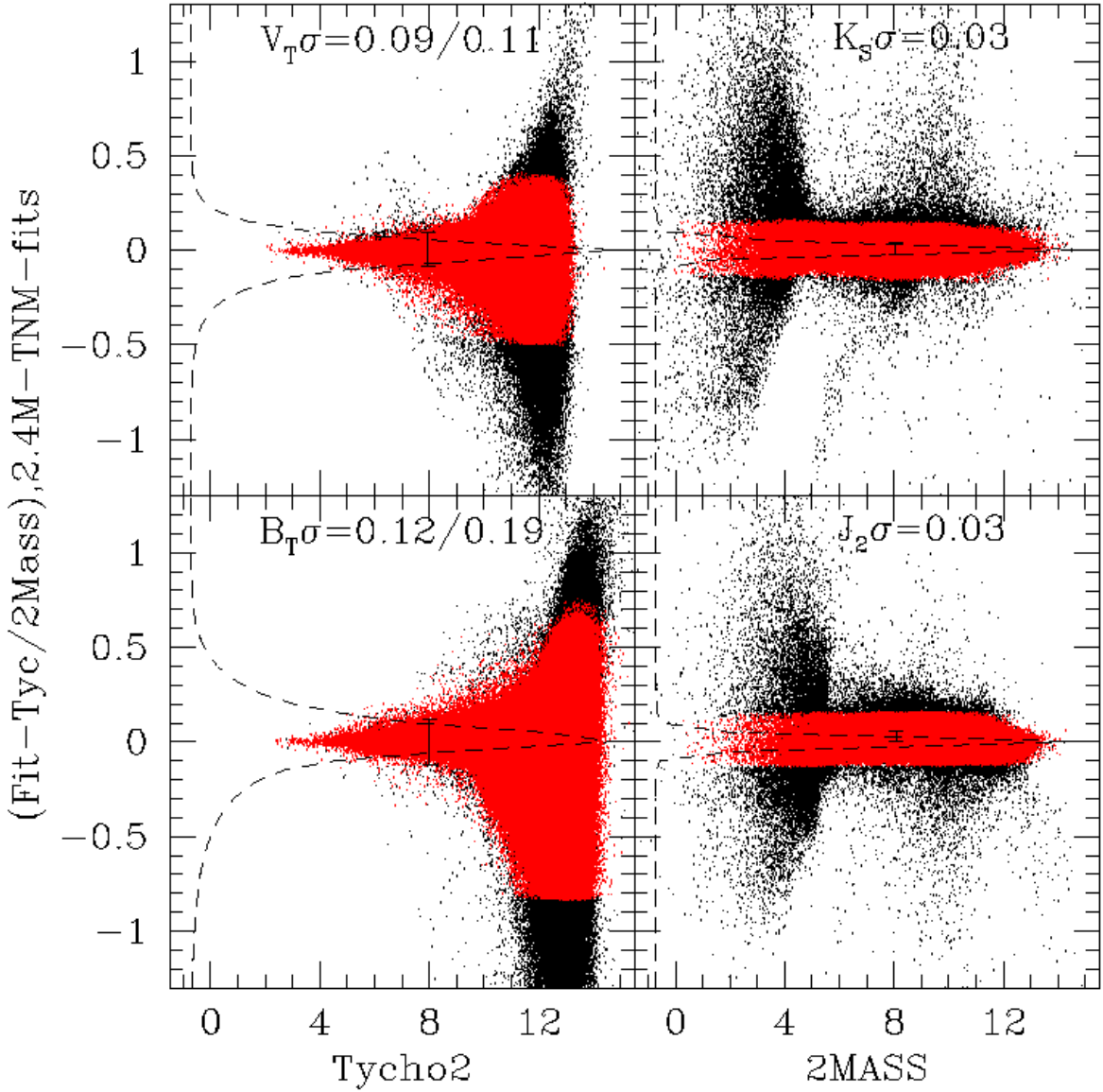}

\caption{TNM fits with six $B_TV_TR_NJ_2H_2K_2$ bands for $\sim$2.4\,M
Tycho2 stars with NOMAD $R_N$ and 2MASS $JHK_2$ magnitudes, plotted as
(Fit-catalog) vs. Tycho/2MASS catalog values on the abscissa.  The
Tycho2 catalog magnitudes vary between $\sim$2.4 to 16.5 mag.  The
2MASS magnitudes reach as bright as -4.5 mag, but with significant
errors there.  Black dots show all the stars rejected by a $3\sigma$
clip to give the fitted dispersions shown. The grey bands through the
middle (red in the electronic version) show 95\%, 96\%, 99\% and 99\%
of all stars left after sigma clipping in $B_TV_TJ_2K_2$ respectively.
The dashed histograms, which show the number distributions of errors
about the zero delta line for all the stars, are quite strongly peaked
around this line. The histograms are only slightly affected by the
rejected points as there are comparatively few of them. The annotated
sigmas are to a limiting magnitude of 13 \& 12.5\,mag in $B_TV_T$ / and
for the full magnitude range in each band.}

\label{BVJK-TNM}
\end{figure}

The fitted optical and infrared magnitudes are compared to the catalog
values in figure \ref{BVJK-TNM}, which illustrates the magnitude range
in the optical (+2.4 to $\sim$15\,mag) and infrared (-4.5 to $\sim$15\,mag).
A $3\sigma$ clip has been applied to all six bands, leaving 95\%,
96\%, 82\%, 99\%, 99\% and 99\% of values in $B_TV_T$, $R_N$ (matched
to fitted Landolt R) and $JHK_2$ bands respectively. The resulting
sigmas are 0.19, 0.11, 0.26, 0.03, 0.03 and 0.03\,mag in
$B_TV_TR_NJHK_2$ respectively, comparable with the quoted errors in
these bands. The sigmas are 0.12 \& 0.09\,mag for $B_TV_T$ to 13 and 12.5\,mag respectively.
Our fitted $B_TV_T$ magnitudes may be more accurate
than the catalog values, as they are tied to the more accurate 2MASS
photometry via the spectral matching process.  Both rejected and
included points are shown, as described in the figure. The fitted
values for $B_TV_T$ are listed in electronic table {\em Tycho2Fit};
other values can be derived by adding the library magnitude from {\em
LibMags} for the appropriate band and spectral type to the fitted V
magnitude.

\begin{figure}
\epsscale{1.0}
\plotone{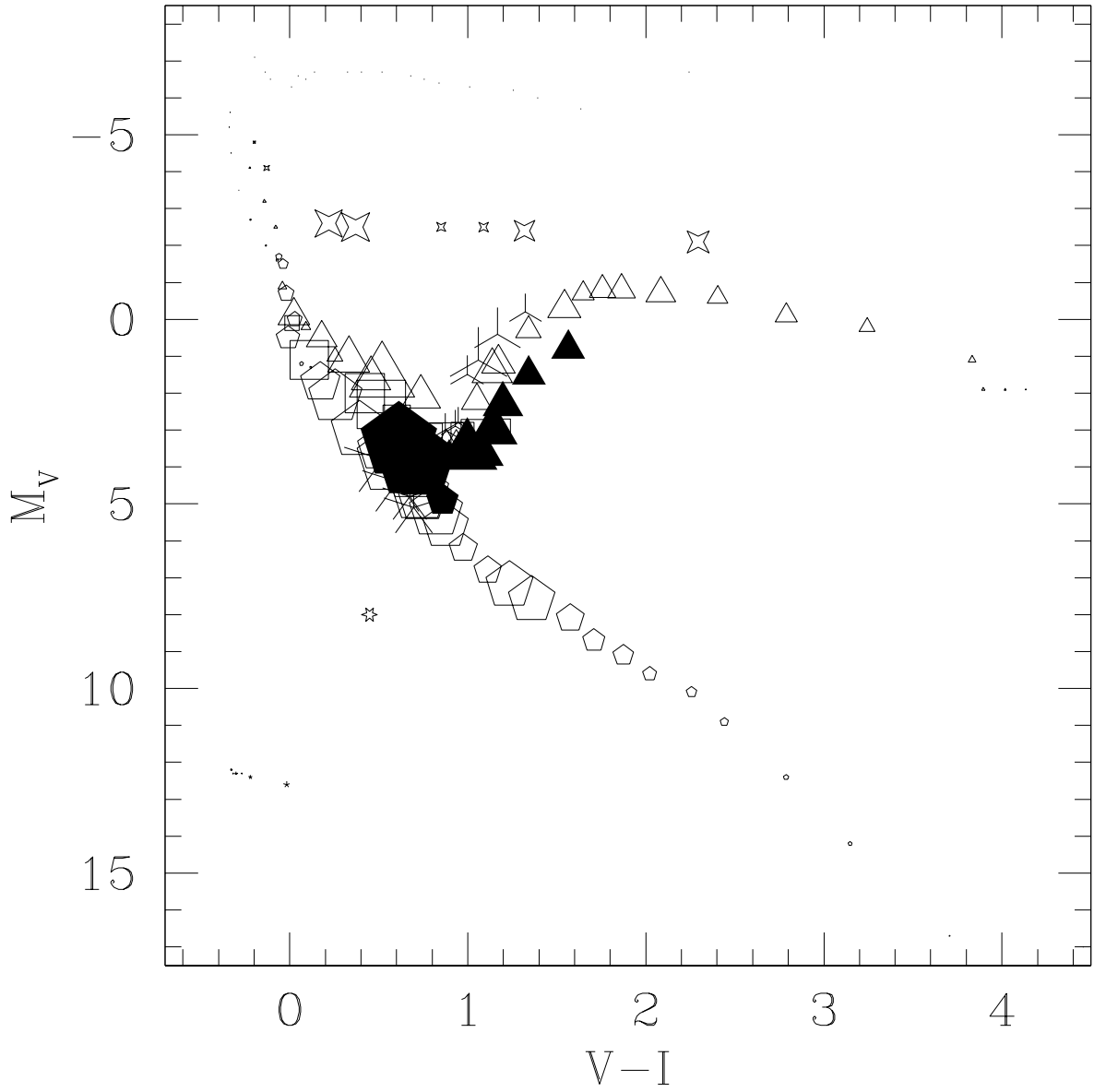}
\caption{HR diagram for $\sim$2.4\,M Tycho2 stars. Same symbols as fig \ref{HRland} }
\label{HRty2}
\end{figure}

Figure \ref{HRty2} shows the HR diagram derived for 2.4\,M Tycho2
stars from these fits, which shows a less populated lower main
sequence and a giant branch more dominated by solar metallicity stars
then the Landolt, SDSS-PT or SDSS-Southern standards, and about 3,340
double star 'DA1/K4V' fits.

\begin{figure}
\epsscale{1.0}
\plotone{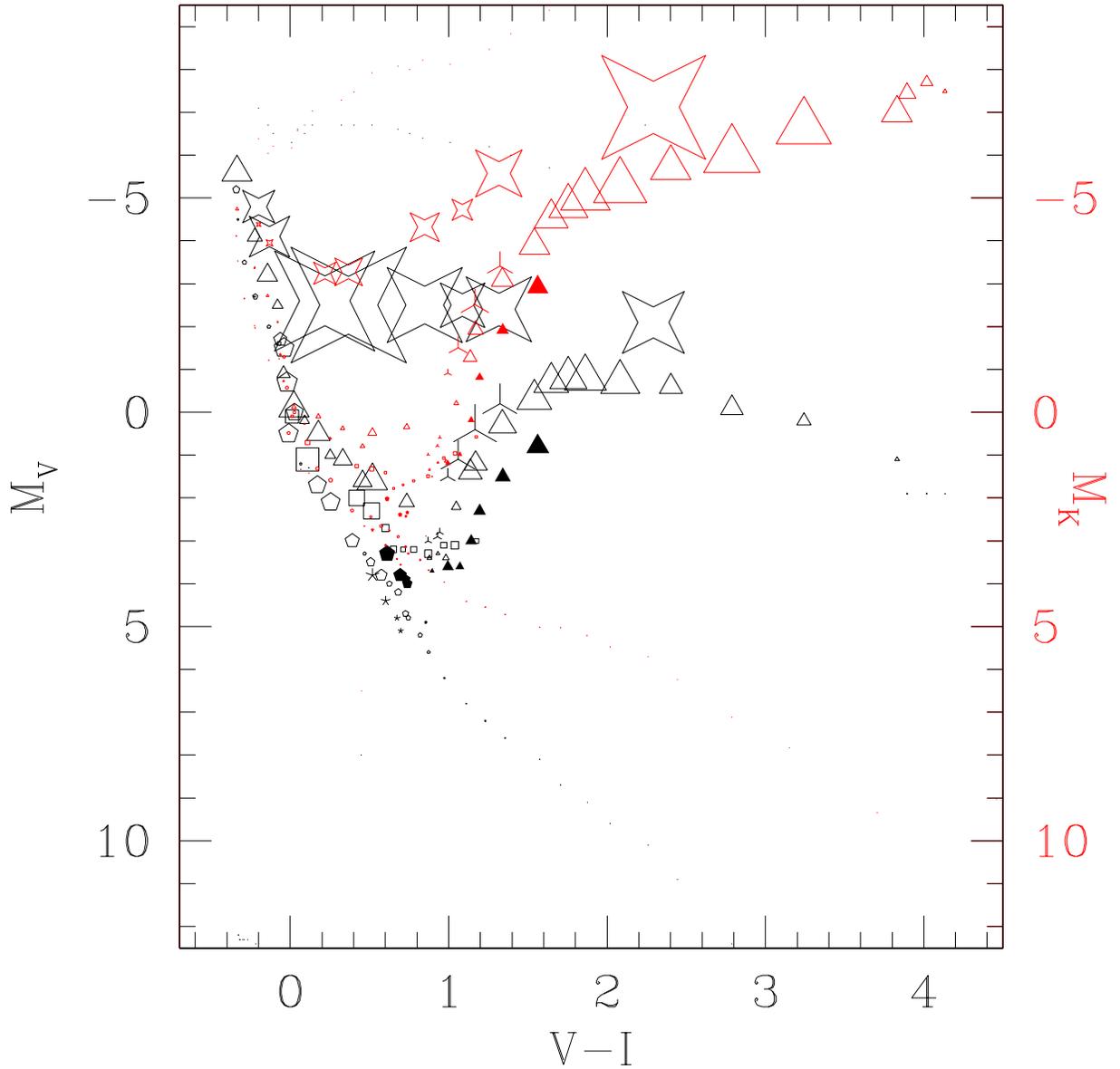}
\caption{HR diagram for $\sim$2.4\,M Tycho2 stars: The black symbols
show the 141 library types plotted as $M_V$ vs. $V-I$, where the area
of the symbols is proportional to the V-light emitted by the
stellar types. The grey symbols (red in the electronic version) occur on the
same vertical line, and are plotted as $M_K$ (on the right ordinate)
vs. $V-I$, and the area of the symbols is proportional to the K-light
emitted by the stellar types. Symbols as before.}
\label{HRty2-vk}
\end{figure}

Figure \ref{HRty2-vk} shows the same data plotted, where now the size of
the symbols represent either the V-light or the K-light of the stellar
types. The Tycho2 distribution is more dominated by solar-abundance
types and by early giant branch types, than is the SDSS-PT standard
population.

\subsection{Spectrally matched catalog of the SDSS survey region}

The catalog of 4.8\,M fitted Sloan survey stars with $g<16$ is
published in electronic table {\em DR7Fit}, which contains the SDSS
DR7 coordinates,
unprimed $ugriz$ (psf) magnitudes and errors for each star, matching 2MASS
$JHK_{2/S}$ magnitudes, errors and quality flag, followed by fitted
values of rank, $chi2$, Ruz, Rgz, R2M, number and type of fitted
spectrum, fitted $UBVRI-ZY$ magnitudes on the {\em vega} system, fitted
$u'g'r'i'z'$ magnitudes, fitted (unprimed) $ugriz$ magnitudes and
Distances in pc.  Other fitted magnitudes can be obtained by reference
to electronic table {\em LibMags}.

The catalog contains about 100 entries with 2MASS quality flag 'X', and
about 43,000 entries with quality flag 'U', modified for magnitude and
error as detailed above.

We found better fits to psf magnitudes and errors than to model
magnitudes and errors, so have spectrally matched the psf
magnitudes. The difference between psf and model magnitudes typically
exceeds what would be expected from the DR7 quoted errors.

\begin{figure}
\epsscale{1.0}
\plotone{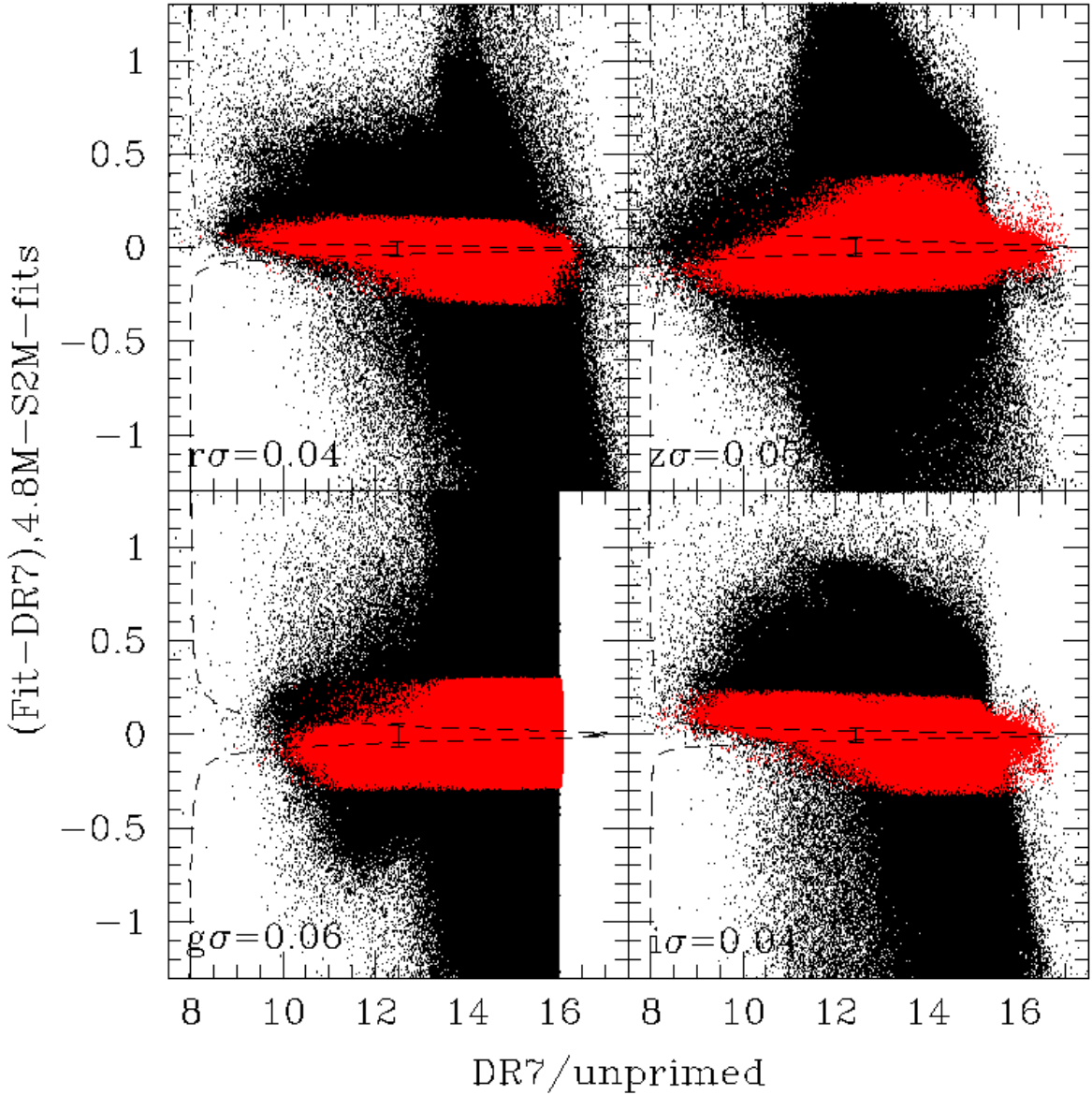}

\caption{S2M fits with eight $ugrizJHK_2$ bands for $\sim$4.8\,M SDSS
stars to $g<16$ with unprimed $ugriz$ and 2MASS $JHK_2$ magnitudes,
plotted as (Fit-DR7) values on the ordinate and DR7 values on the
abscissa.  Black dots show all the stars rejected by an Rgz \& $3\sigma$ clip
to give the fitted dispersions shown. The grey bands through the middle
(red in the electronic version) show about 70\% of stars left
after Rgz \& sigma clipping, The dashed histograms show the number
distributions of errors.  They are reasonably well peaked about the
zero delta line, but there are a surprisingly large number of
discrepant points at relatively bright magnitudes, and near the zero-delta line. }

\label{griz-dr7}
\end{figure}

Figure \ref{griz-dr7} shows the comparison of DR7 $griz$ magnitudes
with our fitted unprimed $griz$ magnitudes.  There are quite a large
number (about 30\%) of discrepant points in one or more DR7 bands. In
this case we have initially rejected points with large values of Rgz,
and then performed the sigma clip to reduce the numbers only slightly.
The sigmas after this process are 0.18, 0.06, 0.04, 0.04 and 0.05 for
$ugriz$ respectively, and by reference to section \ref{consistent},
are inferred to be similar for $UBVRI$.  These S2M fits, for about
70\% of the DR7 survey stars to $g<16$, are what are reported as the
typical $1\sigma$ error per DR7 star per band in the abstract.
 
There are quite a large number of (black) points rejected by the above
process at relatively bright magnitudes and near to the zero-delta
line in each panel.  This is because many of the $\sim$30\% points
rejected as having large Rgz values are seriously discrepant in one or
more color bands, but fit well in other colors.  It is possible, but
not verifiable here, that transparency variations during the drift
scans have affected different color bands for different stars.
Unfortunately the quoted psf errors are often quite low where the
magnitudes appear to be severely discrepant, so our fitting process
has trouble discriminating between accurate and doubtful data.

Note that the number of rejected points per histogram bin away from
the peak are only 1--3\% of those at the peak but, summed over
$\sim$20 such bins contain about 30\% of the (rejected) data
points. The histogram bin size was increased from 0.05 to 0.1\,mag
here to emphasize the outlying values.

The R2M values in {\em DR7Fit} are based on the rms differences between 2MASS
catalog and fitted values. Their sigma clipped values are 0.04, 0.05 \& 0.06 mag for $JHK_2$
respectively.

\begin{figure}
\epsscale{1.0}
\plotone{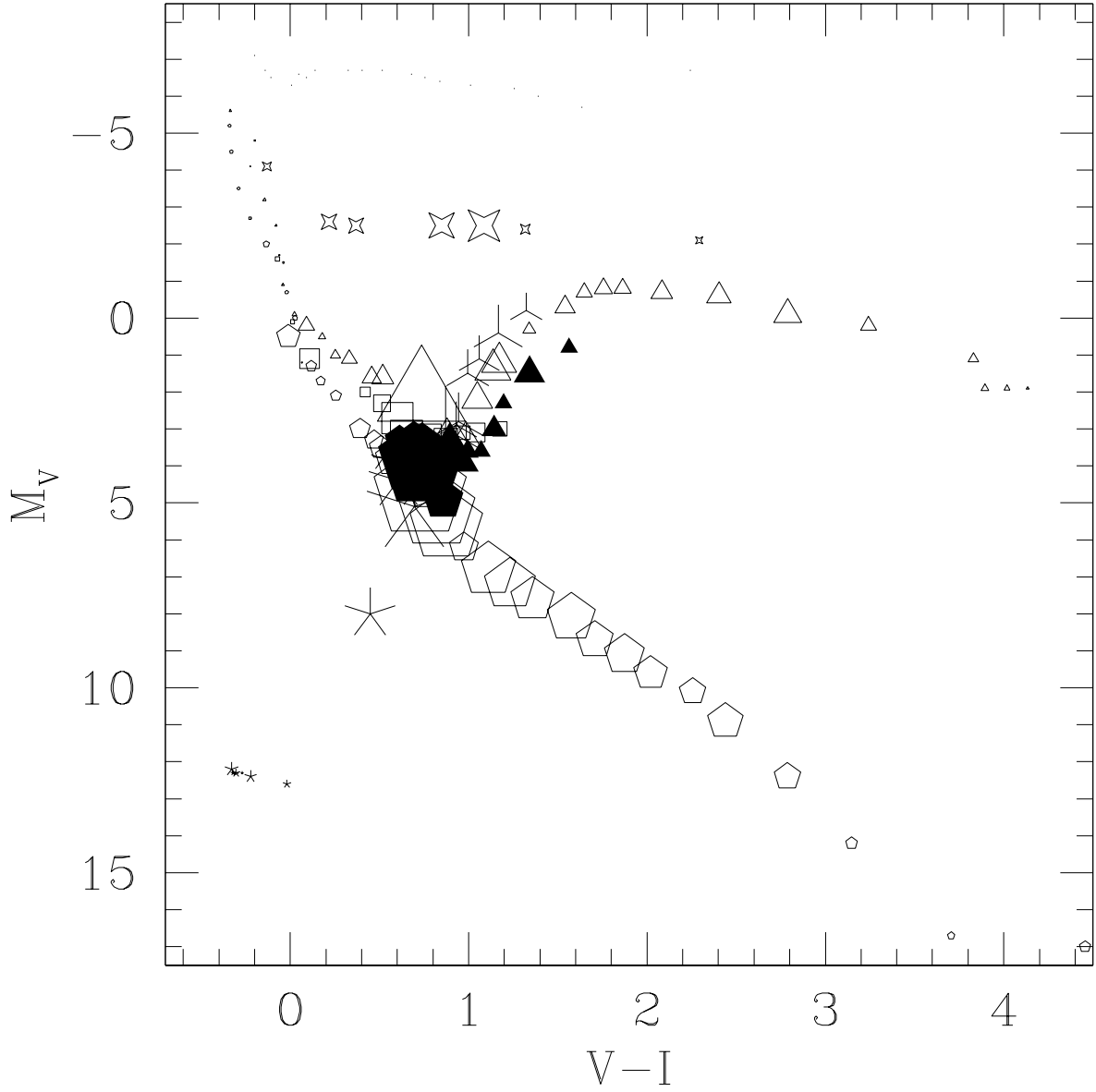}

\caption{HR diagram for $\sim$4.8\,M SDSS stars to $g<16$. Same
symbols as fig \ref{HRland}, with symbol area proportional to number
of occurrences. }

\label{HRdr7}
\end{figure}

Figure \ref{HRdr7} shows the HR diagram derived for 4.8\,M SDSS stars
to $g<16$ from these fits, with symbol area is proportional to number of
occurences of each type.

\begin{figure}
\epsscale{1.0}
\plotone{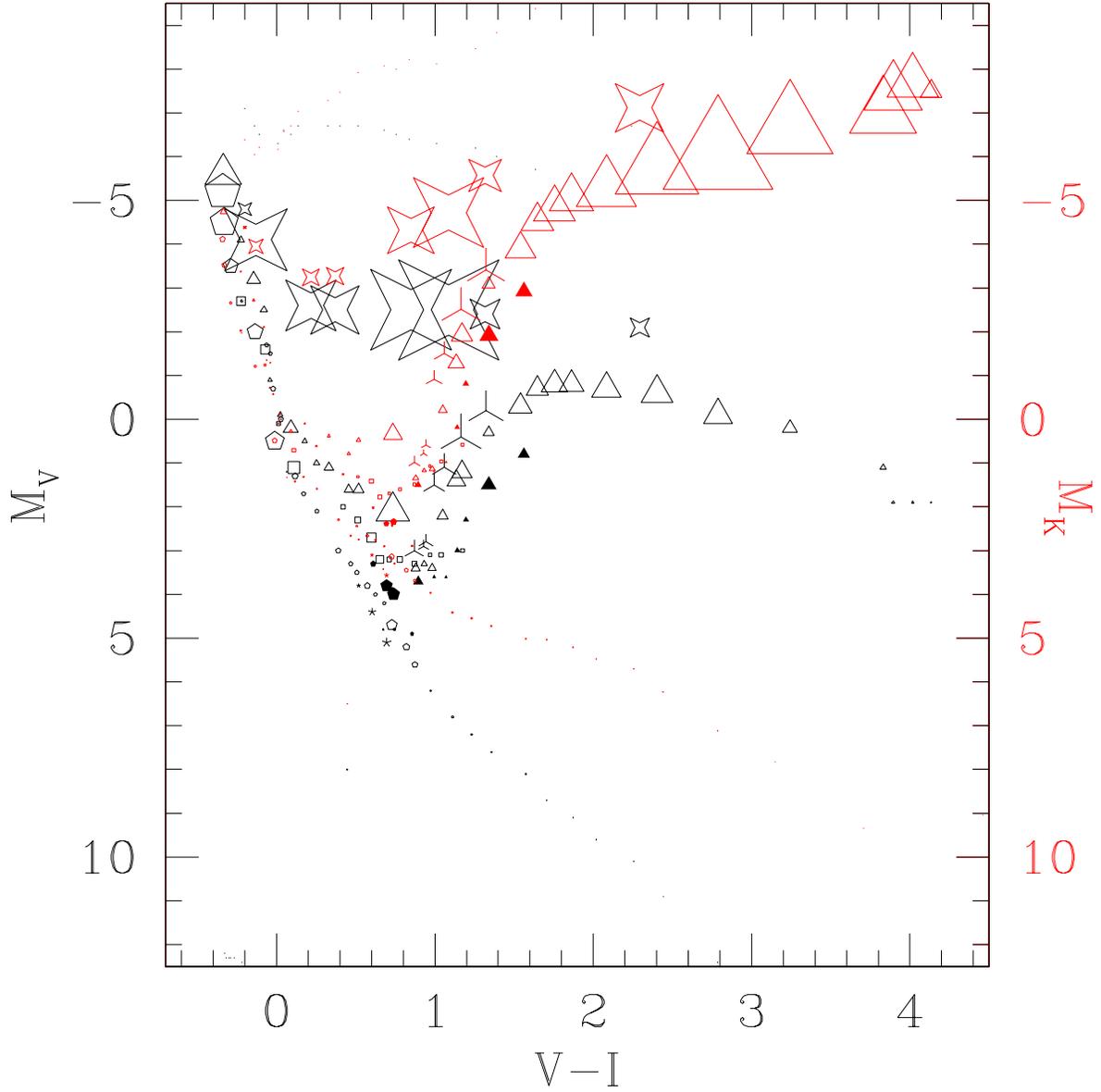}
\caption{HR diagram for $\sim$4.8\,M SDSS stars to $g<16$: Symbols as before, with black symbols
plotted as $M_V$ vs. $V-I$, grey (red) symbols plotted as $M_K$ vs. $V-I$, and symbol area
proportional to the V-light or K-light emitted by the stellar types. }
\label{HRdr7-vk}
\end{figure}

Figure \ref{HRdr7-vk} shows the same data plotted as $M_V$ and
$M_K$ vs. $V-I$, where the symbol area proportional to the V-light
and K-light respectively.

Our fitted magnitudes may improve over discrepant points in
individual filter bands in the input catalog values, since the
spectral matching process associates fitted values in each band with
additional, sometimes more accurate data.

\section{Summary}

We present online catalogs that provide synthetically calibrated \&
fitted magnitudes in several standard filter system bandpasses for
several surveys, including the all-sky Tycho2 survey, and the Sloan
survey region to $g<16$. The errors of the synthesized magnitudes are
mainly limited by the accuracy of the input data, but sufficient stars
should be available in fields of view $\geq$\,30-arcmin to average
enough stars to provide flux calibrations to better than 10\%, and
often to better than 5\%, in most bandpasses and most observing
conditions.  For instance with 9 Tycho2 stars in a typical 30 arcmin
field of view, flux calibration to about 0.1, 0.05, 0.04, 0.03 \&
0.03\,mag should be possible in either $UBVRI$ or $u'g'r'i'z'$
systems. These are typically achievable field-to-field uncertainties
on either standard system. Relative calibration of repeated images of the
same field, with similar or different equipment, should be automatically
possible to 2\% or better, even during variable photometric
conditions, but subject to the averaged systemic uncertainty (above).

The spectral matching technique provides a standard photometric system
check on survey magnitudes and, because of matching to multiple filter
bands, can be more accurate than individual measurements in some
filter bandpasses.  The accuracy of fitted magnitudes improves
substantially with improving survey data accuracy, as does the
reliability of fitted types, luminosity classes and distances.  The
method enables quick statistical analyses of stellar populations
sampled by wide-field surveys.

Substantial improvements can be expected in terms of reliable types,
multi-filter magnitudes and distances with realistic improvements in
the spectral library definition, eg. the NGSL library
\citet{gre06,hea10}, and expected all-sky survey improvements but,
importantly, only a few accurately determined survey bands (spanning
optical and infrared bands) are typically necessary to derive accurate
multi-filter fits.  Spectral matching removes the necessity to
accurately measure {\em all} desired filter bandpasses in all-sky
surveys.

\section{Acknowledgements}

We acknowledge the referee who suggested several useful improvements.
We would like to acknowledge the extensive standard system photometry
referenced here, and the producers and maintainers of the {\em
calspec} database at STScI.  This research has made extensive use of
the Tycho2, 2MASS, SDSS DR7 and NOMAD catalogs, and we gratefully
acknowledge their producers.  We have made use of the SIMBAD database
operated at CDS, Strasbourg, France, and of several data access tools
provided through the cdsclient package.

\clearpage



\clearpage

\clearpage

\clearpage

\clearpage

\clearpage

\clearpage

\clearpage

\clearpage

\clearpage

\clearpage

\clearpage

\clearpage

\clearpage

\clearpage

\clearpage

\clearpage

\clearpage

\clearpage

\clearpage

\clearpage

\clearpage

\clearpage

\clearpage

\clearpage

\clearpage

\clearpage

\clearpage

\clearpage

\clearpage

\clearpage

\topmargin +2cm
\oddsidemargin -2cm
\textwidth 200mm
\begin{deluxetable}{l|rrrrrrrrrrrrr}
\tabletypesize{\scriptsize}
\rotate
\tablecaption{Photometric Data for STIS\_NIC\_003 Calspec Standards  \label{tbl-stda}}
\tablewidth{0pt}
\tablehead{
\colhead{StdName}        &\colhead{G191B2B} &\colhead{GD\,153} &\colhead{GD\,71}  &\colhead{BD+17\,4705} &\colhead{AGK+81\,266}  &\colhead{GRW+70\,5824}  
&\colhead{LDS\,749B} &\colhead{F110}  &\colhead{HD209458} &\colhead{VB8}  &\colhead{P041C} &\colhead{P177D} &\colhead{P330E}  }
\startdata
Type	        &DA0    &DA0	&DA1   &sdF8	   &sdO	       &DA3	    &DBQ4    &D0p   &G0V      &M7V  &G0V   &G0V	  &G0V    \\
B-V		&-0.33	&-0.29	&-0.25 &0.44	   &-0.34      &-0.09       &-0.04   &-0.31 &0.59     &1.92 &0.62  &0.66  &0.64	  \\
V-I		&-0.33	&-0.3:	&-0.3: &0.62	   &-0.35      &-0.21	    &-0.2:   &-0.31 &0.7:     &4.6  &0.7:  &0.7:  &0.7:	  \\
\tableline
{\bf Tycho}	&       &       &      &           &           &            &        &      &         &     &      &      &       \\
$B_T$           &11.354 &       &      &9.969      &11.482     &            &        &11.440&8.334    &     &13.00:&      &       \\
$V_T$           &11.65: &       &      &9.505      &11.89:     &            &        &11.49:&7.699    &     &12.24:&      &       \\
\tableline
{\bf Landolt}   &LU     &HB     &L09   &LU          &LU         &LU          &LU      &LU    &         &B91&cal/lib&cal/lib&cal/lib \\
U               &10.250 &11.860 &11.675&9.724      &10.392     &11.807      &13.717  &10.360&         &      &     &      &       \\
B               &11.455 &13.060 &12.785&9.907      &11.596     &12.682      &14.634  &11.527&         &18.7: &12.62:&14.13:&13.64:\\
V               &11.781 &13.346 &13.033&9.464      &11.936     &12.773      &14.674  &11.832&         &16.78 &12.00:&13.47:&13.00:\\
R               &11.930 &13.484 &13.171&9.166      &12.090     &12.873      &14.675  &11.970&         &14.60 &11.65:&13.12:&12.65:\\
I               &12.108 &13.665 &13.337&8.846      &12.281     &12.979      &14.676  &12.145&         &12.31 &11.28:&12.75:&12.28:\\
\tableline
{\bf 2MASS}	&       &       &      &           &           &            &        &      &         &     &      &      &       \\
$J_2$           &12.543 &14.012 &13.728&8.435      &12.692     &13.248      &14.894  &12.548 &6.591   &9.776&10.864&12.245&11.781 \\
$H_2$           &12.669 &14.209 &13.901&8.108      &12.844     &13.357      &15.050  &12.663 &6.366   &9.201&10.592&11.932&11.453 \\
$K_{2/S}$       &12.764 &14.308 &14.115&8.075      &12.985     &13.451      &15.217  &12.796 &6.308   &8.816&10.526&11.861&11.432 \\
\tableline
{\bf UKIRT}	&       &       &      &           &           &            &        &      &         &     &      &      &       \\
$Z_V $          &       &13.642 &13.398&           &           &            &        &      &         &     &      &      &       \\
$Y_V$           &       &13.929 &13.657&           &           &            &        &      &         &     &      &      &       \\
$J_{MKO}$       &       &14.085 &13.710&           &           &            &        &      &         &9.86 &      &12.212&11.772 \\
$H_{MKO}$       &       &14.165 &13.805&           &           &            &        &      &         &9.22 &      &11.920&11.455 \\
$K_{MKO}$       &       &14.296 &13.899&           &           &            &        &      &         &8.85 &      &11.865&11.419 \\
\tableline
{\bf Stromgren} &       &LF/HM  &W     &           &           &HM          &W/HM    &LF/HM/ &         &     &      &      &       \\
$u_s$           &       &12.881 &12.686&           &           &12.858      &14.750  &11.400&         &     &      &      &       \\
$v_s$           &       &13.194 &      &           &           &12.890      &14.793  &11.650&         &     &      &      &       \\
$b_s$           &       &13.214 &12.897&           &           &12.700      &14.722  &11.720&         &     &      &      &       \\
$y_s$           &       &13.350 &13.120&           &           &12.800      &14.687  &11.860&         &     &      &      &       \\
\tableline
{\bf Sloan-air} &HB     &HB/PT  &HB    &std        &           &            &convrtd &      &         &convrtd&PT  &PT    &PT     \\
u'              &11.033 &12.700 &12.438 &10.560    &           &            &14.507  &      &         &19.287&       &15.112 &14.533 \\
g'              &11.470 &13.047 &12.752 & 9.640    &           &            &14.560  &      &         &17.607&12.260 &13.745 &13.280 \\
r'              &12.007 &13.567 &13.241 & 9.350    &           &            &14.802  &      &         &15.929&11.844 &13.300 &12.841 \\
i'              &12.388 &13.938 &13.612 & 9.250    &           &            &15.034  &      &         &16.106&11.719 &13.158 &12.701 \\
z'              &12.740 &14.287 &13.973 & 9.230    &           &            &15.245  &      &         &11.767&11.703 &13.125 &12.674 \\
\tableline
{\bf Sloan-vac} &convrtd&convrtd&convrtd&convrtd   &           &            &DR7     &      &         &DR7  &convrtd&convrtd&convrtd \\
u               &11.033 &12.700 &12.438 &10.560    &           &            &14.507  &      &         &19.287 &       &15.112 &14.533 \\
g               &11.444 &13.022 &12.729 & 9.664    &           &            &14.551  &      &         &17.714 &12.291 &13.778 &13.313 \\
r               &11.996 &13.557 &13.231 & 9.356    &           &            &14.797  &      &         &16.027 &11.851 &13.308 &12.849 \\
i               &12.364 &13.914 &13.588 & 9.245    &           &            &15.016  &      &         &16.095 &11.716 &13.155 &12.698 \\
z               &12.740 &14.287 &13.973 & 9.230    &           &            &15.245  &      &         &11.767 &11.703 &13.125 &12.674 \\
\tableline
\enddata
\tablecomments{References: B91: \citet{bess91}; HB: \citet{hb06}; HM: \citet{hau98}; LF: \citet{lac81}; L09: \citet{land09}; LU: \citet{lanu07}, W: \citet{weg83};
PT: SDSS-PT secondary standard; DR7: SDSS DR7 release; convrtd: Converted with equations in section \ref{sloan}. \\
Notes: BD\,+17\,4708 is an astrometric binary \citep{lu87}. }
\end{deluxetable}

\clearpage

\topmargin +3cm
\oddsidemargin 0cm
\textwidth 210mm
\begin{deluxetable}{l|rr|rr|rr|r}
\tabletypesize{\scriptsize}
\tablecaption{Photometric Data for NIC\_001, IUE\_004, Oke/Gunn and STIS\_001 Calspec Standards  \label{tbl-stdb}}
\tablewidth{0pt}
\tablehead{
\colhead{StdName}   &\colhead{KF08T3}  &\colhead{KF06T1}  &\colhead{G93-48} &\colhead{GD\,50}  &\colhead{G158-100} &\colhead{BD+26\,2606} &\colhead{F34} }
\startdata
Type	        &K0.5III   &K1.5III &DA3    &DA2   &dG/sdG   &sdF	 &DA \\
B-V		&0.95	   &1.07    &-0.01  &-0.28 &0.68     &0.39	 &-0.34 \\
V-I		&0.98	   &1.09    &-0.2:  &-0.19 &0.84     &0.62	 &-0.28 \\
\tableline
{\bf Tycho}	&          &        &       &      &         &           & \\
$B_T$           &          &        &12.24: &      &         &10.193     &10.834 \\
$V_T$           &          &        &13.20: &      &         &9.779      &11.102 \\
\tableline
{\bf Landolt}   &est     &est     &L09    &L09   &LU       &LU         &LU  \\
U               &        &        &11.942 &12.596&15.511   &9.910      &9.613 \\
B               &14.25:  &14.59:  &12.732 &13.787&15.572   &10.152     &10.838 \\
V               &13.30:  &13.52:  &12.743 &14.063&14.891   &9.714      &11.181 \\
R               &12.80:  &12.96:  &12.839 &14.210&14.467   &9.418      &11.319 \\
I               &12.32:  &12.44:  &12.938 &14.388&14.051   &9.109      &11.464 \\
{\bf 2MASS}	&        &        &       &      &         &           & \\
$J_2$           &11.585  &11.538  &13.203 &14.747&13.488   &8.676      &11.643 \\
$H_2$           &11.090  &10.987  &13.286 &14.863&13.117   &8.934      &11.563 \\
$K_{2/S}$       &10.987  &10.872  &13.397 &15.120&13.016   &8.352      &11.540 \\
\tableline
{\bf UKIRT}	&        &        &       &      &         &           & \\
$Z_V $          &        &        &12.937 &14.396 &13.808  &           & \\
$Y_V$           &        &        &13.143 &14.688 &13.738  &           & \\
$J_{MKO}$       &        &        &13.215 &14.802 &13.427  &           & \\
$H_{MKO}$       &        &        &13.255 &14.878 &13.059  &           & \\
$K_{MKO}$       &        &        &13.330 &14.990 &12.984  &           & \\
\tableline
{\bf Stromgren} &        &        &W/HM    &      &        &        &W/HM \\
$u_s$           &        &        &12.945  &      &        &        &10.585 \\
$v_s$           &        &        &12.930  &      &        &        &10.901 \\
$b_s$           &        &        &12.722  &      &        &        &11.026 \\
$y_s$           &        &        &12.785  &      &        &        &11.191 \\
\tableline
{\bf Sloan-air} &     &        &std     &convrtd&std    &std     &std \\
u'              &     &        &12.760  &13.372 &16.302 &10.761 &10.406 \\
g'              &     &        &12.653  &13.764 &15.201 & 9.891 &10.915 \\
r'              &     &        &12.961  &14.283 &14.691 & 9.604 &11.423 \\
i'              &     &        &13.268  &14.651 &14.469 & 9.503 &11.770 \\
z'              &     &        &13.529  &14.980 &14.377 & 9.486 &12.035 \\
\tableline
{\bf Sloan-vac} &     &        &convrtd &DR6 &convrtd&convrtd&convrtd \\
u               &     &        &12.760 &13.372 &16.302 &10.761 &10.406 \\
g               &     &        &12.641 &13.738 &15.238 & 9.914 &10.891 \\
r               &     &        &12.953 &14.273 &14.701 & 9.610 &11.414 \\
i               &     &        &13.247 &14.628 &14.469 & 9.499 &11.747 \\
z               &     &        &13.529 &14.980 &14.377 & 9.486 &12.035 \\
\tableline
\enddata
\tablecomments{References: HM: \citet{hau98}; L09: \citet{land09}; LU: \citet{lanu07}, W: \citet{weg83}; 
DR6: SDSS DR6 release; convrtd: Converted with equations in section \ref{sloan}. \\
Notes: BD\,+26\,2606 is a spectroscopic binary, possibly variable, but at a level not significant here \citep{lanu07} }
\end{deluxetable}

\clearpage

\topmargin 0cm
\oddsidemargin 0cm
\begin{table}
\begin{center}
\tabletypesize{\scriptsize}
\caption{ZeroPoints, Zero-Mag fluxes and Selected System/Filter bandpasses  \label{tbl-zp}}
\begin{tabular}{c|r|rrr|r|rc|cc}
\tableline\tableline
       &$\lambda_{pivot}$& \multicolumn{4}{c|}{Filter ZeroPoints} & \multicolumn{2}{c|}{Vega STIS\_005} & 0-Mag  &Selected \\ 
Filter &            (nm) & Nstd & Mean    & \multicolumn{2}{c|}{Sigma~~Adopted}           &  Mag & $F_{\nu}$(Jy) &  $F_{\nu}$(Jy)   &BandPass\\ 
\tableline
{\bf Tycho}  &      &      &         &          &                   &      &            &      & \\[-4pt]
$B_T$  &419.6 &  7   & -0.108  & 0.045    & -0.11             &0.046 & 3821       & 3985 & *\\[-4pt]
$V_T$  &530.6 &  5   & -0.030  & 0.020    & -0.03             &0.016 & 3689       & 3746 & *\\[-2pt]
\tableline
{\bf Landolt}  &      &      &         &          &                   &      &            &      & \\[-4pt]
$U_L$  &354.6 &  13   & 0.761  & 0.038    & +0.76             & 0.096 & 1609       & 1758 \\[-4pt]
$B_L$  &432.6 &  17   &-0.103  & 0.076    & -0.10             &-0.004 & 3979       & 3962 \\[-4pt]
$V_L$  &544.5 &  17   &-0.014  & 0.021    & -0.014            & 0.013 & 3646       & 3688 \\[-4pt]
$R_L$  &652.9 &  14   & 0.153  & 0.039    & +0.15             & 0.040 & 3079       & 3195 \\[-4pt]
$I_L$  &810.4 &  14   & 0.404  & 0.066    & +0.40             & 0.050 & 2407       & 2520 \\[-2pt]
\tableline
{\bf UBVRI}  &      &      &         &          &                   &      &            &      & \\[-4pt]
$U_M$  &358.9 &  13   & 0.763  & 0.027    & +0.76             & 0.005 & 1748       & 1755 & *\\[-4pt]
$U_3$  &364.6 &  13   & 0.770  & 0.050    & +0.77             &-0.050 & 1830       & 1748 \\[-4pt]
$B_M$  &437.2 &  17   &-0.116  & 0.020    & -0.12             & 0.012 & 4003       & 4048 & *\\[-4pt]
$B_3$  &440.2 &  17   &-0.124  & 0.024    & -0.12             & 0.012 & 4006       & 4050 \\[-4pt]
$V_M$  &547.9 &  17   &-0.014  & 0.010    & -0.014            & 0.019 & 3626       & 3690 \\[-4pt]
$V_C$  &549.3 &  17   &-0.014  & 0.008    & -0.014            & 0.021 & 3619       & 3691 & * \\[-4pt]
$R_C$  &652.7 &  14   & 0.165  & 0.014    & +0.17             & 0.023 & 3066       & 3131 & * \\[-4pt]
$R_p$  &658.7 &  13   & 0.192  & 0.030    & +0.19             & 0.025 & 2988       & 3058 \\[-4pt]
$I_C$  &789.1 &  14   & 0.386  & 0.016    & +0.39             & 0.031 & 2470       & 2542 & * \\[-2pt]
\tableline
{\bf 2MASS}   &      &      &         &          &                   &      &            &      & \\[-4pt]
$J_2$  &1239.0&  15   & 0.913  & 0.022    & +0.91             &-0.016 & 1601       & 1577 & * \\[-4pt]
$H_2$  &1649.5&  15   & 1.352  & 0.024    & +1.35             & 0.018 & 1034       & 1050 & * \\[-4pt]
$K_{2/S}$  &2163.8 &  15   & 1.830  & 0.018    & +1.83         & 0.008 & 670.3      & 674.9 & * \\[-2pt]
\tableline
{\bf UKIRT}   &      &      &         &          &                   &      &            &      & \\[-4pt]
$Z_V$  &877.6 &   5   & 0.583  & 0.038    & +0.58             &-0.069 & 2268       & 2128 & * \\[-4pt]
$Y_V$  &1020.8&   4   & 0.607  & 0.031    & +0.61             &-0.011 & 2092       & 2072 & * \\[-4pt]
$J_{MKO}$  &1248.8&  4   & 0.936  & 0.020    & +0.94             &-0.027 & 1570       & 1531 & * \\[-4pt]
$H_{MKO}$  &1673.0&  5   & 1.395  & 0.022    & +1.40             &-0.014 & 1019       & 1006 & * \\[-4pt]
$K_{MKO}$  &2200.0&  5   & 1.851  & 0.038    & +1.85             & 0.017 & 653.0      & 663.1 & * \\[-4pt]
$E_{bol}$  &1006.1 &  -   &  -  & -           & +1.40             &-0.194 & 1195       & 1000 &  \\[-2pt]
\tableline
{\bf Stromgren}   &      &      &         &          &                   &      &            &      & \\[-4pt]
$u_s$      &346.1 &   7   & -0.290 & 0.035    & -0.29             & 1.431 & 1267       & 4734 & * \\[-4pt]
$v_s$      &410.7 &   6   & -0.316 & 0.012    & -0.32             & 0.189 & 4094       & 4871 & * \\[-4pt]
$b_s$      &467.0 &   7   & -0.181 & 0.024    & -0.18             & 0.029 & 4175       & 4288 & * \\[-4pt]
$y_s$      &547.6 &   7   & -0.041 & 0.030    & -0.04             & 0.046 & 3612       & 3768 & * \\[-2pt]
\tableline
{\bf Sloan Air}   &      &      &         &          &                   &      &            &      & \\[-4pt]
u'        &355.2 &  12   & -0.033 & 0.021    & -0.03             & 0.974 & 1500       & 3680 & * \\[-4pt]
g'        &476.6 &  14   & -0.009 & 0.027    &  0.00             &-0.093 & 3970       & 3643 & * \\[-4pt]
r'        &622.6 &  14   &  0.004 & 0.020    &  0.00             & 0.148 & 3182       & 36y48 & * \\[-4pt]
i'        &759.8 &  13   &  0.008 & 0.023    &  0.00             & 0.372 & 2587       & 3644 & * \\[-4pt]
z'        &890.6 &  14   &  0.009 & 0.018    &  0.00             & 0.513 & 2263       & 3631 & * \\[-2pt]
\tableline
{\bf Sloan Vacuum}   &      &      &         &          &                   &      &            &      & \\[-4pt]
u         &355.7 &  12   & -0.034 & 0.019    & -0.03             & 0.961 & 1517       & 3676 & * \\[-4pt]
g         &470.3 &  14   & -0.002 & 0.036    &  0.00             &-0.101 & 3994       & 3640 & * \\[-4pt]
r         &617.6 &  14   &  0.003 & 0.019    &  0.00             & 0.142 & 3197       & 3645 & * \\[-4pt]
i         &749.0 &  13   &  0.011 & 0.022    &  0.00             & 0.356 & 2624       & 3641 & * \\[-4pt]
z         &889.2 &  14   &  0.007 & 0.019    &  0.00             & 0.513 & 2263       & 3631 & * \\[-2pt]
\tableline
\end{tabular}
\end{center}
\end{table}




\end{document}